\def\h{\eta}

\def\l{\lambda}
\def\m{\mu}
\def\n{\nu}

\def\L{\Lambda}

\def\ha{{1\over 2}}

\def\be{\begin{equation}}
\def\te{\end{equation}}
\def\bea{\begin{eqnarray}}
\def\nn{\nonumber}
\def\tea{\end{eqnarray}}

\newskip\humongous \humongous=0pt plus 1000pt minus 1000pt

\newif\ifdtup

\documentstyle[prd,aps]{revtex}

\makeatletter                    
\@addtoreset{equation}{section}  
\makeatother                     

\begin{document}

\title{Coarse-Grained Effective Action and  Renormalization Group Theory
in Semiclassical Gravity and Cosmology}

\author{E. A. Calzetta\footnote{Email: 
calzetta@df.uba.ar}}

\address{{\it
Departamento de F\'\i sica and IAFE, 
Facultad de Ciencias Exactas y Naturales\\ 
Universidad de Buenos Aires- Ciudad Universitaria, Pabell\' on I\\ 
1428 Buenos Aires, Argentina}}

\author{B. L. Hu\footnote{Email: hub@physics.umd.edu}}
\address{{\it  Department of Physics, University of Maryland\\
College Park, MD 20742, USA}}

\author{Francisco D.\ Mazzitelli 
\footnote{Email: fmazzi@df.uba.ar}}

\address{{\it
Departamento de F\'\i sica\\
 Facultad de Ciencias Exactas y Naturales\\ 
Universidad de Buenos Aires- Ciudad Universitaria, Pabell\' on I\\ 
1428 Buenos Aires, Argentina\\}}

\date{\today}
\maketitle

\begin{abstract}
In this report we introduce the basic techniques (of the closed-time-path
coarse-grained  effective action) and ideas (scaling, coarse-graining and 
backreaction)
behind the treatment of quantum processes in dynamical background
spacetimes and fields. We show how they are useful for the
construction of renormalization group (RG) theories for studying 
these nonequilibrium processes and discuss the underlying
issues. Examples are drawn from quantum field processes in 
an inflationary 
universe, 
semiclassical cosmology and stochastic gravity.
In Part I we begin by establishing a relation between scaling and inflation,
and show how eternal inflation (where the scale factor of the universe 
grows exponentially)  can be treated as static critical phenomena, 
while a `slow-roll' 
or power-law inflation can be treated as dynamical critical phenomena. 
In Part II  we introduce the key concepts in open systems
and discuss the relation of coarse-graining and backreaction. We recount 
how  the (in-out, or Schwinger-DeWitt) coarse-grained effective 
action (CGEA) devised by Hu and Zhang
can be used to treat  
some aspects of the effects of the environment on the system. 
This is illustrated by the { stochastic inflation} model where 
quantum fluctuations appearing as noise
backreact on the inflaton field. We show how  RG 
techniques can be usefully applied to obtain the running of coupling 
constants in the inflaton field, followed by a discussion of the 
cosmological and  theoretical 
implications. In Part III we present  the Closed-Time-Path (CTP, in-in,
or Schwinger-Keldysh)  CGEA 
introduced  by Hu and Sinha. We show how to calculate  perturbatively the  
CTP CGEA 
for the $\lambda \Phi^4$  model. We mention how it is useful for calculating 
the backreaction of environmental  fields on the system field 
(e.g. light on heavy, fast on slow) or one
sector of a field on another (e.g. high momentum modes on low, 
inhomogeneous modes on homogeneous), and problems in other areas of physics 
where this method can be usefully applied. 
This is followed by an introduction to the influence functional in the 
(Feynman-Vernon) formulation 
of quantum open systems, illustrated by the quantum Brownian motion models. 
We show 
its relation to the CTP CGEA, and indicate how to identify the 
noise and dissipation kernels therein.
We derive the master and Langevin  equations for interacting quantum fields,
represented in the works of  Lombardo and Mazzitelli
and indicate how they can be applied to the problem of 
coarse-graining,  decoherence 
and structure formation in de Sitter universe. 
We perform a nonperturbative 
evaluation of the CTP CGEA and show how to derive the renormalization 
group equations 
under an adiabatic approximation adopted for the modes  
by Dalvit and Mazzitelli. 
We assert that this approximation is incomplete as the effect of noise is 
suppressed. We then discuss why 
noise is expected in the RG equations for nonequilibrium processes.  
In Part IV, 
following  Lombardo and Mazzitelli, we use the RG equations to derive  
the Einstein-Langevin equation in  {stochastic semiclassical  gravity}.
As an example, we calculate the quantum correction to 
the Newtonian potential.  We end with a discussion on why a  
stochastic component
of RG equations is expected for nonequilibrium processes. 
\end{abstract}

\section{Aim and Scope}

We discuss  how the concepts of open systems and the techniques of 
the coarse-grained effective action and influence functionals can be applied to 
nonequilibrium quantum processes  in the early universe and in semiclassical
gravity,  leading to renormalization group (RG) theories for the description
of the interaction dynamics of these theories. 

Our wish  is that by examining a sample class of problems of 
fully dynamic nature --as different from equilibrium (finite temperature) or 
near-equilibrium (linear response theory) -- 
we can lay out the issues and 
approaches useful for the construction of a RG theory for 
nonequilibrium (NEq)  processes  involving quantum fields. 

Description of phase transitions involves a  scale 
that measures the behavior of the order parameter field in
the critical region. 
The energetics of the system is characterized by its quantum dynamical
and statistical mechanical properties. At the heart of NEq statistical
mechanics is the interplay of the dynamical scales
of the system (from the time-dependent order-parameter field) and
some background (e.g.,  all physical processes in an expanding universe
are measured against the time-dependent metric function such as the scale factor $a(t)$).
In equilibrium treatment, this is usually captured by a finite temperature effective potential. 
But  since the order parameter field is generally time dependent, and there
may not be a thermal equilibrium environment present
in these dynamical processes,  one should really be working  with an effective action
or a free energy density functional. 
Unlike scattering problems commonly found in particle physics where one can
determine the transition amplitude between the in and out states based on the 
in-out or Schwinger-DeWitt effective actions, in evolutionary problems frequently
encountered in statistical mechanics, the development of the expectation 
value (of an operator associated with some physical variable, such as the
energy momentum tensor) need be obtained from the in-in or Schwinger-Keldysh
effective actions. When open system concepts like coarse-graining and backreaction
of the system and environment are applied, the CTP CGEA or the  influence action
are more appropriate.  (For the development and application
of these ideas applied to problems in gravitation and cosmology, see, e.g.,
\cite{CH87,CH89,HuPhysica,cgea,jr,HPZBelgium,CH94,Banff}).

This paper is in the nature of a report rather than a review  -- in that 
we will present or develop only those works which are useful for the 
construction of the conceptual and technical frameworks to treat this broad 
class of problems,  taking specific examples from semiclassical gravity and inflationary cosmology as illustrations. 

RG concepts and techniques have been used in other areas of gravitation and cosmology.
We mention some representative works:

1) RG in gravitational collapse:  The use of universality and scaling ideas in classical 
gravitational collapse
first discussed by Choptuik \cite{chop93} has grown since
then into an interesting area of classical graviational research.
Fractal structure and scaling laws in a self-gravitating gas have 
been investigated
by de Vega, Sanchez and Combes \cite{fract}

2)  RG in quantum field theory in curved spacetime: notable work since 
the 80's by Calzetta,
Hu, O'Connor, Jack, Parker, Toms and others as well as the 
Tomsk group can be found in 
the book of Buchbinder, Odintsov and Shapiro \cite{buchetal92}. 
For more recent works  see  \cite {Kiretal93} and \cite {bon95}.

3) Scaling in quantum gravity has been pursued by Ambj\/orn in a 
simplicial gravity approach \cite{ambj} and by Antoniadis, Mazur and Mottola 
\cite{antomazmot},   to mention just a few notable avenues of inquiry.

The organization of this paper is as follows:
In Sec. 2  we first give a descriptive summary  of the nature of  the problems 
encountered in phase transitions in the early universe, focussing on the inflationary 
cosmology.
We distinguish the case of eternal inflation from that of slow roll and indicate
why they can be viewed as static and dynamic critical phenomena 
respectively. 
In Sec. 3 we begin a discussion of the relation of
scaling,  coarse-graining and backreaction with the example of 
stochastic inflation, thus bringing out the basic concepts
of open systems. In Sec. 4 we introduce the (`in-out') coarse-grained 
effective action (CGEA) for a $\lambda \phi^4$ field to incorporate the 
backreaction effect of the short wavelength modes viewed as the environment,
on the long wavelength modes of the system. In Sec. 5 we perform a
rescaling of the modes and the field in the spirit  of RG transformations
and derive the correspondng RG equations for the coupling constants of the system field.
We briefly discuss the theoretical and cosmological implications. 
In Sec. 6, we present  the `in-in' or closed-time-path  
(CTP) CGEA and show how it is
useful for the derivation of real and causal dynamical
equations for expectation values of operators of the system fields 
with backreaction from the environment field.
We carry out a perturbative evaluation of the CTP CGEA
and show how it could be useful for the consideration of backreaction
of environment field (modes) on the system field (modes).
In Sec. 7 we introduce the influence functional formulation via the quantum
Brownian motion model and show its relation  with the CTP CGEA. 
In this process we obtain the  master and semiclassical Langevin equations for interacting 
quantum fields. 
We show how the noise kernels can be identified  in this equation and how their
behavior can be used as a measure of  decoherence. 
In Sec. 8 we carry out a nonperturbative evaluation of the CTP CGEA, and  
derive the RG equations under different approximations.
We indicate how our real-time CTP Coarse-Grained Effective Action (CGEA) 
approach differs from  the Euclidean Averaged  Effective Action (EAEA) 
approach.  
In Sec. 9 we use  RG theory to derive the Einstein-Langevin equation in stochastic 
semiclassical 
gravity. 
In Sec. 10 we show how the RG equations change the Newtonian potential. 
In  Sec. 11, following a short summary, we discuss the salient features of RG theory 
for nonequilibrium processes and argue why there is a stochastic  component 
in the RG equations for such systems.

Those readers who only wish to learn the 
{\bf methodology} which underlies nonequilibrium  quantum field  processes, 
but have no
special interest in the specifics of such {\bf  processes} in gravitation and cosmology,  can do
just read Sections 4, 6, 8 and 11, from which they will be able 
to apply the methods 
to  their own areas of research  (e.g., nuclear/particle, 
atomic/optical physics).

\vskip 1cm 

{\Large{\bf PART ONE: Scaling and Inflation}}

\section{Phase Transitions  in the Early Universe }


We use phase transitions in the early
universe (from the Planck time to the GUT time, or even at the later electroweak era)
as  working examples to illustrate the basic issues and  methods 
involved. This is because phase transitions in the early universe
are mediated by and involve many NEq  processes of 
fundamental interest (e.g., nucleation, spinodal decomposition, 
particle creation, decoherence)
that lead to many important physical consequences
such as entropy generation and structure formation. 
We will only dwell on  those aspects which illustrate the
applications of  the RG theory and suggest its extension to dynamical
processes. The development of a RG theory for  NEq processes here -- from 
the construction of the CGEA to the 
derivation of the RG equations --- is based 
on the concepts (scaling, coarse-graining, backreaction) and techniques
(CTPEA and influence functional) of 
nonequilibrium mechanics and quantum field theory. 
These ideas and techniques are generally applicable also to problems
outside of gravitation and cosmology, such as the quark-gluon plasma 
and atoms interacting with a Bose-Einstein condensate. 
On the other hand this parallel presentation of
RG theory in the context of NEq processes may be helpful to
cosmologists who wish to find a more solid theoretical anchor 
for the discussion of phase transitions in the early universe.

\subsection{Effective Action for Dynamic Order-Parameter Fields}

As noted earlier knowledge of the exact form  of  the CTP CGEA
holds the key to a complete description of a phase transition.
One can deduce not only the qualitative features
(first or second order) but also the quantitative details
(mechanisms and processes). Therefore the construction of the effective action 
for the order parameter  field in different cosmological spacetimes is
usually the necessary first step towards a description of phase transitions in 
the very early universe.

For the purpose of illustration it is useful to establish the connection with
ideas of RG theory at the outset. The two themes established in \cite{cgea,jr}, which 
we follow here are:
1) Eternal  inflation can be described equivalently as an exponential scale transformation,
thus rendering this special class of dynamics as effectively static. 
2) The class of `slow-roll' inflation can  be treated as a dynamical 
perturbation off the effectively static class of exponential inflation and 
be understood as a dynamical critical phenomenon in cosmology. 
\footnote{A comment on the meaning of the
term `critical dynamics' as used in the context of cosmological
phase transitions is in order. By it we
refer to studies of phase transitions mediated by a time-dependent
order parameter field in contrast to static critical phenomena
where the order-parameter field is  constant in time. We are using this term
in a general sense, not necessarily referring to the specific conditions of
critical phenomena as discussed in condensed matter systems \cite{CriDyn}.
For example,
critical phenomena usually deals with the change of the order parameter
field near the critical point as a function of temperature.
In cosmology, temperature $T$ is a parameter usually (e.g. under the assumption
of adiabatic expansion) tied in with the scale factor $a$ and does not play
the same role as in critical phenomena. In the new
inflationary scenario, the critical temperature $T_{c}$ is defined as the
temperature at which a global ground state (the true vacuum) first appears.
The stage when vacuum energy begins to dominate and inflation starts is the
beginning of the phase transition.  The stage when the system begins to 
enter the true vacuum
and reheat can be regarded for practical purposes as the end of the phase transition.
Throughout the process of inflation the system is in a `critical' state.
The progression of a cosmological phase transition is measured not by
temperature, but by
the change of field configurations in time. Criticality corresponds to
the physical condition that the correlation length
$\xi  = m^{-1}_{eff} \rightarrow  \infty $, or
$ m^2_{eff} = {d^2V / {d {\phi}^{2}}}|_{ \phi=0}\rightarrow  0$
(which  may or may not be possible).
In critical dynamics studies of  condensed matter systems
one usually analyses the time-dependent Landau-Ginzburg equation, with a
noise term representing the effect of a thermal bath, and studies how the
system (order-parameter field) settles into equilibrium as it approaches the
critical point. We are not concerned with the corresponding cosmological
problem here.
An attempt to describe this aspect of the inflationary transition
was made in \cite{CorBru}. See also \cite{HPR}.}

Let us concentrate on situations where the order parameter field changes
either with space or time. (A familiar example in condensed matter physics
is anisotropic superconductivity where one can use a gradient expansion
in the Landau-Ginzburg-Wilson effective potential to account for the
differences coming from the next-to-nearest neighbor interactions.)
For cosmological problems,
it is the time-dependence of the background field which one needs to deal with.
\footnote{Strictly speaking, phase transition studies usually carried out
assuming a constant field in the de Sitter universe
are unrealistic, in that they only address the situation after the universe
has entered the inflationary stage and inflates indefinitely. This model cannot
be used to answer questions raised concerning the likelihood that
the universe will still inflate
if it had started from a more general, less symmetric initial state,  such
as the mixmaster universe \cite{Wald,BGH,SheHuOC}. Nor can one use this
model to study the actual process of phase transition (e.g., slow roll-over),
and the problem of exit (graceful or not) to the
`true' Friedmann phase. To do this, as is well-known, one
usually assumes that the potential is not exactly flat, but has a downward
slope which enables the inflaton field
to gradually (so as to give sufficient inflation)
settle into a global ground state. The cosmological
solution is, of course, no longer
a de Sitter universe.}
Thus for a realistic description
of many inflationary transitions one needs to treat the case of a 
dynamical field and a nonflat or even quasi-static 
potential. The form of the potential and the
metric of the background spacetime together determine
the behavior of the scalar field in the Laplace-Beltrami equation, but the
field in turn provides the source of the Einstein equation which determines
the behavior of the background spacetime metric. Hence they ought to be solved
self-consistently. (One usually considers  only the homogeneous mode
of the scalar field for the dynamics of inflation and the
inhomogeneous modes of quantum fluctuations for processes like structure
formation.)

At the classical level, the wave equation
for the background scalar field (assumed homogeneous)
with self-interaction potential $V(\phi )$ in a
spatially-flat Robertson-Walker (RW) spacetime is given by
\begin{equation}
\ddot \phi + 3H(t) \dot \phi + V'(\phi ) = 0
\label{1phieq}
\end{equation}
and the Einstein equations read
\begin{equation}
\dot H + 3 H^2 = 8 \pi G V (\phi), ~~~
\dot H = - 4 \pi G \dot \phi ^2
\end{equation}

\noindent where $H(t) \equiv { \dot a / a }$ is the Hubble rate,
a dot denotes derivative with respect to cosmic time $t$, and a prime
denotes a  derivative taken with respect to its argument.
A trivial but important solution to these equations is obtained by assuming
that
$V(\phi) = V_0 =constant$,
$ \phi =  \phi _0 =constant$ and $H=H_0=constant$,
which is  the de Sitter universe $a= e^{H_0t}$
with a constant field. A less trivial but useful solution is the so-called
`power-law' inflation models \cite{PowInf}, with an exponential potential
and a slowly-varying inflaton field. One can carry out a derivative expansion
of the background field to obtain a quasi-local effective action for the description
of such classes of spacetimes and fields.  (An example of  `slow-roll' transition.)

\subsection{Inflation as Scaling: Static Critical Phenomena}

This idea arose from the work of Hu and Zhang  \cite{cgea}
on coarse-graining and backreaction in stochastic inflation. 
In trying to compare the
inflationary universe with phase transitions in the Landau-Ginzburg model,
using a $\lambda \phi^4$ field as example, they realized that the exponential
expansion of the scale factor can be viewed as the system undergoing
a Kadanoff-Migdal scale transformation \cite{scaling}
(this is explained in detail in Sec. 3.B).  
In other words, time in this case plays the 
role of
a scaling parameter. It does not have to be viewed  as a dynamical parameter.
Thus for this special class of expansion, the dynamics of spacetime can be
replaced equivalently by a scaling transformation. In so doing one renders
eternal inflation into a static setting. By contrast, the larger class of
power-law expansion $a=t^\gamma$ does not possess this scaling property.
A useful parameter which marks the difference between these two classes of
dynamics is
$\zeta  = |\dot H|/ {H^2} = \ddot\alpha/\dot\alpha^2$,
where $\alpha \equiv ln a$, which can be regarded
as a `nonadiabaticity parameter' of dynamics:
the de Sitter exponential behavior with $\zeta= 0$ is `static',
the slow-roll with small $\zeta << 1$ is `adiabatic',
while the RW low-power-law with  $\zeta \approx 1 $ is `nonadiabatic'.

\subsection{Quasi-Static Field, `Slow-Roll' as Dynamical Critical Phenomena}

The effective potential $V(\phi)$ gives a well defined description of
phase transition only for a constant background (order-parameter) field.
If the order-parameter field is dynamic, the effective potential is ill-
defined and a host of problems will arise. Indeed,
the very meaning of phase transition can become questionable.
This is because as the field changes the effective action functional changes,
and the location of the minima changes also. The notion of symmetry breaking
and
restoration is meaningful only when there exist well-defined global and local
minima which do not change much in the time scale of the phase transition.
Changing background fields will also
engender particle creation, which affects
the nature and energetics of phase transition as well.
Therefore, in the context of phase transitions
involving dynamic fields, short of creating a new framework,
one can at best discuss the problem in a perturbative
sense, where the background field is nearly constant (quasi-static),
so that an effective quasi-potential can still be defined \cite{qEffPot,SinHu}.
An effective Lagrangian for a slowly-varying background field can be obtained
by carrying out a quasilocal expansion in derivatives of the field,
the leading term being the effective potential \cite{HuOC84}.
\begin{equation}
\cal L =
\cal L (   \phi, \partial_ \mu   \phi, \partial_\mu \partial_\nu  \phi, ...)
\end{equation}

One can
use this method to derive effective quasi-potentials for scalar fields in
flat space (for an example of its application to electroweak finite 
temperature
transition see, e.g., Moss et al \cite{MTW}) or (in conformal time)
for the conformally-flat Robertson-Walker spacetimes.
This is useful for studying cosmological phase transitions where the background
spacetime changes only gradually, as in the Friedmann (low-power law) solutions
$a= t^\gamma , \gamma <1$. (For a description see \cite{SinHu}.)
However, for the inflationary universe where the scale factor
undergoes rapid expansion following
either an exponential $a=e^{Ht}$ or a high power-law behavior,
the quasilocal expansion which assumes that the background field varies
slowly is usually inadequate.

Using the conceptual framework introduced above, one can understand why
the particular subclass of high-power-law expansion
associated with an exponential potential can hence be viewed as quasi-static.
It is in this context that one can introduce the quasi-`static'
approximation to derive the effective action for scalar fields to depict this
more realistic `slow-roll' inflation, now carried out as
a quasilocal perturbation from the de Sitter space, which is viewed as
effectively static.
For slowly-varying background fields one can use the method of derivative
expansion to derive the quasilocal effective Lagrangian. Usually this makes
sense only for static (or conformally-static, like the RW) spacetimes.
However, one can view the special class of exponential expansion as
effectively `static'. This can be understood with  the ideas of `dynamical
finite size effect' 
\cite{Edmonton}  and implemented by treating inflation as
`scaling' transformations \cite{cgea}. The `slow roll-over' type of phase transition
used in many inflationary models
can be viewed as a quasi-static case, and derived as a dynamic perturbation
from the de Sitter universe \cite{jr}. 
This view reveals a close analogy of this case with
dynamical critical phenomena where the scaling parameter $s$
plays the role of inflation and the time parameter $t$ measures how much
the system departs from the exponential expansion solution.
The main points proposed in \cite{jr} can be  summarized 
by the following schematic diagram:

\vskip .5cm
\noindent Constant Field in Static or ~~~~~SCALING~~~~~Exponential Expansion
$a=   e^{Ht}$\\
Conformally Static Spacetimes ----------------$>$   `Eternal Inflation' \\
(Finite Size Effect)~~~~~~~~~~~~~~~~~~~~~~~~~~~~~~~~~(Dynamical Finite Size Effect)   \\

$~~~~~~~~~~~~~~~~~~~~~~~~ |~~~~~~~~~~~~~~~~~~~~~~~~~~~~~~~~~~~~~~~~~~~~~~~~|~~~~~~~~~~~~$\\

{\it Quasilocal~~ Approximation~~~~~~~~~~~~~~ Derivative~~ Expansion}\\

$~~~~~~~~~~~~~~~~~~~~~~~~ |~~~~~~~~~~~~~~~~~~~~~~~~~~~~~~~~~~~~~~~~~~~~~~~~|~~~~~~~~~~~~$\\

$~~~~~~~~~~~~~~~~~~~~~~~~{\rm v}~~~~~~~~~~~~~~~~~~~~~~~~~~~~~~~~~~~~~~~~~~~~~~~{\rm v}~~~~~~~~~~~~$\\

\noindent Slowly Varying Field in~~~~~~~~~~~~~~~~~~~~~~~~High Power-Law
expansion   $a=t^\gamma$\\
RW Universe (Low Power-Law) ----------------$>$   `Slow Roll-Over'\\


The inadequacy of the finite temperature  effective potential 
for the description of phase transitions in the early universe was what 
motivated
us to look for more general methods useful for dynamic and nonequilibrium processes,
especially involving quantum fields.
We will trace out this pathway with two themes, one for the exposition  of
methods  to treat NEq processes, and the other for the development of 
RG theories for such processes. This will lead us from the effective action 
to the influence functional methods, and  for RG theory, from CTP CGEA
to RG equations.  

It is better to start with a physical example to  motivate, so 
we first discuss a  simple conceptual point which will enable us  to see 
inflation in the light of scaling. \footnote{Sec. 3, 4, 5 are excerpted from Lectures I and II of \cite{cgea}}.

\section{Coarse-Graining, Scaling and Inflation}
\subsection  { Inflation}

Consider a massive (m), self-coupled $(\lambda )$ scalar field 
$\phi (\vec{x},t)$ 
in a background spacetime with action 
\begin{equation}
S[\phi] = \int d^4x \sqrt{-g} 
\left [- \frac{1}{2} \phi \Box \phi - \frac{1}{2} (m^2+\xi R)\phi^2 -
\frac{1}{4!}\lambda\phi^4\right]  
\label{3phiact}
\end{equation}                                          
where
$ \Box = \frac{1}{\sqrt{-g}} \frac{\partial}{\partial x^{\mu}}
g^{\mu \nu} \sqrt{-g} \frac{\partial}{\partial x^{\nu}}$  
is the Laplace-Beltrami operator and $\xi  = 0, \frac{1}{6}$  
correspond respectively to minimal and conformal coupling of the field
with the background spacetime with scalar curvature $R$. We will consider 
Robertson-Walker (RW) and de Sitter (DS) spacetimes with line element
\be
ds^{2} = dt^{2} - a^{2} d\Omega ^{2}
\label{3rw}
\te
where $d\Omega ^{2}$ is that of the 3-space
$(d\Omega ^{2} = d\vec{x}^{2}$ for spatially flat).  We
restrict our attention here only to the spatially homogeneous and isotropic
RW, or the spacetime- homogeneous and isotropic DS cases, where a single
isotropic scaling parameter is applicable. For the RW universe
     $$a(t)=t^\gamma,$$
where $\gamma=1/2$ for a radiation-dominated source
and $\gamma=2/3$ for a pressureless dust source.
For a de Sitter universe (in the spatially-flat RW coordinatization)
\be
a(t)= exp(Ht).
\label{3.3}
\te
Spatially homogeneous but anisotropic spacetimes are more complicated as
they require different scaling parameters in different directions.
 
In a spatially-homogeneous spacetime, the scalar wave function
$\Phi (\vec{x},t)$
\be
\Phi (\vec{x},t) = \sum  \phi _{k}(t) u_{k}(\vec{x})
\te
is the product of a function of time  $\phi _{k}(t)$ and 
a function of space
$u_{k}(x) [ = e^{i\vec{k}\cdot \vec{x}}$ for spatially-flat, and
$ = Y_{lmn}(\chi ,\theta ,\phi ), k=(l,m,n)$ for spatially-closed cases.]
The wave equation for the time-dependent amplitude function $\phi _{k}(t)$
associated with the kth mode of a scalar field $\Phi (\vec{x},t)$ with
self-interaction potential $V(\Phi )$
in a RW spatially-flat spacetime is given by
\be
{d^{2}\phi _{k}\over dt^{2}} + 3H(t) {d\phi _{k}\over dt} 
+ ({k^2\over a^2} + m^2 +\xi R)\phi_k
+ V_k'(\phi ) 
= 0
\label{3modes}
\te
where $H(t) \equiv { \dot a / a } = \dot\alpha $ , or
$\alpha  = \ln a = \int  Hdt $.  (Here a dot denotes derivative with
respect to cosmic time $t$).

The solution to \ref{3modes} 
depends on $V'(\phi)=dV/d\phi$.  For inflation we are
interested in cases where i) $ V'(\phi ) << 3H \dot\phi $ .  This would 
correspond to considering only the relatively flat portion of the potential, 
where inflation takes place for an extended period of time (usually assumed 
$\Delta \alpha = \int^{t_{1}}_{t{0}} Hdt > 68~ e$-folding time to engender
sufficient entropy corresponding to the  observed universe).  In new 
inflation, 
this corresponds to assuming a flat plateau; and in chaotic inflation, 
a gradual downward slope.  In the realistic de Sitter case, the order 
parameter 
field $\phi $ changes very slowly in this inflation regime (slow roll-over), 
its rate controlled by $3H d\phi /dt$, which can be regarded as a viscous term 
in the dynamics.  One can safely  also assume that
ii) $ \ddot\phi << 3H\dot\phi $
in this regime. (By contrast the $ \ddot\phi $  term dominates the dynamics in
the reheating regime).  A further simplification is to assume that $\phi $ = 
constant in this regime throughout, corresponding physically
to eternal inflation, i.e., $a = e^{Ht}$ for all time.
This idealized case is of course
unphysical in that the universe will never roll-over (in new and chaotic
inflation),
or tunnel (in old inflation) to the true vacuum.  Nevertheless it captures
the salient features of inflation.  We shall consider this idealized case in
detail here in the light of static scaling.  In summary one can distinguish
the following cases:\\
a)Minkowski field theory (static) \ \ \ \ \ \ \ \ \  H  = 0 \\
b)Eternal inflation (stationary)\ \ \ \ \ \ \ \ \ \  H  = constant,
  $ \dot\phi= 0$ \\
c)Slow-roll inflation (quasi-stationary)\ \  H  $\simeq$ 
constant,$ \dot\phi\neq 0,
    \ddot\phi = 0 $\\
d)Realistic inflation (dynamic)\ \ \ \ \ \ \ \ \ \ \ H $\simeq$ constant,
  $ \dot\phi\neq 0,      \ddot\phi \neq 0 $\\
In both cases a) and b), the term $3H\dot\phi$ in \ref{3modes} is ignored,
while in cases b) and c) the term $ \ddot\phi $  is ignored.
 
We now discuss the effect of (eternal) inflation on the scalar field.
We will show that {\sl a scalar field with $\lambda \phi ^{4} $
 self-interaction in an isotropically expanding spacetime undergoing
 eternal inflation behaves exactly like the same field in a static
spacetime undergoing a  scaling transformation}.  The physics of inflation
(we refer specifically to the initial stage only, but not the subsequent slow
roll-over nor the reheating stages) can thus be understood in terms of
scaling completely.
 
\subsection   { Scaling}
 
To fix ideas, consider the spatially-flat RW metric Eq.\ref{3rw} with $a$ =
constant. This is just the Minkowski spacetime.  Let us consider an ordered
sequence of such static hyperspaces (foliation) with scales $a_{0}, a_{1},
a_{2}$, 
etc, parameterized by $t_{n} = t_{0} + n \Delta t, n = 0, 1, 2, \ldots $ .  
These spacetimes have the same geometry and topology but differ only in the 
physical {\sl scale} in space.  One can always redefine the physical scale 
length $x'_{(n)} = a_{n} x$ to render them equivalent.  If each copy has scale 
length magnified by a fixed factor $e^{H\Delta t}$ over the previous one in 
the 
sequence, i.e. ${a_{n+1}/a_{n}}= e^{H\Delta t}$, we get exactly the
physical 
picture as in an eternal inflation.  After $n$-iterations i.e. 
$a_{n}/a_{0} = e^{n(H\Delta t)}$, or, in terms of a continuous parameter
$a(t) = a_{0}e^{Ht}$.  It is
important to recognize here that $t$ can be any real parameter not necessarily
describing the dynamics.  Time in this depiction plays the role of a
scaling parameter, and dynamics is nothing other than scaling.
 
To see this simple point in another light, let us describe a totally
different physical situation (which we will see at the end is exactly
identical) where scaling plays the dominant role but one would
otherwise not be inclined to associate with the physics of inflationary
cosmology.  This is the Kadanoff-Migdal transformation 
\cite{scaling,CriPhe}
 widely applied to condensed matter systems for the study of critical
 behavior.  Consider a 3-dimensional cubic lattice with lattice length $L$,
and an order parameter field $\phi $ describing the magnetization of the
system. The Ginsburg-Landau-Wilson Hamiltonian for a Heisenberg magnet
bears similarity with the $\lambda \phi ^{4}$ theory in a well-known manner
 \cite{CriPheQFT}.
 
The Kadanoff-Migdal transform is an artificial procedure for relating
the microscopic and macroscopic properties of a system based on the
existence of scaling properties in the system in the infrared limit.  It
involves taking a certain number of fixed spin sites (e.g. 4 in 2 dimensions,
8 in 3 dimensions)  and replacing them by one block spin with
adjusted (e.g. doubled bond strength) interaction strength between nearest
neighbors.  Carrying out this transformation $n$ times leads to a
coarse-grained
system.  If the system (ferromagnet) manifests scaling properties near
the critical point, as had been observed before the renormalization group
theory was invented for its description, then the resulting rescaled
macroscopic system should preserve the same properties as the original
microscopic system.  The appearance of long range order near the critical
point makes this procedure a viable one, permitting enormous simplification
while preserving the salient features of the system.
 
Let us examine a two-dimensional example of this process, sometimes
also called decimation.  Denote by $a$ the original length and $a_{n}$
the scale length after carrying out the $K-M$ transformation $n$ times.
If 4-spins are combined into 1 block spin in 2 dimensions, $a_{1} = 2a_{0}$,
and $a_{n}/a_{0} = 2^{n}$.  This gives the same result as inflation, 
except that the scaling factor $s \equiv  {a_{n+1}/ a_{n}}$ here is 2, 
instead of $s = e^{(H\Delta t)}$.  Defining
\be
s \equiv  {a_{n+1}\over a_{n}} \equiv  e^{\sigma } ,
\te
$\sigma $ for inflation is $H\Delta t$ , a constant, while for decimation
in the above example is $\ln 2.$
We shall see that it is precisely this scaling property in inflation which
imparts all the distinct and desirable physical features, from the
scale-invariant Zel'dovich-Harrison perturbation spectrum to the Hawking
effect.
 
While the above real-space scaling explains well the physical idea, the
dual transformation in
momentum space is easier to calculate.  This is the original Wilson-Fisher
RG transformation \cite{CriPhe} \cite{CriPheQFT}.
In momentum space, the lattice
spacing acts like an
ultraviolet cut off $\Lambda $ of wave numbers $k$ ($\Lambda  = \infty $ for
 continuous systems).  The block spin transformation corresponds to a) 
eliminating the higher wave number modes (e.g. $k > {\Lambda / s})$ and 
b) rescaling (e.g., $k \rightarrow  k'= ks)$.  The real
parameter $s > 0$ acts in a) like a coarse-graining (cut parameter) and in b)
like a rescaling parameter.  In our example above, $ s =2.$ In every iteration,
wavemodes with $k > {\Lambda /2}$ are integrated out first, then the
remaining modes from $0 < k < {\Lambda / 2}$
are rescaled such that the new $k'$  space covers the full range with the same
$UV$ cutoff.  For definiteness, we will refer to these two steps as
{\sl coarse-graining} (separation) and {\sl rescaling}.  Together these two 
steps
constitute the renormalization group transformation.  
 
Thus in this example, at any iteration, wave numbers ${\Lambda / s} < k <
\Lambda$ are being integrated away, and only the lower wave mode sector
is kept. For the long wavelength, low $k$ cutoff, we assume the horizon
size $H^{-1}$, because those with
lower $k$ having wavelengths greater than the horizon size are of no
physical significance to the system, at least during the inflationary stage.
(Some of these wavelength will reenter the horizon in the reheating phase
and influence later events.) In the language of subdynamics one
can call the former the environment (irrelevant sector), and the latter the
system (relevant sector).  The division of these two sectors changes with
the order of iteration (in the language of Kadanoff-Migdal transformation) or
time-evolution (for inflationary cosmology) of the system.  The coupling of
the environment with the system is determined by $V(\phi )$.  In addition, 
for a
self-interacting field there are low-low and also high-high mode couplings.
A useful way to keep track of these interactions, in particular, is by way
of the so-called coarse-grained effective action which we will discuss in
the next section.

\subsection { Coarse-graining and Stochastic Inflation}

In the above discussion we have shown how one can transcribe the
dynamical process of {\sl inflation} to the static transformation of {\sl 
scaling.}  
We also showed how the evolution of the universe during inflation can be viewed
as a succession of {\sl coarse-grainings}, with the higher wave number modes
cast
away and their effect on the remaining (low $k$ )modes accounted for by a
coarse-grained effective action.  The resulting RG 
equations describe the flow of the coupling constants.
Using this equivalence, we can conceptually (and technically) replace the
{\sl inflationary expansion} process in time by a running of the field towards
 the infrared regime.  

Let us compare the coarse-graining scheme used in the critical phenomena example above 
(CP, or Case A) with that of an important class of inflationary cosmologies called 
stochastic inflation (SI, or Case B), because there are close similarities and basic
differences.  Starobinsky's scheme \cite{StoInf} is to divide a free
scalar field  into two parts:  the system  contains
long-wavelength modes outside the horizon 
$(p = {k / a}< \epsilon H, \ell  = p^{-1}$ being the
physical wavelength and $\epsilon  < 1 $ a small parameter), which are treated
classically; and the bath contains quantum fluctuations of wavelengths $\ell$
shorter than the horizon.  Thus the bath in stochastic inflation is obtained
from the full spectrum by a window function $\theta (k - \epsilon aH)$, 
and changes at each
moment.  Many authors have claimed that these high $k$ modes with a
dynamic cutoff
which obey the wave equation (\ref{3modes})  (with first order or with both first and
second order time derivative terms) give rise to a white noise.  They
then proceed to set up and solve a Langevin equation for the classical field
$\phi $
driven by the de Sitter dynamics and with this white noise source.  Thus the
name stochastic inflation. 
\footnote{Authors of \cite{cgea,HPZBelgium,CH95}
question how a white noise can be deduced by this dynamic partitioning of a
{\sl free} field. They suggested instead how colored noise can be generated
from coarse-graining an interacting field without assuming any dynamic
partitioning  
(see Section VII B).
This provides a more general scheme for stochastic inflation,
where one can discuss issues like non-Gaussian galaxy distribution from
vacuum fluctuations.}
 
The physical setup of Starobinsky's stochastic inflation differs from the
example we used above to illustrate the relation of scaling
and inflation in how the system is defined. Technically
there exists a simple transformation between coarse-graining
in critical phenomena (CP, Case A) and that in stochastic inflation
 (SI, Case B).  Let us explain:
 
Compare the composition of the system (S) and the environment (E) at
the beginning and end of inflation for critical phenomena (Case A) and for
stochastic inflation (Case B).   For Case B, at the start $t_{0}$,
assume $a(t_{0}) = 1$ and $\epsilon  = 1$ for simplicity.  At any later time 
the system consists of
all modes $k <  Ha$ or $\ell  > H^{-1}$.  As the universe inflates
$a = e^{Ht}$, more $k$
modes are being red-shifted out of the horizon into the system.  Thus $S$ of
stochastic inflation increases (from the $k$ region $H_{0}K$ at $t_{0}$ to 
$HK$ at $t^{H}_{*}$ to $GK$ at $t^{G}_{*}$ , see Fig. 1).
The reason why in the stochastic inflationary program
the system is taken to consist of modes with wavelengths {\sl greater} than the
horizon $H^{-1}$ in the inflation regime is because these are the
modes which would reenter the RW horizon in the radiation-dominated RW
phase; and
only those higher  $(k > k_{G})$ modes (here treated as environment ) which
reenter the horizon {\sl before} galaxy formation  can physically influence
$k_{G}$ in
the RW phase.   So the system in stochastic inflation is really designed for
and chosen by the physics in the later radiation-dominated $RW$ regime, rather
than during the inflationary regime.  Note that modes with wavelengths greater
than the horizon size at the onset of inflation have not yet reentered today's
horizon, because in, say, the new inflationary scenario, today's observable
universe is believed to be only a small portion $(l_{H})$ of a much larger
bubble inflated from a fluctuation domain $(l_D) $ at $t_{0}$. The mode
$ k_{H}$ exits the horizon at
time $t^{H}_{*}$.  The wave mode $k_{H}$ of physical wavelength $\ell _{H} 
($today's Hubble radius)
which is just entering the RW horizon today [and which disappeared outside
the de Sitter horizon earlier than any observable physical scale $(HR_{H}
H_{\hbox{now}}$ 
in Fig. 1)] forms the upper limit of $\ell _{\hbox{phys}}$.
Of course more wave modes
with $\ell  > \ell _{H}$ will be entering into the RW horizon tomorrow.
These modes have not yet
come into contact with our universe so they should be considered as
unphysical by today's observers.  However, they had interacted previously
in the de Sitter phase at times $t < t^{H}_{*}$ with all other k modes
$< k_{H_{0}}$ within the de Sitter
horizon.  In the inflationary phase, it is this interaction which one should
address.
 
By contrast, in critical phenomema (Case A) depicted earlier, both the system and the bath
are contained in the horizon throughout the period of inflation from $t_{0}$
to $t_{1}$ (the OLH$_{1}$ portion in Fig. 1). $ H^{-1}$ is the lower (IR)
cutoff. The partition is defined by the incremental fraction
 $s = e^{H\Delta t} = a(t+\Delta t)/a(t)$ of $k$ modes
slipping into the bath at each moment $\Delta t ($or each iteration) by
coarse-graining.  For example, one possible partition $k_{L} ($point $L$ in
Fig.1) is determined by the last wave mode which left the horizon at the end
of inflation at $t_{1}$.  All modes with $k > k_{L}$ never made it out and
always stayed inside of the DS horizon.
These will be the common "bath" to all systems at all times in the DS phase.
Another choice more geared towards the galaxy formation problem
is defined by the $k$ modes of interest (e.g. galaxy scale $k_{G})$
and made up of $k < k_{G}$.  The environment then contains modes $k > k_{G}$,
including the common portion
$k > k_{L}$.  In this designati
on we make sure that:
1)   The system consists of all scales within today's Hubble radius 
(RW horizon).
2)   Their interactions in the inflationary phase with the higher and
lower modes are incorporated.
3)   The environment is always within the de Sitter horizon.
We are interested in two effects:  1)  The effect of the higher $k$ modes
field $\phi_>$ on the system field made up of lower k modes
$\phi_<$, and 2)  How the
field parameters (such as mass $m$ and coupling constants $\lambda $)
change in this process of inflation (or scaling).
These are answered respectively by 1) deriving the effective action and the
associated effective equations for $ \phi_< $ , and 2) deriving the 
renormalization group equation for these field parameters.
 
Despite the difference in the aims of the two programs the
coarse-graining technique used in both cases are related in a simple
manner.  By
replacing $s \rightarrow  s^{-1}$, or equivalently $(\Delta t) \rightarrow  
-(\Delta t)$ in Case A, one gets the
procedure in Case $B$ for stochastic inflation.  This amounts to changing
inflation to deflation (time-reversal), or interchanging system and bath.
One can also view the difference as representing a passive versus an active
view.  By this we mean inflating the background (spacetime) has, for a local
observer in a scale-invariant system, the same effect as
coarse-graining the system in a fixed background.
 
A figurative description which encapsulates the main ideas in this
comparison is the effect of a zoom lens.  The physics of a scale-independent
system under inflation is equivalent to holding the system static and
viewing it through a zoom lens.  Here zooming (scaling) replaces dynamics
(inflation).  Features of an evolving system undergoing inflation (the
passive interpretation) can be depicted equally by the magnifying  view
through a zoom lens (scaling) of a fixed system (the active interpretation).
Obviously this relation holds only if the structure is homogeneous and the
lens has uniform magnification. (This corresponds to homogeneity in space
and uniform scaling.)

In the next section we present an effective
action formalism for carrying out the coarse-graining.  We shall keep the
formalism general and non-specific to the system-environment stipulation
so that it can be applied to different situations in a variety of problems.

\vskip 1cm 
{\Large {\bf PART TWO: Coarse-Graining and Renormalization Group}}

\section{ Coarse-Grained Effective Action }

\subsection{Coarse Graining and Backreaction}

In quantum field theory, when the background field or spacetime 
dynamics follows some  simple discernable (e.g., classical) behavior one commonly performs 
what is called  a 
background field decomposition, $$ \phi = \phi_c + \phi_q $$ on a   (e.g., self-interacting)
scalar field $\phi$. One then calculates the effect of the fluctuation 
field $\phi_q$ on the
background field $\phi_c$ (as well as the background geometry 
$g_{\m\n}$ when necessary)
using a loop expansion in 
the in-out effective action.
For many problems such as those
encountered in statistical mechanics where one is
interested in the detailed behavior of only a part of the overall system
(call it the system) interacting with its surrounding 
(call it the environment),
one can decompose the field describing the overall system $\phi$ according to
$$   \phi= \phi_S +\phi_E  $$
where $\phi_S$ denotes the system field and $\phi_E$ the environment field.
One can calculate the backreaction of the latter on the former with the
`coarse-grained effective action' (CGEA) first introduced by Hu and Zhang \cite{cgea}, 
where one does not integrate over  the quantum fluctuations, 
as in the conventional definition (Schwinger-DeWitt) of
effective action in orders of $\hbar$,
but rather with respect to a small parameter $\epsilon$ which
can generally be expressed as the ratio of two time, length, 
mass or momentum scales
characteristic of the system and environment discrepancy (examples are
slow-fast time scale, long-short length scale, light-heavy mass scale or 
soft-hard momentum scale).  In the {\bf stochastic  inflation} scheme  
\cite{StoInf}, one regards the system field as containing only 
the lower modes and the environmental field as containing the higher modes 
with the 
division provided  by the event horizon in de Sitter spacetime. 
In {\bf post-inflationary reheating}, the quantum (inhomogeneous) 
fluctuation fields 
are parametrically amplified by the time-dependent (homogeneous) 
inflaton  (background) field,
resulting in particle creation. Its backreaction effect is to damp 
away the oscillations of
the inflaton field  and reheat the universe. 
(See, e.g., \cite{RamseyHu,Hasi}).  Similar processes of particle creation 
at the Planck time leading to {\bf damping of anisotropy} of the background
spacetime were earlier treated by means of a closed-time-path (CTP), or in-in
effective action. (See, e.g.,  \cite{CH87}).  
In {\bf quantum cosmology} one usually makes the so-called 
`minisuperspace approximation' 
of quantizing only the homogeneous universes and ignoring the inhomogeneous
ones. The validity of this truncation scheme can be examined as a
backreaction problem. One can view the homogeneous and inhomogeneous universes
as the low and high lying modes of excitation of spacetime (this notion can
be made more precise in terms of gravitational perturbations off a
background spacetime as described by the Lifshitz wave operator), treat 
the former as constituting the system field (minisuperspace) and the 
latter as the
environmental field. The magnitude of the backreaction of the latter on
the former will give a measure of the validity of the minisuperspace
approximation. This approach was taken  by Hu and Sinha who
introduced the CTP CGEA \cite{SinhaHu}. The same idea and method should
be applicable to backreaction effects of hard thermal loops 
on the soft sector in quark-gluon plasma  \cite{BraPis}.

In the high-low mode division, one can think of
the system as the average of the field over a spatial volume 
$\Lambda_c^{-3}$, $\Lambda_c={\L\over s}$ being the 
inverse of the critical wavelength that
separates the system from the environment. 
The  {\it Euclidean } version of this
coarse-graining leads to
the  so-called average effective action (AEA), introduced
by Wetterich  \cite{wette1}.
These two formalisms, {
\it Euclidean} and {\it real time} CGEA, 
though sharing similar ideas have rather different 
ranges of applications.  Wetterich's  average effective action 
averages the  fields 
over a {\it space-time} Euclidean region, while Hu's \cite{cgea} 
coarse-grained effective action 
averages field over a {\it spatial} region.
The Euclidean AEA  is more suitable for near-equilibrium systems,  
and has indeed been
applied to QCD problems like quark-gluon plasma in a quasi-stationary state,
while the real-time CTP CGEA gives causal evolution equations for
expectation values of quantum operator and is more suitable for  
the analysis  of dynamical problems. 
We will invoke the AEA in Section VIII A.

\subsection{`In-Out' Coarse Grained Effective Action}

We now introduce the {\it in-out} version of the CGEA. This is 
because is pedagogically simpler, and is closer to what 
one learns from textbooks, i.e., 
the Schwinger-DeWitt effective action. 
Functionally it is also 
 adequate for the introduction of ideas on coarse-graining and 
backreaction to the extent of
seeing the running of coupling constants leading to a RG theory.
But the effective equation of motion with backreaction obtained 
from an {\it in-out} 
version contains non real and
non causal terms. This was first noticed by DeWitt and Jordan 
\cite{DeWJor,Jordan} and
Calzetta and Hu  \cite{CH87,CH89} in a study
of the backreaction of particle creation on a background spacetime or field.  
For a  correct treatment of  the backreaction of the environment 
on the system one needs the {\it in-in} or CTP version, which we will 
discuss  in Sec. 6.

Consider the $ \lambda \phi ^{4}$ model with action (\ref{3phiact})
in a spatially-flat Robertson-Walker
universe.
The scale factor $a(t) = e^{\alpha }$ is left unspecified for now.
Rather than viewing $a(t)$ as a dynamical function determined from Einstein's
equations we will, in the spirit of scaling as explained in an earlier section, regard
$a$ as a constant and $t$ as a parameter.  Although the space can have
different scales  $a(t)$ at different times $t$, spacetime remains flat
at all times. (One can rescale $\vec {x}$ at different times to make
them all equal in value.) Thus we can simply think of doing Minkowski
field theory here, leaving $ a(t) $ as a parameter while carrying
out coarse-graining in the 3-dimensional space. The content of this subsection
can thus be used without reference to cosmology (simply set $a(t)= 1$ ).
However we do want to tag the scale factor  $a(t)$ along so that later
we can perform scaling transformations and discuss  inflationary cosmology.

For the RW
 spacetime, the action (\ref{3phiact}) reads
\begin{equation}
S[\phi] = \int d^3x \int dt a(t) [- \frac{1}{2}
\phi(\hat{d}_2+\hat{d}_1-\hat{\Delta} +\tilde{m}^2)\phi - \frac{1}{4!}
\tilde{\lambda}\phi^4]   
\end{equation}
where
\be
\hat{d} _2 = a^2(t) \partial ^2/ \partial t^2, \ \ \ \
\hat{d} _1 = a^2(t) 3H \partial/ \partial t, \ \ \ \
\hat{\Delta}=a^2(t) \Delta
\te
are the second and first order time derivative operators and $\Delta$ is the
spatial Helmholtz operator respectively. Here
$H \equiv \dot{a} /a = \dot{\alpha}$ and
\be \tilde{m}^2(t) = a^2(t)(m^2+ \xi R) \te
\begin{equation}
 \tilde{\lambda}(t) = a^2(t)\lambda  
\end{equation}
are the conformally-related generalized mass and field self-coupling
parameter respectively.  We will now assume $\hat{d}_{1} = \hat{d}_{2} = 0$,
i.e., consider the case of eternal inflation. In
this `static' limit, one can introduce the momentum space
representation on the spatial section
\be
\phi(\vec{x},t) =(2\pi)^{-3/2} \int d^3k e^{i\vec{k}\cdot\vec{x}}
\phi ( \vec{k},t)
\te    
Then
\[
S[\phi] = \int dt a(t) \int d^3k [- \frac{1}{2} (k^2 +
\tilde{m}^2(t))\phi( \vec{k}) \phi(- \vec{k})]
\]
\[ 
+(2\pi)^{-3} \int dta(t) \int
d^3 \vec{k} _1d^3 \vec{k} _2d^3 \vec{k} _3d^3 \vec{k} _4
[- \frac{1}{4!}
\tilde{\lambda}(t)\delta ( \vec{k}_1+ \vec{k}_2+ \vec{k}_3+
\vec{k} _4) 
\]
\begin{equation}
\phi (\vec{k} _1,t) \phi ( \vec{k} _2,t) \phi ( \vec{k} _3,t)
\phi( \vec{k}_4,t)]
\label{4sphi}
\end{equation}    
For interacting fields in a static spacetime mode-mixing is manifest.
If $\hat{d}_{1} \neq 0$ "time" translation-invariance is lost and there will
be frequency mixing. This becomes relevant in the context of critical
dynamics, as mentioned in Sec. 2.
 
Now we separate the field $\phi (x,t)$ into two parts, $\phi_<$ and 
$\phi_> $, which one can refer to as the system and the environment,
\[
\phi = \phi_< + \phi_>
\]
We assume that $\phi_<$ contains the lower $k$ wave modes and
$\phi_>$ the higher $k$ modes and consider these two cases: \\
Case A (critical phenomena)
\begin{equation}
\phi_<: \mid \vec{k} \mid < \Lambda /s,~~~~~~
\phi_>: \Lambda/s < \mid \vec{k} \mid < \Lambda,
\label{casea}
\end{equation}
Case B  (stochastic inflation)
\begin{equation}
\phi_<: \mid \vec{k} \mid<  \epsilon Ha, ~~~~~~
\phi_>: \mid \vec{k} \mid > \epsilon Ha,~~~~~
\epsilon \approx 1.
\label{caseb}
\end{equation}
Here $\Lambda $ is the ultraviolet cutoff and $ s > 1$ is the
coarse-graining parameter which gives the fraction of total $k$ modes counted 
in the environment.  The separation of 
$\phi $ can also be made in other manners, depending on the physical setup 
of the problem and the questions one asks.  The formalism we present here
can be easily adapted to other types of
decomposition.

$S[\phi ]$ in (\ref{4sphi}) can be written as
\begin{eqnarray}
S[\phi] & = & \int dta(t) \int d^3 \vec{k}
(-\frac{1}{2}) (\vec{k}^2+\tilde{m}^2(t) 
\phi_< (\vec{k},t) \phi_<(-\vec{k},t)) \nn\\
& & 
+ \int dta(t) \int d^3 \vec{q}
(-\frac{1}{2}) (\vec{q}^2 + \tilde{m}^2(t))
\phi_>(\vec{q},t) \phi_>(-\vec{q},t) \nn\\ 
& & 
+(2\pi)^{-3} \int dt a(t)
\int d^3 \vec{k_1} d^3 \vec{k_2} d^3 \vec{k_3} d^3 \vec{k_4}
(- \frac{1}{4!} \tilde{\lambda}(t)) \nn\\             
& & 
\times 
\phi_<(\vec{k}_1,t) \phi_<(\vec{k}_2,t)
\phi_<(\vec{k}_3,t) \phi_<(\vec{k}_4,t)  
\delta(\vec{k}_1+ \vec{k}_2+ \vec{k}_3 + \vec{k}_4) \nn\\   
 & &
+(2\pi)^{-3} \int dta(t)
\int d^3\vec{k_1} d^3\vec{k_2} d^3\vec{k_3} d^3\vec{q_1}
(-\frac{1}{4!} \tilde{\lambda}(t)) \nn\\
& &
\times 4 
\phi_<(\vec{k}_1,t) \phi_<(\vec{k}_2,t)
\phi_<(\vec{k}_3,t) \phi_>(\vec{q_1},t)
\delta(\vec{k}_1 + \vec{k}_2+ \vec{k}_3 + \vec{q}_1) \nn\\   
& &
+(2\pi)^{-3}\int dt a(t)
\int d^3\vec{k_1} d^3\vec{k_2} d^3\vec{q_1} d^3\vec{q_2}
(-\frac{1}{4!} \tilde{\lambda}(t)) \nn\\
& &
\times 6 \phi_<(\vec{k}_1,t)    \phi_<(\vec{k}_2,t)
\phi_>(\vec{q}_1,t) \phi_>(\vec{q}_2,t) 
\delta (\vec{k}_1 + \vec{k}_2 + \vec{q}_1 + \vec{q}_2) \nn\\ 
& & 
+(2\pi)^{-3} \int dt a(t)
\int d^3\vec{k_1} d^3\vec{q_1} d^3\vec{q_2} d^3\vec{q_3} 
(-\frac{1}{4!} \tilde{\lambda} (t)) \nn\\ 
& &
\times  4 
\phi_< (\bar{k}_1,t) \phi_> (\vec{q}_1,t) 
\phi_>(\vec{q}_2,t) \phi_>(\vec{q}_3) 
\delta (\vec{k}_1 + \vec{q}_1 + \vec{q}_2 + \vec{q}_3) \nn\\ 
& & 
+(2\pi)^{-3} \int dt a(t)
\int d^3\vec{q_1} d^3\vec{q_2} d^3\vec{q_3} d^3\vec{q_4}
(-\frac{1}{4!} \tilde{\lambda} (t)) \nn\\  
& &
\times 
\phi_> (\vec{q}_1,t) \phi_> (\vec{q}_2,t) 
\phi_> (\vec{q}_3,t) \phi_> (\vec{q}_4,t) 
\delta (\vec{q}_1 + \vec{q}_2 + \vec{q}_3 + \vec{q}_4)
\label{4sphi2}
\end{eqnarray}
where all integrals of $k$ and $ q $ have the limits indicated
in (\ref{casea}) or (\ref{caseb}) 
depending on the cases in question. Henceforth we shall use the 
shorthand notation
\begin{equation}
\phi_<(1) \equiv \phi_< (\vec{k}_1,t), ~~~~~
\phi_>(\tilde{1}) \equiv \phi_> (\vec{q}_1,t)
\end{equation}
and denote
\begin{equation}
G_<(1)       =  [\vec{k}^2 + \tilde{m}^2(t) -i\epsilon]^{-1}, ~~~~
G_>(\tilde{1}) =  [\vec{q}^2 + \tilde{m}^2(t) -i\epsilon]^{-1}.
\end{equation}  
We can separate terms in (\ref{4sphi2}) into three groups
\begin{equation}
S[\phi] 
= S[\phi_<] 
+ S_0 [\phi_>] 
+ S_I [\phi_<, \phi_>]
\end{equation}   
where
$$
S[\phi_<]=
- \frac{1}{2} G_<^{-1}(1) \phi_< (1) \phi_<(-1) \\
- \frac{1}{4!} \tilde{\lambda} (t) \phi_< (1) \phi_< (2)
\phi_< (3) \phi_< (4) \delta (1+2+3+4)   
$$
$$
S_0[\phi_>]
= - \frac{1}{2} G_>^{-1} (\tilde{1})\phi_>(\tilde{1}) 
\phi_> (-\tilde{1}) 
$$
and
\begin{eqnarray}
S_I[\phi_<,\phi_>]
& = & 
- \frac{1}{4!} \tilde{\lambda} (t)[ 4 
\phi_<(1) \phi_<(2) 
\phi_<(3) \phi_>(\tilde{1})
\delta (1+2+3+\tilde{1}) \nn \\
& & 
+ 6\phi_<(1) \phi_<(2)
\phi_>(\tilde{1}) \phi_>(\tilde{2}) 
\delta(1+2+\tilde{1}+\tilde{2}) \nn \\
& & 
+ 4\phi_<(1) \phi_>(\tilde{1})
\phi_>(\tilde{2}) \phi_>(\tilde{3})
\delta(1+\tilde{1}+\tilde{2}+\tilde{3}) \nn \\
& & 
+ \phi_>(\tilde{1}) \phi_>(\tilde{2})
  \phi_>(\tilde{3}) \phi_>(\tilde{4})
  \delta(\tilde{1}+\tilde{2}+\tilde{3}+\tilde{4})] \nn \\
& &
\equiv S_1 + S_2 + S_3 + S_4
\label{3.14}
\end{eqnarray}
Here integration over k's and q's with the respective ranges
is understood, and we have used the simplified notation of
$1, \tilde 1$ to denote $\vec k_1, \vec q_1$, etc.
 It is easier to use Feynman diagrams for the propagators and
vertices corresponding to the two fields $\phi_<$ and $\phi_>$.  
They are listed in Figure 2.
 
The functional integral formalism for quantum fields is set up in the
usual way.  Assume that at $t = \pm  \infty $, the interaction is turned 
off so that the  in-vacuum $ | 0_{-} \rangle $ and the out-vacuum
$|0_{+}  \rangle $
are defined.  The vacuum persistence amplitude on the generating functional
$  \langle 0_{+}|0_{-}  \rangle $ gives the probability amplitude of the
field remaining in the vacuum after the interacting system evolves in time.
 
With the splitting $\phi  = \phi_< + \phi_>$, the Schwinger-DeWitt
(in-out) generating functional becomes,
\begin{equation}
Z[\phi] =
N \int D\phi e^{iS[\phi ]} =
N \int D\phi_< \int D\phi_>
e^{\{iS[\phi_<]+S_0[\phi_>]
+S_I[\phi_<,\phi_>] \}}
\end{equation}
Introducing an integral over $ \phi_> $ of
$\exp{\hbox{iS}_{0}[\phi_>]}$ as the norm for the functional
average in the denominator and changing the constant $N$ to $N'$  
accordingly, we get
\begin{eqnarray}
Z[\phi] 
& = &
N' \int D\phi_<~ e^{iS[\phi_<]} 
(\int D\phi_>~ e^{iS_0[\phi_>]}
e^{iS_I[\phi_<,\phi_>]} \large{/}
\int D\phi_>e^{iS_0[\phi_>]})  \nn  \\
& = & 
N' \int D\phi_<~e^{iS[\phi_<]} 
\times \langle  e^{iS_I[\phi_<,\phi_>]} 
\rangle _{\phi_>}
\end{eqnarray}
Here $ \langle ~~~~ \rangle _{\phi_> }$ defines averaging
over the  $\phi_>$ field under the free action $S_{0}[\phi_>]$
(We write $ \langle ~~~~ \rangle _0 $  interchangeably to emphasize that
the averaging is with the free action $S_{0}[\phi_>]$.)
 
 Denoting
\begin{equation}
\langle e^{iS_I[\phi_<, \phi_>]}  \rangle _{\phi_>} 
= \exp {i \Delta S[\phi_<]}    
\end{equation}
\noindent we get
\begin{equation}
Z[\phi_<] 
= N' \int D \phi_< ~ \exp{iS_{eff}[\phi_<]}
\end{equation}
The coarse-grained effective action is given by
\begin{equation}
S_{eff}[\phi_<] 
= S[\phi_<] + \Delta S[\phi_<],   
\end{equation}
where
\be
\Delta S [\phi_<] = - iln
\langle  \exp{iS_I} [\phi_<, \phi_>]  
\rangle _{\phi_>} 
\te
If the interaction between the $\phi_<$ and $\phi_>$ 
fields is small (e.g. $\lambda \ll 1$ in a $ \lambda \phi ^{4}$ theory),
one can expand this in a Dyson-Feynman series
\begin{eqnarray}
\Delta S[\phi_<] 
& = &
- i ln \sum^{\infty}_{n=0} \frac{i^n}{n!} \;
\langle S^n_I  \; [\phi_<, \phi_>] \rangle_{\phi_>} \nn \\
& =  &
\langle S_I[\phi_<,\phi_>]  \rangle_{\phi_>} 
+ \frac{i}{2}
[\langle S_I[\phi_<,\phi_>]^2 \rangle _{\phi_>}
-\langle S_I[\phi_<,\phi_>] \rangle _{\phi_>}^2 ]
+  ...                                 
\label{3.20}
\end{eqnarray}
where in the second line terms up to the second order in
$S_I$ are written out explicitly.
 
Let us now examine each term in (\ref{3.20}) in detail, 
starting with the       
first order term $ < S_{I}[\phi_<,\phi_>] >_{\phi_>}$.  
We see that terms in (\ref{3.14}) involve the average of products of 
$\phi_>$ fields, but those containing an odd number of fields in
the product average to zero. Also any product containing four
$\phi_> $ fields (last term in (\ref{3.14})) averages out to a quantity
which is independent of $\phi_<$, i.e.,
$\langle  \phi_> (\tilde{1})  \rangle _0
= \langle \phi_> (\tilde{1}) \phi_>(\tilde{2}) 
\phi_> (\tilde{3}) \rangle _0 = 0$
and $\langle \phi_> (\tilde{1}) \phi_> (\tilde{2})
\phi_>(\tilde{3}) \phi_>(\tilde{4}) \rangle_0$
is independent  of $ \phi_< (\vec{k},t)$. Thus the only contribution to
$ \langle S_{I}[\phi_<,\phi_>]  \rangle _{0}$
is the quadratic product
\begin{equation}
\langle \phi_>(\tilde{1}) \phi_>(\tilde{2}) \rangle _0 
=-ia^{-1}(t)G_>(\tilde{1}) \delta (\tilde{1}+\tilde{2}) \delta(t_1-t_2).
\label{3.21}
\end{equation}                                           
We denote by $S_{I}[\phi_<, \phi_>]_j $  the terms
containing $ j $ $\phi_>$ fields, i.e. 
$ S_{I\; 1} \sim \phi_<\phi_<\phi_<\phi_>$, 
$ S_{I\; 2} \sim \phi_<\phi_<\phi_>\phi_>$, etc.  
In the Feynman diagram depiction (Fig.2c), the averaging over $\phi_>$
amounts to closing the ~ lines representing the environment field propagators  
on themselves, and it is easy to see 
how the above conclusion is reached.  
 Explicitly, the only contributing first order term is
\begin{eqnarray*}
\langle  S_I[\phi_<, \phi_>]_2  \rangle _0 
& = & 
(2\pi)^{-3}\int a(t)dt \int d^3 \vec{k_1} d^3 \vec{k_2}
\int d^3 \vec{q_1} d^3 \vec{q_2}
6 (- \frac{1}{4!}) a(t) \tilde{\lambda}(t) \\
& &
\times \phi_<(\vec{k}_1,t) \phi_<(\vec{k}_2,t)
\langle  \phi_>(\vec{q}_1,t) \phi_>(\vec{q}_2,t) \rangle _0
\delta(\vec{k}_1+\vec{k}_2+\vec{q}_1+\vec{q}_2) 
\end{eqnarray*}
Using (\ref{3.21}), we get
\begin{equation}
\langle S_I[\phi_<,\phi_> ]_2 \rangle _0
= \int a(t)dt \int d^3 \vec{k} (-\frac{1}{2})
\phi_<(\vec{k},t) \ \delta \tilde{m}^2(t,s)_1 \
\phi_<(\vec{k},t)
\end{equation}
where
\begin{equation}
\delta \tilde{m}^2(t,s)_1
=- \frac{i}{2} \delta(0)\tilde{\lambda}(t)a^{-1}(t)
(2\pi)^{-3}\int d^3\vec{q}~[\vec{q}^2+\tilde{m}^2(t)-i\epsilon ]^{-1}.
\label{4.23}
\end{equation}
This is depicted as (1) in Fig. 2c.  The subscript 1
on $\delta\tilde{m}^2$ denotes that it is the first order correction to
$\tilde{m}^2$.
The second order correction is given by (5) and (6) of Fig. 2c and
computed below. 
 
Now we examine the second order terms in (\ref{3.20}). The second term
in the square bracket 
$ \langle S_{I}[\phi_<,\phi_>]  \rangle ^{2}_{0}$  
gives three non-contributing disconnected graphs.
The first term $ \langle (S_{I}[\phi_<,\phi_>])^{2}  
\rangle _{0} $ has the following components:
\begin{eqnarray*}
\langle S^2_I \rangle _0
& = &
   \langle S^2_1 \rangle _0  +  \langle S^2_2  \rangle _0  
+  \langle S^2_3 \rangle _0  +  \langle S^2_4  \rangle _0 \\
& &
+ 2[ \langle S_1S_2 \rangle _0  +  \langle S_1S_3  \rangle _0
+     \langle S_1S_4 \rangle _0  \\
& & 
+  \langle S_2S_3 \rangle _0 + \langle S_2S_4 \rangle_0
+  \langle S_3S_4 \rangle _0 ]           
\end{eqnarray*}
where we have written in short
\[
S_j = S_I [\phi_<, \phi_>]_j ~~~~~~ (j = 1,2,3,4)
\]
as in Eq. (\ref{3.14}) . In terms of the Feynman diagrams in Fig. 2c
constructed from Fig. 2b,
there are four contributing connected
diagrams of the second order.  They are denoted by (3)-(6) in Fig. 2c.
We see that diagram (1) is the first order and (5) (6) are the  second order
corrections to the mass $ \tilde{m}^2$, (4) is a second order correction to
the coupling constant $\tilde{\lambda}$, while (3) generates $\phi^{6}$
correction.
 
Adding these contributions together we obtain finally for the
coarse-grained effective action up to second order
in $\lambda$ :
\begin{eqnarray}
S_{eff}[\phi_<] 
& = & \int a(t)dt \int d^3\vec{k}(- \frac{1}{2})
\phi_<(\vec{k},t)[k^2+\tilde{m}^2(t,s)]\phi_<(-\vec{k},t) \nn \\
& &
+ \int a(t)dt 
\int d^3\vec{k}_1 d^3\vec{k}_2 d^3\vec{k}_3 d^3\vec{k}_4
(- \frac{1}{4!}) \tilde{\lambda} (t,s, \{ \vec{k}\} ) \nn \\
& &
\times
\phi_< (\vec{k}_1,t) \phi_< (\vec{k}_2,t) 
\phi_< (\vec{k}_3,t) \phi_< (\vec{k}_4,t) 
\delta (\vec{k}_1 + \vec{k}_2 + \vec{k}_3 + \vec{k}_4) \nn \\
& &
+ \int a(t)dt \int d^3\vec{k}_1...d^3\vec{k}_6 
(- \frac{1}{6!}) \nu_6 (t,s, \{ \vec{k}\} ) \nn \\
& &
\times
\phi_< (\vec{k}_1,t)...\phi_<
(\vec{k}_6,t) \delta (\vec{k}_1 + ... + \vec{k}_6).
\label{3.24}
\end{eqnarray}
where the integration range of $\vec{k}_i$ is as indicated in  (\ref{casea}) and (\ref{caseb}):
$\mid \vec{k} \mid \leq \Lambda / s $ for the case of critical phenomena and
$\mid \vec{k} \mid < \epsilon Ha $ for stochastic inflation.\\
Here,
\[
\tilde{m}^2(t,s) = \tilde{m}^2(t)
+ \delta \tilde{m}^2(t,s)_1
+ \delta \tilde{m}^2(t,s)_{2a}
+ \delta \tilde{m}^2(t,s)_{2b}
\]
\begin{equation}
\tilde {\lambda} (t,s, \{\vec{k}\})
= \tilde{\lambda} (t) + \delta \tilde{\lambda} (t,s,\{\vec{k}\})_1
\label{4.25}
\end{equation}
The subscripts under $\delta \tilde{m}^2$ and $\delta \tilde{\lambda}$
denote the order of
correction to the mass and the coupling constants
arising from averaging over $\phi_>$ fields. They are
given by: 
\begin{eqnarray}
\delta \tilde{m}^2(t,s)_{2a} & = &
(-\frac{1}{4})[i\delta (0)]^2 [\tilde{\lambda} (t)/a(t)]^2
(2\pi)^{-3}\int d^3\vec{q}_1 [\vec{q}^2_1 + \tilde{m}^2(t) -i\epsilon ]^{-2}
\nonumber\\
& &
(2\pi)^{-3}\int d^3\vec{q}_2 [\vec{q}^2_2 + \tilde{m}^2(t) -i\epsilon ]^{-1}
\end{eqnarray}
\[
\delta\tilde {m}^2(t,s)_{2b} =
(-\frac{1}{6})[i\delta (0)]^2 [\tilde{\lambda} (t)/a(t)]^2
(2\pi)^{-3}\int d^3\vec{q}_1 (2\pi)^{-3}\int d^3\vec{q}_2
[\vec{q}^2_1 + \tilde{m}^2(t) -i\epsilon ]^{-1}
\]
\be
[\vec{q}^2_2 + \tilde{m}^2(t) -i\epsilon ]^{-1}
[(\vec{k}-\vec{q}_1-\vec{q}_2)^2+ \tilde{m}^2(t) -i\epsilon ]^{-1}
\te
\[
\delta \tilde {\lambda} (t,s, \{\vec{k}\})=
[i\delta (0)]~\frac{3}{2} a(t)^{-1} \tilde{\lambda} ^2(t)
(2\pi)^{-3}\int d^3\vec{q} [\vec{q}^2+ \tilde{m}^2(t) - i\epsilon]^{-1}
\]
\begin{equation}
[(\vec{k}_1 + \vec{k}_2 - \vec{q})^2
+\tilde{m}^2(t) + i\epsilon]^{-1}    
\end{equation}
\begin{equation}
\nu_6(t,s,\{\vec{k}\})=
-10 \tilde{\lambda} ^2 (t) [(\vec{k}_1 + \vec{k}_2 + \vec{k}_3)^2
\tilde{m}^2(t) -i \epsilon]^{-1}
\label{nu6}
\end{equation}
 


\section{Backreaction in the Inflationary Universe:
Renormalization group equations and the running of coupling constants}

We have so far discussed the separation and averaging processes leading to a
coarse-grained effective action which contains the averaged effect of the
environment on the system.  Now we will introduce a rescaling of the k-space
(truncated in critical phenomena,  enlarged in stochastic inflation) 
to bring it back to the original size (but with less content in CP, more in SI).  
By requiring that this transformed
Lagrangian has the same form as the one we started out with we obtain a set of
renormalization group equations for the field parameters.  At this point our earlier
observation  on the equivalence of scaling and inflation for the
field and spacetime in question becomes relevant.  That is, rescaling in a
static picture serves the function of inflation in a dynamic picture. Our
derivation here of the differential renormalization group equations
is similar to the Wilson-Fisher
\cite{CriPhe}  or the Wegner-Houghton methods \cite{weghou}.
 
After introducing a rescaling of the k-space (spatial dimension D=3),
\begin{equation}
\vec{k}' = s \vec{k}, \; \; \; \phi '(\vec{k}',t) = s^{- \frac{D+2}{2}} 
\phi_< (\vec{k},t)
\end{equation}
the effective action (\ref{3.24}) becomes
\[ S_{eff} [\phi '] =  \int dt a(t) \int d^3 \vec{k}'(-\frac{1}{2}) \phi '
(\vec{k}',t) [ \vec{k}'^2 + s^2\tilde{m}^2(t,s)] \phi '(\vec{k}',t) \]
\[+ \int dt a(t) \int d \vec{k}'_1 \cdots  d \vec{k}'_4 (-\frac{1}{4!})
s^{4-D} \tilde{\lambda} (t,s,\{\vec{k}'\}) \phi ' (\vec{k}'_1,t) \cdots
 \phi'(\vec{k}'_4,t)
\delta (\vec{k}'_1 + \vec{k}'_2 + \vec{k}'_3 + \vec{k}'_4) \]
\be
+\int dt a(t) \int d \vec{k}'_1 \cdots d \vec{k}'_6 (-\frac{1}{6!})s ^{6-2D}
\nu_6 (t,s,\{ \vec{k}'\}) \phi' (\vec{k}'_1,t) \cdots \phi' (\vec{k}'_6,t)
\delta(\vec{k}'_1 + \cdots + \vec{k}'_6)
\te
Note that the $k'$-integrations now resume the full range as k in 
(\ref{casea}) or (\ref{caseb}).  In
terms of the rescaled variables $\phi'\;,\rm{S_{eff}}$ will have the same 
form up
to a certain order as the original S$\rm{_{eff}}$ in terms of $\phi$ provided 
that we identify
\[\tilde{ m}'^2(t,s) = s^2\tilde{m}^2(t,s) \]
\begin{equation}
\tilde{\lambda} '(t,s) = s^{4-D} \tilde{\lambda} (t,s)
\end{equation}
The effective mass and coupling constants are given in 
(\ref{4.25}-\ref{nu6}).

We can now proceed to tackle the coarse-grained integrals therein.
At this point one
needs to stipulate the system and bath separation such as the Cases A and B
given as examples in Section 3. We will work out the details for critical phenomena (Case A) here.
The case for stochastic inflation (Case B) can be obtained via the simple relation between Case A and B also
mentioned in Section 3. For small changes in $s\simeq1+ d\sigma$, one can
derive a set of differential renormalization group equations for the mass
and the coupling constant as follows:
 
\[ dx/d\sigma = 2x - \frac{1}{2} y/(1+x)  \]
\begin{equation}
dy/d\sigma = \epsilon y + \frac{3}{2} y^2/(1+x)^2
\label{rgephi4}
\end{equation}
where
\[ x \equiv \tilde{m}^2(t,\sigma)\Lambda^{-2} \]
\begin{equation}
y \equiv \tilde{\lambda} (t,\sigma) \Lambda ^{-\epsilon} \Omega a^{-1}(t)
\end{equation}
Here, $ \epsilon= 4-D $, and $\Omega$ is the solid angle integration in
D-dimension.
\par
As expected, these equations have the same form as in the Ginzburg-Landau-
Wilson model \cite{CriPhe}\cite{CriPheQFT}\cite{Wilson} .
They govern the flow of the field parameters $(\tilde{m}^2,\tilde{\lambda})$
as the scaling changes.  There exist certain points in this parameter space
 known
as the fixed points, where further application of the RG transformation leads
to an invariant result.  The fixed points are thus the steady-state solutions
to the RG equations $dx/d\sigma = 0$  and $dy/d\sigma$ = 0.  For the
 $\phi^4$
theory, it is well-known that there is a trivial fixed point at
\begin{equation} x_f^0 = 0,~~~~~ \ y_f^0 = 0
\end{equation}
and a nontrivial fixed point at
\begin{equation}
x^*_f = - \epsilon/6,~~~~~ \ y^*_f = - 2 \epsilon/3
\end{equation}
Near the trivial fixed point  the solution to (\ref{rgephi4}) is 
\begin{equation}
\left(
\begin{array}{c}
x\\
y
\end{array} \right)\\
=A \left(
\begin{array}{c}
1\\
0
\end{array} \right)
e^{2\sigma} + B
\left(
\begin{array}{c}
1\\
2(2-\epsilon)
\end{array} \right) e^{\epsilon \sigma}
\end{equation}
The critical point of interest to us is the nontrivial fixed point.  We want to
find out how in its neighborhood the field parameters flow towards this fixed
point.  By setting $x = x^{*}_f + \Delta x$ and $y = y^{*}_f + \Delta y$
we get
\[ d(\Delta x)/d\sigma = (2 - \frac{1}{3} \epsilon ) \Delta x -
\frac{1}{2} (1 + \frac{1}{6} \epsilon ) \Delta y \]
and
\begin{equation}
d(\Delta y)/d\sigma = - \epsilon \Delta y
\end{equation}
with the solution
 
\begin{equation}
\left(
\begin{array}{c}
\Delta x\\
\Delta y
\end{array} \right) = A
\left( \begin{array}{c}
1\\
0
\end{array} \right) e^{(2 - \epsilon /3)\sigma} + B \left(
\begin{array}{c}
1\\
4(1+\epsilon /6)
\end{array} \right) e^{-\epsilon \sigma}
\label{3.38}
\end{equation}

\subsection{Cosmological Consequences}
\par
As we explained earlier, in eternal inflation the renormalization group
transformation parameterized by $s$ is equivalent to time evolution
parametrized by $t$.  Thus under such conditions we can treat the coupling
constants like $\lambda (t,s)$ as depending on either $s$ (flow)
or $t$ (evolution).
The fixed point corresponds to the extreme infrared limit which in the case of
eternal inflation lies at infinitely late time $t_{f}$.  In practice it
corresponds to the time the inflation regime ends $t_{1}\simeq t_{f}$ before 
reheating
sets in.  Thus in (\ref{3.38}) $\Delta y$ is a measure of how $y$ changes from $t$
to $t_{1}$.
Explicitly, with $\epsilon  = 1$ and $\sigma = \int H dt$, the
$O(\epsilon )$ expansion result gives
\begin{equation}
y(t) = y(t_{f}) + {14B\over 3} e^{-H(t_{f}-t)}.
\end{equation}
At the fixed point $t_{f} $, $ y(t_{f}) = - {2\over 3} \epsilon $.
Recall that $y = \tilde{\lambda(t)}\Lambda ^{-\epsilon } \Omega a^{-1}(t)$ and
$\tilde{\lambda }(t) = \lambda (t) a^2(t)$ ,this yields
\be
\lambda (t) = - {2\over 3} e^{-H(t-t_{f})}\Bigl[1 - 7B e^{H(t-t_{f})}\Bigr].
\label{5lambda2}
\te
\noindent Comparing $\lambda (t)$ at two times $t > t_{0}$,
\be
{\lambda (t)\over \lambda (t_0)}
\simeq  e^{-H(t-t_{0})}\Bigl[1 - 7B(e^{H(t-t_{f})} - e^{H(t_{0}-t_{f})})
\Bigr].
\te
Assuming $t_{f} >> t, t_{0}$ (which is clearly satisfied for
realistic inflation $t_{f} > t_{0}e^{68}$) the second term in the
 square bracket can be neglected.  We get,

\be
\lambda (t)/\lambda (t_{0}) \simeq  e^{-H(t-t_{0})}
\label{5lambda}
\te
\noindent Thus the coupling constant $\lambda $ of the effective field
theory (in a system which is depleting in content) actually  decreases
with time. 
\footnote{Of course, even though the running coupling goes like
$\lambda (t)/\lambda (t_{0}) \simeq  e^{-H(t-t_{0})}$,
a loop expansion of a quantity such as the density-density
correlation function will show that in the same 
limit each loop appears with a factor of $ e^{H(t-t_{0})}$.
The self-consistent cancellation of these two factors is 
associated with a crossover phenomenon and in principle
could be accounted for by using an H-dependent (environmental-friendly)
RG\cite{cspriv}. We thank C. R. Stephens for this remark.}
This is a new effect which could have important consequence
for the galaxy formation problem.

As we know from standard calculations \cite{GalForIU} the density contrast
$\delta \rho /\rho$ depends strongly on $ \lambda$. For GUT processes one
needs to assume an unnaturally weak $\lambda \approx 10^{-12}$ in order
that $ \delta \rho /\rho \approx 10^{-4}$  at the time the fluctuations
reenter the (RW) horizon. What (\ref{5lambda}) implies is that the strength of
$\lambda$ at time $t^*_G$ when the fluctuation mode corresponding to galaxy
scale left the de Sitter horizon is much weaker than $\lambda$ at $t^*_H$,
the time the Hubble size left the de Sitter horizon. The running of the
coupling constants seems to provide a way to reduce their strength. The
reason behind this phenomena is, as we recall, due to the backreaction
of the environment of high k modes on the system of lower modes,
which in this particular choice is diminishing in time.

In contrast, for stochastic inflation where the system (consisting of
modes with physical
wavelength greater than the de Sitter horizon) increases in content, the
coupling constants will, from this analysis,  actually increase exponentially
with time during the inflationary regime (era from $t_{0}$ to $t_{1})$.
Because a formal analogy exists between this case (Case B) of stochastic inflation (SI)  and 
critical phenomena  (CP, Case A) one can indeed
just read off the result from the results we obtained earlier.  The transcription is simply i)
Changing $\Lambda $ in  CP to $\epsilon H$ in  SI.
Recall that $\Lambda $ refers to the
ultraviolet cutoff in CP, whereas $H^{-1}$ is the horizon size.  The
ultraviolet cutoff for SI is assumed to be $\infty $.  ii)  Changing
$s$ in CP to $s^{-1}$ in SI. Recall that the coarse-graining parameter $s$ plays the
role of the scale function $a(t)$.  Thus $s^{-1}$ in CP is now acting like
$a(t)$ in SI [see Eq. (\ref{caseb})].  Note that $s>1$ in CP but $s<1$ in
SI. Note also that transformation (ii) is equivalent to a time reversal,
i.e., changing $\Delta t \rightarrow  - \Delta t$.
The way $\lambda$ runs in stochastic inflation is given by Eq. (\ref{5lambda2}),
with $-t$ replacing $t$ there.  The ensuing discussion
of the implications on galaxy formation is similar, except that, of course,
one reaches the opposite conclusion, i.e., the strengthening of $\lambda $ 
in time increases the density contrast as each mode reenters the RW horizon.

The opposite results arising from these two cases should not be viewed 
as unsettling.  They arise from very different stipulations of the system
(CP decreasing in time, SI increasing).
The cosmological consequences will depend
on whether and how $\lambda$ runs, but the theoretical issues this
investigation brings out should be valid.
What is shown here is that if one
takes into account the interaction between the environment and the
system, the running of coupling parameters in the theory
as {\sl observed in the system} is an unavoidable consequence.
How they run (with time or scale or energy) will depend sensitively
on how the system is selected with respect to the environment, and,
of course, the nature of the interaction (gauge fields would
presumably run differently from $\lambda \phi ^{4}$, as their $UV$
and $IR$ behavior differ).  

On this point, let us reexamine the physical criteria for the 
choice of the system
versus the environment in our examples. The cutoff is determined by
the highest  $k$ mode which left the de Sitter horizon, which in turn
depends on the duration of inflation. At the start of inflation of
course there is no knowledge when inflation will end and the modes do not
know which ones among them are to be counted
in (the system) and which are to be out (the environment). A more
straightforward division would be to have a fixed cutoff at the beginning
of inflation, e.g., those with k greater than the galaxy size ( $k_G$ )
constitute the environment and those smaller than $k_G$ but greater than
$ H^{-1} $ be the system. This would give rise to running, but presumably
at a different rate. Indeed, if one is interested in the behavior of a
particular mode, say, that of the galaxy scale  $k_G$, one can regard this
one mode as the system and the rest as the environment. At all times this
one mode is interacting with the rest of the
whole spectrum, some of these leave the DS horizon earlier, some
later. The behavior of this particular mode determines at a
critical time  $t^*_G$ (when it leaves the DS horizon) its own amplitude
when it reenters the RW horizon, but it is influenced by all the
lower and higher modes at earlier times.
The effect of coarse-graining for this choice of the system (single mode) is
similar to the minisuperspace problem in quantum cosmology. It
is expected that the coupling constants will not run in these circumstances. 
Even in more general cases it may also be that the changing
$\lambda$ as experienced by
modes in the system may not exert any effect on galaxy formation,
because at the moment any kth mode departs the de Sitter horizon it will
have the same amplitude no matter what history of interaction
it has experienced, and it is this amplitude which determines the
density contrast for galaxy formation in the RW era. This point, however,
raises an interesting theoretical issue.

\subsection{Theoretical Implications}

We see that different choices of
the integration  range in the calculation of the CGEA give rise to
very different behavior of the coupling constants. In critical phenomena we stipulated
     the system to be in the same way as the real space renormalization
     group treatment of critical phenomena, i.e., that the ultraviolet
     cutoff is fixed at the start of inflation-- this corresponds to
     the inverse of the lattice size, which is a fixed value for
     condensed matter considerations. In each iteration the UV cutoff is
     redshifted, making the effective range of frequency smaller at
     successive periods. In critical phenomena scaling is an
     artificial procedure to facilitate the approach to the infrared
     limit. However, in inflation every different time interval
     corresponds to a realistic physical situation. If one adopts
     the coarse-graining scheme similar to critical phenomena, one would
     get a  diminishing range for the system as the universe expands.
     This is  the cause of the changing coupling constant, and is also
     the source of a serious problem. Imagine that if one
     turns the problem around by demanding that the physics observed in
     each of the subsequent moments should be identical, specifically,
     if it possesses the full range from the Planck scale to infinity (k
     from 0 to inverse Planck length)-- this is certainly what one usually
     assumes, otherwise the physical range we observe in the
     post-inflationary era such as in today's
     universe would only consists of the low k modes--if one makes this
     reasonable assertion, then we would not have the running coupling
     constants problem. This is equivalent to assuming that the physical
     wavemodes  $p=k/a$ rather than the intrinsic wavemodes k has the
     integration range between  zero and the inverse Planck length.
     However, making such an adjustment in order to make the range
     identical at each subsequent moment would  require a mechanism to
     `replenish' the high frequency modes
     between the redshifted UV limit and the Planck length, which would
     open up another serious problem. Notice that
     this problem is not particular to inflation nor curved spacetime.
     It is already there for any theory in a dynamical setting, e.g., in
     cosmology. It only becomes more serious in inflation because
     the redshifting is exponential. Similar questions would arise in
     black holes when one tries to compare what is observed between
     different observers from the event horizon to asymptotic infinity.
     The relation between black hole and inflation can be easily
     understood by viewing them in terms of
     exponential scaling, as in the dynamical finite size effects
 \cite{Edmonton}. 

One can analyze this problem in the context of the renormalization group
theory by using an explicit cutoff (defined at one instant as the Planck
scale). By demanding that this cutoff be the same at all instants and at
all energies, one would presumably not get any flow. This would in turn 
demand some theory on the physical consequence of  transPlanckian frequencies
\footnote{This issue raised originally in the context of inflationary cosmology
in \cite{cgea} has a parallel in black holes, which was pursued independently by Jacobson,
Unruh and others. For a discussion of trans-Planckian modes problems and explorations 
for its implication for quantum gravity, see, 
e.g., \cite{Jacobson}.}

\vskip 1cm 
{\Large{\bf PART THREE.  Backreaction and Noise from CTP- CGEA}}

\section{The Closed-Time-Path Coarse- Grained Effective Action}

As we introduced the CGEA in the `in-out' version \cite{cgea} we remarked that
for a correct treatment of backreaction problems, the `in-in' 
or closed-time-path CTP version of CGEA is the right way to proceed, 
as it gives a real and causal 
equation of motion for the effective dynamics of the open system.  
The CTP CGEA was first  introduced by  Hu and Sinha \cite{SinhaHu} 
to analyze the validity of the
minisuperspace approximation in quantum cosmology. 
The influence functional method \cite{if} for interacting quantum fields  
(based on quantum Brownian motion \cite{qbm}) was introduced  by 
Hu, Paz and Zhang in  \cite{HPZBelgium,ZhangPhD,Banff} 
and Calzetta and Hu \cite{CH94},
and applied to a range of problems in semiclassical gravity and 
cosmology such as
decoherence, structure formation, entropy generation and reheating  
\cite{CH95,RamseyHu,Matacz,Koks}.
Lombardo and Mazzitelli \cite{LomMaz}  computed perturbatively 
the CTP CGEA for a 
self-interacting theory in a Robertson-Walker spacetime  to discuss
the quantum to classical transition of the low modes of the field, and
to put on firmer grounds the quantum theory of structure formation in inflationary
models. Greiner and M\"uller \cite{GreMul} also used the 
CTP CGEA to analyze
the classical limit of the soft modes of a  
quantum field when the hard modes are 
in thermal equilibrium.
Dalvit and Mazzitelli \cite{DalMaz}  showed that it is possible to 
write an exact 
equation for the dependence of the CGEA on the scale that separates the
system and the environment. This is expected to be useful for a 
nonperturbative calculation of the CGEA, and to discuss the 
appearance of noise in the 
renormalization group equations.

In this part
we will review the above mentioned works. We will start 
with the necessary definitions and the perturbative evaluation of the 
CTP CGEA for a
${\lambda \phi^{4}}$ scalar field in a Friedmann-Robertson-Walker 
(FRW) background spacetime. We will describe how to use the CTP CGEA
to analyze the validity of the minisuperspace approximation in 
quantum cosmology
and the issue of quantum to classical transition of  fluctuations 
which give rise to structure formation in inflationary models. We will
finally describe  attempts to compute the CTP CGEA using
non perturbative approximations.

\subsection{Perturbative evaluation of the CTP CGEA }

We consider again the scalar field action given by
Eq.(\ref{3phiact}).
In a flat Robertson-Walker
spacetime with metric $ds^2=dt^2-a^2(t)d\vec x^2=a^2(\eta)[d\eta^2-
d\vec x^2]$, the
action can be written as
\begin{equation}
S(a,\chi)=\int d^4x [{1\over 2}\eta^{\mu\nu}\partial_{\mu}\chi
\partial_{\nu}\chi-{1\over 2}m^2 a^2\chi^2-{1\over 2}(\xi-{1\over 6})R a^2 
\chi^2-{\lambda\over 4!}\chi^4]
\label{chiaction}
\end{equation} 
where $\chi =a\phi$. From now on, $d^4x$ denotes $d^3x\,\,d\eta$.
To see the flat space results simply set $a=1$ and $R=0$.
Let us make a system-environment field splitting 
\begin{equation}
\chi(x) = \chi_<(x) + 
\chi_>(x),\label{splitting}
\end{equation} where we define the system by 
\begin{equation}
\chi_<(\vec x, t) = \int_{\vert \vec k\vert < \Lambda_c} 
{d^3\vec k\over{(2 \pi)^3}} \phi(\vec k, \eta) \exp{i \vec k . \vec 
x},\label{sys}
\end{equation} 
and the environment by 
\begin{equation}
\chi_>(\vec x, \eta) = \int_{\vert \vec k\vert > \Lambda_c} 
{d^3\vec k\over{(2 \pi)^3}} \phi(\vec k, \eta) \exp{i \vec k . \vec 
x}.
\label{env}
\end{equation} 
The system-field contains the modes with 
wavelengths longer than the critical value $\Lambda_c^{-1}$, while the bath or 
environment-field contains wavelengths shorter than $\Lambda_c^{-1}$.
$\L_c$ corresponds to ${\L\over s}$ of previous sections.      

After the splitting, the total action (\ref{chiaction}) can be written as
\begin{equation}
S[a,\chi] = S_0[\chi_<] + S_0[\chi_>] + S_{int}[a,\chi_<, 
\chi_>],
\label{actions}
\end{equation}
where $S_0$ denotes the kinetic term and the interaction part 
is given by
\begin{eqnarray}
&& S_{int}[a,\chi_<, \chi_>] = - \int d^4x 
\{ {M^2\over 2}\chi_<^2 + {\lambda\over{4!}}\chi_<^4 
 +{M^2\over 2}\chi_>^2 + {\lambda\over{4!}}\chi_>^4 + 
\nonumber\\ && {\lambda\over{4}} \chi_<^2(x) \chi_>
^2 + {\lambda\over{6}} \chi_<^3
\chi_> + {\lambda\over{6}} \chi_< 
\chi_>^3\}.
\label{inter}
\end{eqnarray}
with $M^2 = m^2 a^2 + (\xi -{1\over 6}) R a^2$.

We are interested in the influence of the environment 
on the evolution of the 
system. Therefore the CTP CGEA $S_{\L}[a,\chi_<,a',\chi_<']$  is 
the object of relevance. It is defined by 
\begin{eqnarray}
&&
\exp {i S_{\Lambda_c}[a,\chi_<,a',\chi_<']}
=\exp {i (S_0[\chi_<] - S_0[\chi_<'])}\int d\chi_{>f}
\int^{\chi_{>f}}{\cal D}\chi_>\int^{\chi_{>f}}{\cal 
D}\chi_>'\nonumber \\
&&\times 
\exp {i \{S_0[\chi_>] + S_{int}[a,\chi_<,\chi_>] - 
S_0[\chi'_>] - S_{int}[a',\chi'_< , \chi'_>]\} }.
\end{eqnarray}
The integration is over all fields $\chi_>$ (and $\chi_>'$) with
positive (and negative) frequency modes in the remote past
that coincide at the final time $\chi_>=\chi_>'=\chi_{>f}$

We will calculate the CTP CGEA perturbatively in $\lambda$
and $M^2$, up to quadratic order in both quantities.
After a simple calculation we obtain 
\begin{eqnarray}
&& S_{\L_c}[a,\chi_<,a',\chi_<'] =  S_0[\chi_<] - S_0[\chi_<'] + \{\langle 
S_{int}[a,\chi_<,\chi_>]\rangle_0 - \langle 
S_{int}[a',\chi_<',\chi_>']\rangle_0\}\nonumber \\
&&+{i\over{2}}\{\langle S_{int}^2[a,\chi_<,\chi_>]\rangle_0 - \big[\langle 
S_{int}[a,\chi_<,\chi_>]\rangle_0\big]^2\}\nonumber \\
&&- i\{\langle S_{int}[a,\chi_<,\chi_>] S_{int}[a',\chi_<',\chi_>']\rangle_0 - 
\langle S_{int}[a,\chi_<,\chi_>]\rangle_0\langle 
S_{int}[a',\chi_<',\chi_>']\rangle_0\}  \nonumber\\
&&+{i\over{2}}\{\langle S^2_{int}[a',\chi_<',\chi_>']
\rangle_0 - 
\big[\langle 
S_{int}[a'\chi_<',\chi_>']\rangle_0\big]^2\},\label{inflac}
\end{eqnarray} 
where the quantum average of a functional of the fields is defined 
with respect to the free action $S_0$
\begin{equation}\langle B[\chi_>,\chi_>'] \rangle_0= 
\int d\chi_{>f} 
\int^ {\chi_{>f}}{\cal D}\chi_>\int^{\chi_{>f}}{\cal 
D}\chi_>' \exp {\{S_0[\chi_>] - S_0[\chi_>']\}} 
B.\label{averag}
\end{equation} 
Equation (\ref{inflac}) is the {\it in-in} version of the
Dyson-Feynman series (\ref{3.20}).
   
We define the propagators of the environment field as
\begin{equation}
\langle \chi_>(x)\chi_>(y)\rangle_0 = i 
G_{++}^{\Lambda_c}(x-y),\label{feyn}
\end{equation}
\begin{equation}
\langle \chi_>(x)\chi_>'(y)\rangle_0 = - i G_{+-}^{\Lambda_c} 
(x-y),\label{frecmas}
\end{equation} 
\begin{equation}
\langle 
\chi_>'(x)\chi_>'(y)\rangle_0 = - i G_{--
}^{\Lambda_c}(x-y).\label{dyson}
\end{equation} 
These propagators are not the usual Feynman, positive-frequency Wightman, and 
Dyson propagators of the scalar field since, in this case, the momentum 
integration is restricted by the presence of the (infrared) 
cutoff $\Lambda_c$. 
The explicit expressions are  
\begin{equation}G_{++}^{\Lambda_c} (x-y)= \int_{\vert \vec p\vert > \Lambda_c} 
{d^4p\over{(2 \pi)^4}} e^{i p (x - y)} {1\over{p^2 + i 
\epsilon}},\label{feypro}\end{equation} 
\begin{equation}G_{+-}^{\Lambda_c} (x-y) = -\int_{\vert \vec p\vert > 
\Lambda_c} 
{d^4p\over{(2 \pi)^4}} e^{i p (x - y)} 2 \pi i 
\delta (p^2) \Theta(p^0),\label{frecmaspro}\end{equation}
\begin{equation}G_{--}^{\Lambda_c} (x-y) = \int_{\vert \vec p\vert > 
\Lambda_c} 
{d^4p\over{(2 \pi)^4}} e^{i p (x - y)} {1\over{p^2  - i 
\epsilon}}.\label{dysonprop}\end{equation} 
As an example,  we show the 
expression for the propagator $G^{\Lambda_c}_{++}$.
The usual Feynman propagator is
\begin{equation}G_{++}(x)={i\over{8 \pi^2}}{1\over{\sigma}} - 
{1\over{8 \pi}} 
\delta (\sigma),\nonumber \end{equation}
while 
\begin{eqnarray}
&& G^{\Lambda_c}_{++}(x) = {i\over{8 \pi^2}}
\Big[{cos[\Lambda_c (r - x^0)]\over{r 
(r - 
x^0)}} + {cos[\Lambda_c (r + x^0)]\over{r(r+x^0)}}\Big] 
-{1\over{8 \pi^2}}\Big[{sin[\Lambda_c (r - x^0)]\over{r 
(r - x^0)}} - {sin[\Lambda_c 
(r + x^0)]\over{r (r + x^0)}}\Big]\nonumber \\ && \equiv G_{++}(x) 
- G_{++}^{\vert 
\vec p\vert <\Lambda_c}(x), \end{eqnarray} 
where $\sigma = {1\over{2}}x^2$. 

The CTP CGEA can be computed from Eqs. (\ref{inflac})-(\ref{dyson}) 
using standard techniques. After some algebra we 
find 
\begin{eqnarray} 
&& S_{\L_c}[a,\chi_<, a',\chi'_<]= S_0(\chi_<) - S_0(\chi'_<)
-{1\over 2}\int d^4x (M^2(x)\chi_<^2 - M^{'2}\chi_<^{'2}) \nn\\ &&
 -\lambda \int d^4x 
\Big[{1\over{24}} (\chi^4_<(x) - \chi_<^{'4}(x)) 
+ {1\over{2}} i 
G_{++}^{\Lambda_c}(0) (\tilde M^2(x) - \tilde M^{'2}(x))\Big]\nonumber \\ &&+ 
\int d^4x \int d^4y \Big[-{\lambda^2\over{72}} \chi_<^3(x) G_{++}^ 
{\Lambda_c}(x-y) \chi_<^3(y) - {\lambda^2\over{36}} \chi^3_<(x) G_{+-}
^{\Lambda_c}(x-y) 
\chi_<^{'3}(y)\nonumber \\ && + {\lambda^2\over{72}} \chi_<^{'3}(x) G_{--
}^{\Lambda_c}(x-y) \chi_<^{'3}(y) - {1\over{4}} 
\tilde M^2(x) i G_{++}^{\Lambda_c 
2}(x-y) \tilde M^2(y) \nonumber \\ && + {1\over{2}}\tilde 
M^2(x)i G_{+-}^{\Lambda_c 
2}(x-y) \tilde M^{'2}(y)-{1\over{4}}\tilde M^{'2}(x) i G_{--}^{\Lambda_c 2} 
(x-y)\tilde M^{'2}(y)\nonumber \\ && + {\lambda^2\over{18}} 
\chi_<(x) G_{++}^{\Lambda_c 
3}(x-y)\chi_<(y) + {\lambda^2\over{9}} \chi_<(x) G_{+-}^{\Lambda_c 
3}(x-y)\chi_<'(y)\nonumber \\ && - {\lambda^2\over{18}}\chi_<'(x) 
G_{--}^{\Lambda_c 
3}(x-y) \chi'_<(y) \Big],\end{eqnarray} 
where we introduced the notation $\tilde M^2 =M^2 +{\lambda\over 2}\chi_<^2$.

Defining $$P_\pm ={1\over{2}}(\chi^4_< 
\pm \chi'^4_<) ~~~;~~~ R_\pm ={1\over{2}}(\chi_<^3 \pm \chi'^3_<)$$ 
$$\lambda Q_\pm 
={1\over{2}}(\tilde M^2 \pm \tilde M^{'2}) ~~~;~~~ \chi_\pm ={1\over{2}}
(\chi_< \pm 
\chi'_<),$$ 
and using simple identities for the propagators, the 
real and imaginary parts of the 
CTP CGEA can be written as 
\begin{eqnarray} 
&& Re S_{\L_c} = S_0(\chi_<) - S_0(\chi'_<)  - \lambda \int 
d^4x\{{1\over{12}}P_-(x) + {i\over{2}}G_{++}^{\Lambda_c}(0) 
Q_-(x)\}\nonumber \\ 
&&+ \lambda^2 \int d^4x\int d^4y\Theta(y^0-x^0) 
\{-{1\over{18}} R_+(x) Re G^{\Lambda_c}_{++}(x-y) 
R_-(y) + \nn\\ &&{1\over{4}} Q_+(x) Im G^{\Lambda_c 2}_{++}(x-y) Q_-(y)+ 
{1\over{3}} \chi_+(x) Re G^{\Lambda_c 3}_{++}(x-y) \chi_-(y)\},
\label{inff}\\
&& Im S_{\L_c}  =
\lambda^2 \int 
d^4x \int d^4y \{-{1\over{18}} R_-(x) Im G^{\Lambda_c}_{++}(x-y) R_-(y)
\nonumber 
\\ &&-{1\over{4}} Q_-(x) Re G^{\Lambda_c 2}_{++}(x-y) Q_-(y) + {1\over{3}} 
\chi_-(x) Im G^{\Lambda_c 3}_{++}(x-y) \chi_-(y)\}. 
\label{inff2}
\end{eqnarray} 

The real part of the 
CTP CGEA in Eq.\ (\ref{inff}) contains divergences and must
be renormalized. As the propagators in Eqs.\ (\ref{feyn})-(\ref{dyson})
differ from the usual ones only by the presence of the infrared
cutoff, the ultraviolet divergences coincide with those of
the usual $\lambda\chi^4$-theory. The 
effective action can therefore 
be renormalized using the standard procedure.

Consider  the square of the Feynman propagator.
Using dimensional regularization we find
\begin{equation}G^{\Lambda_c 2}_{++}(x) = G_{++}^2(x) + G_{++}^{(\vert \vec 
p\vert <\Lambda_c) 2}(x) - 2 G_{++}(x) G_{++}^{(\vert \vec p\vert 
<\Lambda_c)}(x),\end{equation}
where
\begin{equation}G_{++}^2(x)={i\over{16 \pi^2}}[{1\over{n-4}}+\psi(1) - 4 
\pi]\delta^4(x)+i\Sigma(x) - \eta(x) - Log[4 \pi 
\mu^2],
\label{g2}
\end{equation}
\begin{equation}\Sigma (x)={1\over{(2 \pi)^4}}\int d^4p e^{i p x} Log \vert 
p^2\vert,\nonumber\end{equation}
\begin{equation}\eta (x)={\pi\over{(2 \pi)^4}}\int d^4p e^{i p x} \Theta 
(p^2).\nonumber\end{equation}
Note that the divergence is the usual one, i.e., proportional
to ${1\over n-4}\delta^4(x-y)$ and independent of $\Lambda_c$.
Consequently, the term $Im G_{++}^{\Lambda 2}(x-y)Q_+(x)Q_-(y)$
in Eq.\ (\ref{inff}) is also divergent and renormalizes the coupling constant
$\lambda$ and the constants that appear in the gravitational action
(as usual, in order to renormalize the theory of a quantum field in
a curved space, it is necessary
to include in the gravitational action the Einstein Hilbert term,
a cosmological constant, and terms quadratic in the curvature tensor).
The other divergences can be treated in a similar way.
One can also check that the imaginary part of the effective
action does not contain divergences.\footnote{Of course, a
successful ultraviolet renormalization does not
guarantee that an approximation scheme such as RG-improved
perturbation theory will be well behaved. A good example
is in Section V, where loops depend on a factor
$ e^{H(t-t_{0})}$ which would invalidate perturbation theory.
Further `infrared' $H$-dependent (environmentally friendly)
renormalization of $\lambda$ is needed.}

As  we will show with the help of QBM models, the (nonlocal) 
real and imaginary parts of $ S_{\L_c}[a,\chi_<,a',\chi_<']$  
can be associated 
with the dissipation and  noise respectively, and can be
related by an integral equation known as the 
fluctuation-dissipation relation. 

Similar perturbative results for the CTP CGEA have been obtained
by Greiner and M\"uller in Ref. \cite{GreMul}. In order to derive
effective field equations for the soft modes of the scalar field,
they computed the CTP CGEA assuming that the modes in the environment
are at thermal equilibrium at a temperature $T$ such that
$\L_c\ll T$. Their result for the effective action is essentially
given by expressions (\ref{inff}) and (\ref{inff2}), for the particular
case $a(t)=1$ and replacing the vacuum propagators
by thermal propagators in order to take
into account the state of the environment. 

In Section VII.B we will see that the field equations
derived from the CTP CGEA are real and causal.

\subsection{Applications: Dynamics of  System Fields/Modes with
Backreaction of Environment Fields/Modes}

The coarse-grained effective action is a very useful method to treat 
coarse-graining and  backreaction problems.
Examples include stochastic inflation \cite{HPZBelgium,CH95,LomMaz}
and reheating  in inflationary cosmology \cite{RamseyHu,BoyDeV}, 
effect of hard  thermal loops in  QCD plasma, and
 the effect of individual atoms on a BEC condensate. 
The first instances
the CTP effective action was applied to were for backreaction of 
quantum fluctuations
and particle creation on the background spacetime \cite{CH87,Jordan} 
or 
for interacting quantum fields \cite{CH89,Paz90}. The CTP CGEA was first 
introduced
\cite{SinhaHu} to address the validity of the so-called minisuperspace 
approximation. 
Although  it was introduced in the context of quantum cosmology, 
the method
has a wide  range of applications. The relevant issues are
the { backreaction} effect of the  inhomogeneous modes in an  
interacting  quantum field on the homogeneous mode; and 
the validity of quantizing  a truncated theory: does it preserve the 
salient features of a fully quantized theory?
These issues underlie  many problems in physics, especially  when we view 
effective theories
as playing a more fundamental  role in the description of Nature  
\cite{Weinberg}.  

\vskip 1cm 


\section{Master and Langevin Equations in Quantum Field Theory}

We now  show how  to use the open system concepts and techniques to give a first principles 
derivation of the evolution equation for the reduced density matrix that describes
the system under the influence of the environment. We will
use the influence functional method to extract the noise and dissipation
kernels. 
The open system framework is best illustrated by the quantum mechanical
problem of Brownian motion (QBM). Feynman and Vernon \cite{if} first treated
this problem with the influence functional method. Using this method, 
the master equation was derived by Caldeira and Leggett  
\cite{qbm} for  Markovian processes
(Ohmic environment at high temperature), and by  Hu Paz and Zhang \cite{HPZ} 
for a general environment including  nonMarkovian processes. 
(See also \cite{Ingold}
and \cite{UnrZur}). As we will soon find out, the influence functional 
method is intimately related
to  the CTP  coarse-grained effective action  \cite{Chen,CH94}.

Hu, Paz and Zhang \cite{ZhangPhD,HPZBelgium} first  generalized the
quantum mechanical problem of Brownian motion to quantum fields,
taking  the system and environment as two independent fields. Lombardo and
Mazzitelli \cite{LomMaz} treated the same problem, taking as system and environment 
the low and high frequency modes of a single, self-interacting scalar field. 
We will follow their treatment here.
 
We begin with a  brief review of the problem of quantum Brownian motion.
Denote by $x$ the coordinate
of the Brownian particle, by $\Omega$ its frequency,
and by $q_i$ the coordinates of the oscillators in the 
environment. The influence of the environment on the 
Brownian particle can be described by the reduced density matrix  
$\rho_r(x,x',t)$ that is obtained from the full density matrix
by integrating out the environmental degrees of freedom $q_i$.

For a linear coupling $x q_i$,
the  master equation for the reduced
density matrix $\rho_r(x,x',t)$ is of the form \cite{jppshort}
\begin{eqnarray}
&& i\hbar \partial_t \rho_r(x,x',t)=\langle x\vert[H,\rho_r]\vert x'\rangle
-i\gamma(t)(x-x')(\partial_x - \partial_{x'})\rho_r(x,x',t)\nonumber\\
&&+ f(t)(x-x')(\partial_x + \partial_{x'})
\rho_r(x,x',t) - iD(t)(x-x')^2\rho_r(x,x',t),
\label{meqbm}
\end{eqnarray}
where the coefficients $\gamma (t), D(t)$ and $f(t)$ depend 
on the properties of the environment (temperature $\beta^{-1}$
and spectral density $I(\omega)$). The first term on the RHS of 
Eq. (\ref{meqbm}) gives the usual Liouvillian
evolution. The second is a dissipative term with a time dependent  dissipative coefficient $\gamma (t)$. The last two are 
diffusive terms. The one proportional to the anomalous diffusion coefficient
$f(t)$ does not produce decoherence, i.e., the off-diagonal terms
in the density matrix are not suppressed in time. To see this in a simple
example, assume that $\rho_r=\rho_0 e^{-B(t)(x-x')^2}$. Inserting
this into Eq. (\ref{meqbm}) it is easy to prove that the $f-$term produces
an oscillating function $B(t)$ but no damping. 
(We refer the reader to Refs. \cite{jppshort,diana} for a more
detailed justification). The last term with  diffusion coefficient $D(t)$ 
\begin{equation}
D(t)=\int_0^t ds~~cos (\Omega s)~~\int_0^{\infty}d\omega
~~I(\omega)coth({1\over 2}\beta\hbar\omega)~~cos(\omega s),
\label{dt}
\end{equation}
gives the main contribution
to decoherence. Indeed, an approximate solution
of Eq.\ (\ref{meqbm}) is \cite{HPZ,jppshort}
\begin{equation}
\rho_r[x,x';t] \approx \rho_r[x,x',0] \exp{\Big[-(x - 
x')^2\int_0^t D(s) ds}\Big],\label{qbmdecay}\end{equation}
and we see that the off-diagonal terms of the density matrix
are suppressed as long as $\int_0^t D(s) ds$ is large enough.
For non-linear couplings like $x^nq_i^m$, one expects 
the master equation to contain  
terms of the form
$iD^{(n,m)}(t)(x^n-x'^n)^2\rho_r$.

We now proceed to quantum field theory.
The total density matrix (for the system and environment fields) is defined
by 
\begin{equation} \rho[\chi_<,\chi_>,\chi_<',\chi_>',\h]=\langle\chi_< 
\chi_>\vert {\hat\rho} \vert \chi_<' 
\chi_>'\rangle,\label{matrix}
\end{equation}
where $\vert \chi_<\rangle$ and $\vert \chi_>\rangle$ are the eigenstates of 
the field operators ${\hat\chi}_<$ and ${\hat\chi}_>$, respectively.
For simplicity, we will assume that 
the interaction is turned on
at the initial time $\h_0$ and that, at this time, the system and the environment are 
not correlated. (The physical consequences of such a choice is elaborated in \cite{HPZ}. 
More general initial conditions can be introduced by a stipulated preparation function 
\cite{Pazpre}.)
As such, the total density matrix can be written as the product of 
the density matrix operator for the system and for the bath 
\begin{equation}{\hat\rho}[\h_0] = {\hat\rho}_{<}[\h_0] 
{\hat\rho}_{>}[\h_0].\label{sincorr}\end{equation} 
We will further assume that 
the initial state of the environment is the vacuum. 

The reduced density matrix is defined by 
\begin{equation}\rho_{red}[\chi_<,\chi'_<,\h] = \int {\cal D}\chi_> 
\rho[\chi_<,\chi_>,\chi_<',\chi_>,\h].\label{red}
\end{equation}
and evolves in time according to
\begin{equation}\rho_r[\chi_{<f},\chi_{<f}',\h] = \int d\chi_{<i}\int 
d\chi_{<i}' J_r[\chi_{<f},\chi_{<f}',\h\vert \chi_{<i},\chi_{<i}',\h_0] 
\rho_r[\chi_{<i}\chi_{<i}',\h_0],\label{evol}
\end{equation} 
where $J_r[t,\h_0]$ is the reduced evolution operator 
\begin{equation}J_r[\chi_{<f},\chi_{<f}',\h\vert \chi_{<i},\chi_{<i}',\h_0] = 
\int_{\chi_{<i}}^{\chi_{<f}}{\cal D}\chi_< \int_{\chi'_{<i}}^{\chi_{<f}'}{\cal 
D}\chi_<' \exp{{i\over{\hbar}}\{S[\chi_<] - S[\chi_<']\}} 
F[\chi_<,\chi_<'].\label{evolred}
\end{equation}
The  influence functional (or Feynman-Vernon functional) 
$F[\chi_<,\chi_<']$ is defined as 
\begin{eqnarray}
&& F[\chi_<,\chi_<']= \int d\chi_{>i} \int 
d\chi_{>i}' \rho_{>}[\chi_{>i},\chi_{>i}',\h_0] \int d\chi_{>f}
\int_{\chi_{>i}}^{\chi_{>f}}{\cal D}\chi_>\int_{\chi_{>i}'}^{\chi_{>f}}{\cal 
D}\chi_>'\nonumber \\
&&\times \exp {{i\over{\hbar}}\{S[\chi_>] + S_{int}[\chi_<,\chi_>] - 
S[\chi'_>] - S_{int}[\chi'_< , \chi'_>]} \}.
\end{eqnarray}
We see that when the environment is initially in its vacuum state, 
the influence 
functional coincides with the CTP CGEA \cite{Chen,CH94}. Treatment of
two self-interacting fields via the influence functional leading to noise
is contained in \cite{Banff}

As the propagator $J_r$ is defined 
through a path integral, we can obtain an estimation using the saddle point 
approximation: 
\begin{equation}
J_r[\chi_{<f},\chi'_{<f},\h\vert\chi_{<i},\chi'_{<i},\h_0] 
\approx \exp{ {i\over{\hbar}} S_{\L_c}[\chi_<^{cl},\chi_<^{'cl}]},
\label{prosadle}
\end{equation}
where $\chi_<^{cl} (\chi_<^{' cl})$ is the solution of the equation of motion
${\delta Re S_{\L_c}\over\delta\chi_<}\vert_{\chi_<=\chi'_<}=0$
with boundary conditions $\chi_<^{cl}(\h_0)=\chi_{<i}(\chi'_{<i})$
and $\chi_<^{cl}(\h)=\chi_{<f} (\chi'_{<f})$.
This formula enables us to analyze the quantum to classical transition
of the system field using the perturbative evaluation of the CTP CGEA
described in the previous section.

\subsection{ Master equation and decoherence of long wavelengths}

The decoherence effects are contained in the imaginary part 
of the CTP CGEA, which is already of order $\lambda^2$.
As a consequence, in the evaluation of $Im S_{\L_c}$ 
we can approximate $\chi_<^{cl}$ 
by the solution of the free field equation satisfying the appropriate boundary 
conditions. For simplicity we will consider massless, 
conformally coupled fields. Therefore the classical solution reads 
\begin{equation}\chi_<^{cl}(\vec x, s) =  [\chi_{<f} {sin(k_0 s)\over {sin(k_0 
\h)}} + \chi_{<i} {sin[k_0 (\h - s)]\over{sin(k_0 \h)}}] cos(\vec k_0 . \vec 
x)\equiv \chi_<^{cl}(s) cos(\vec k_0 . \vec 
x)
,\label{classicphi}\end{equation} where we assumed that the system-field 
contains only one Fourier mode with $\vec k = \vec k_0$. This is a sort of 
``minisuperspace" approximation for the system-field that will greatly 
simplify the calculations. 

As in the QBM problem, we can derive the diffusion coefficients
from the master equation, which would provide relevant information 
about decoherence.  Hu Paz and Zhang \cite{ZhangPhD,HPZBelgium} derived
the noise kernel of two interacting $\phi^4$  quantum fields  in de 
Sitter space
and analyzed its behavior in relation to decoherence.  
Lombardo and Mazzitelli \cite{LomMaz} did the same,  with one self-interacting field split into two mode sectors. They  followed the method proposed in \cite{jppshort}, 
and computed the time derivative of the 
propagator $J_r$, eliminating the dependence on the initial field 
configurations $\chi_{<i}$ and $\chi'_{<i}$ that enters through 
$\chi_{<}^{cl}$ and $\chi_<^{'cl}$. 
The master equation is of the form
\begin{eqnarray}
&& i \hbar \partial_\h \rho_r  [\chi_{<f},\chi'_{<f},\h] = 
\langle \chi_{<f}\vert\Big[{\hat H}_{ren}, 
{\hat\rho}_r\Big]\vert\chi'_{<f}\rangle 
- i \lambda^2  [{(\chi_{<f}^3 
- \chi_{<f}^{'3})^2 V\over{1152}}D_1(k_0;\h )\nonumber \\
 && + {(\chi_{<f}^2 - 
\chi_{<f}^{'2})^2 V\over{32}}D_2(k_0;\h ) - {(\chi_{<f} -
\chi'_{<f})^2 V\over{6}}D_3(k_0;\h )] \rho_r[\chi_{<f},\chi'_{<f},\h ] + .... 
\label{master}
\end{eqnarray} 
Due to the complexity of the  equation,
we only show the correction to the usual unitary evolution term 
coming from the noise kernels.

This equation contains three time-dependent 
diffusion coefficients $D_i(\h )$. Up to one loop, 
only $D_1$ and $D_2$ survive 
and are given by 
\begin{eqnarray}&& D_1(k_0;\h )= \int_0^t ds ~~ cos^3(k_0 s)Im 
G_{++}^{\Lambda_c}(3k_0;\h -s)\nonumber \\ &=& {1\over{6 k_0}}\int_0^t ds ~~ 
cos^3(k_0 s) ~~ cos(3 k_0 s) ~~ \theta(3 k_0 - \Lambda_c)\nonumber \\  &=& {2 
k_0 \h + 3 sin(2 k_0 \h) + {3\over{2}} sin(4 k_0 \h ) + {1\over{3}} sin(6 k_0 
\h )\over{576 k_0^2}},~~~~ {\Lambda_c\over{3}}<k_0<\Lambda_c\label{d1}
\end{eqnarray} 

\begin{equation}D_2(k_0;\h ) = \int_0^\h ds ~~ cos^2(k_0 s)(Re G_{++}^
{\Lambda_c 2}(2k_0;\h -s)+ 
2Re G_{++}^{\Lambda_c 2}(0;\h -s)).\label{d2prev}\end{equation} 
Using that
\begin{eqnarray}Re G_{++}^{\Lambda_c 2}(2k_0;\h -s)&=& 
{\pi\over{k_0}}\bigg\{ \int_{\Lambda_c}^{2k_0+\Lambda_c}
dp\int_{\Lambda_c}^{2k_0 + p}dz 
cos[(p + z)s]\nonumber \\
&& + \int_{2k_0+\Lambda_c}^{\infty}dp \int_{p-2k_0}^{p+2k_0}dz 
cos[(p+z)s]\bigg\},\end{eqnarray} 
\begin{equation}Re G_{++}^{\Lambda_c 2}(0;\h -s)= 
\pi\bigg\{2\pi \delta(s) - 2 {sin(2 \Lambda_c 
s)\over{s}}\bigg\},\end{equation} 
the $D_2$ diffusion coefficient can be written as 
\begin{eqnarray}D_2(k_0;\h ) &=& {\pi\over{4}}\bigg\{3 \pi - ({3\over{2}} - 
{\Lambda_c\over{2k_0}})Si[2 \h (\Lambda_c - k_0)]\nonumber \\ && - (2 - 
{\Lambda_c\over{2 k_0}}) Si[2 \Lambda_c \h ] - ({3\over{2}} + 
{\Lambda_c\over{2k_0}})Si[2 \h (\Lambda_c + k_0)] -(1 +
{\Lambda_c\over{2k_0}})Si[2\h (2k_0 + \Lambda_c)]\nonumber 
\\ &&+ {cos[2\Lambda_c 
\h ]\over{4k_0t}} - {cos[2 \h (\Lambda_c + k_0)]\over{4k_0\h }}
+{cos[2 \h (\Lambda_c - 
k_0)]\over{4k_0\h }}-{cos[2 
\h (2k_0+\Lambda_c)]\over{4k_0\h }}\bigg\},\label{d2}\end{eqnarray} 
where $Si[z]$ denotes  the sine-integral function \cite{abramo}.

Eq.\ (\ref{master}) is the field-theoretical version of the QBM master 
equation we were looking for.
In our case, the system is coupled in a nonlinear form. Owing 
to the existence of three interaction terms ($\chi_<^3 \chi_>$, 
$\chi_<^2 \chi_>^2$, and $\chi_< \chi_>^3$) there are three diffusion 
coefficients in the master equation. The form of the 
coefficients is fixed by these couplings and by the particular
choice of the quantum state
of the environment.

Our results are valid in the single-mode approximation of 
Eq.\ (\ref{classicphi}). In this approximation one obtains a reduced density 
matrix for each mode $\vec{k}_0 $, and neglects the interaction 
between different system-modes. Due to this interaction, in the general case, 
$\rho_r$ will be different from $\prod_{\vec k_0} \rho_r(\vec k_0)$. 
This  point deserves further study.  

A detailed
analysis of the quantum to classical transition in the
model we are considering is a very complicated task.
One should analyze in detail the master equation and 
see whether the off-diagonal elements of the reduced density 
matrix are suppressed or not. One should also
study the form of the Wigner function, and see
whether it predicts classical correlations or not \cite{HabLaf,HalYu}
 Having in mind the analogy with the QBM, here we will only be concerned 
with the diffusive terms of the master equation.
By examining how the value of the diffusion coefficients change in time,
one can get an indication of how effective  decoherence is 
\cite{HPZBelgium,LomMaz}.  
This simplified method  can only provide a rough approximation.

A mode in the system will decohere if the diffusion coefficients are different 
from zero during an appreciable period of time. Therefore we ask the 
following question: which is the maximum
value of the cutoff $\Lambda_c$ such that, a few e-foldings after the
initial time, all modes with $k_0\leq\Lambda_c$ still suffer the diffusive
effects? The value of the diffusion coefficients at a given
time depends on the value of the adimensional quantity $l=\L_c\h$.
For the particular case of a de Sitter spacetime we have
$a(t)=\exp (Ht)$ and
$$l=\Lambda_c\eta={\L_c\over H}(1 - {1\over a})$$
which, a few e-foldings after the initial time, is
approximately given by $l=\Lambda_c/H$. We can distinguish two
possibilities. When $l\sim 1$ both coefficients
$D_1$ and $D_2$ are appreciably different from zero
for all values of $k_0$. On the other hand,
when $l\gg 1$ the situation is completely different:
$D_1$ is very small everywhere while $D_2$ is also very small in the
infrared sector. The diffusive effects are important
only
for $k_0\sim\Lambda_c$. The conclusion is that, in order
to have decoherence for all modes with $k_0<\Lambda_c$, we can include
in our system only those modes with wavelength larger than 
$H^{-1}$. This is consistent with Starobinsky's original suggestion. If we 
include wavelengths shorter than $H^{-1}$, the frequency threshold
in the environment increases, the infrared sector of the system 
cannot excite the environment and therefore does not decohere.

A more
realistic calculation should include a time dependent cutoff \cite{habib},
since the system should contain, at each time, the modes with
$k_{ph}={k_0\over a}<H$. More importantly, the scalar
field should be minimally coupled ($\xi =0$) to the curvature.
In spite of this, we think that our example 
illustrates the main aspects of the problem. Indeed, one could repeat
the calculations for a given mode of the minimally coupled scalar field,
and obtain a master equation, similar to our Eq. (\ref{master}). The main
difference would be that, for $\xi =0$, the propagators in De Sitter
spacetime would differ from their flat-spacetime counteparts
Eqs. (\ref{feypro}-\ref{dysonprop}) (See Zhang's thesis \cite{ZhangPhD}). 
The evaluation of the diffusion coefficients using
curved spacetime propagators would be more realistic 
for discussions of structure formation in inflationary cosmology.

\subsection{The Langevin equation}

We now show how  to derive a stochastic (Langevin)
 equation for the system field from the CTP CGEA. This equation takes into account
 the three fundamental effects of the environment on the system:
 renormalization, dissipation and noise.

The real and imaginary parts of the CTP CGEA for our model are
 given in Eqs. (\ref{inff}) and (\ref{inff2}).                   
One can regard
the imaginary part of $S_{\L_c}$ as coming from three noise 
sources $\nu(x)$, $\xi(x)$, and $\eta(x)$ 
with a Gaussian functional probability distribution given by 
\begin{eqnarray}
&& P[\nu(x), \xi(x), \eta(x)]= N_\nu N_\xi N_\eta \exp\bigg\{
-{1\over{2}}\int d^4x\int d^4y \nu(x)\Big[{\lambda^2\over{9}} 
Im G_{++}^{\Lambda_c}\Big]^{-1}\nu(y)\bigg\}\nonumber \\
&&\times \exp\bigg\{-{1\over{2}}\int d^4x\int d^4y \xi(x)
\Big[{\lambda^2\over{2}} Re G_{++}^{\Lambda_c 2}\Big]^{-1}\Big]
\xi(y)\bigg\}\nonumber \\
&&\times \exp\bigg\{-{1\over{2}}\int d^4x\int d^4y \eta(x)\Big[{-2\lambda^2
\over{3}} 
Im G_{++}^{\Lambda_c 3}\Big]^{-1}\eta(y)\bigg\},\end{eqnarray}
where $N_\nu$, $N_\xi$, and $N_\eta$ are normalization factors. 
Indeed, we can write the imaginary part of the 
influence action as three functional integrals over the Gaussian fields 
$\nu(x)$, $\xi(x)$, and $\eta(x)$: 
\begin{eqnarray}
&& \int {\cal D}\nu(x)\int {\cal D}\xi (x) \int {\cal D}\eta(x) 
P[\nu,\xi,\eta] \exp{ -{i\over{\hbar}} \bigg\{R_-(x)\nu(x) + Q_-(x) \xi(x) 
+ \chi_-(x) \eta(x)}\bigg\}\nonumber \\
&&= \exp\bigg\{-{i\over{\hbar}}\int d^4x\int d^4y \
\Big[{\lambda^2\over{18}}R_-(x) ImG_{++}^\Lambda(x,y)R_-(y)\nonumber \\
&&+ {\lambda^2\over{4}}Q_-(x) Re G_{++}^{\Lambda 2}(x,y)Q_-(y) - 
{\lambda^2\over{3}}\chi_-(x) Im G_{++}^{\Lambda 3}(x,y)\chi_-(y)\
\Big]\bigg\}.\end{eqnarray}
Therefore, the CTP-CGEA can be rewritten as
\begin{equation}
S_{\L_c}[\chi_<,\chi_<']=-{1\over{i}} ln \int {\cal D}
\nu P[\nu]\int {\cal D}\xi P[\xi]\int {\cal D}\eta P[\eta] 
\exp\bigg\{i S_{eff}[\chi_<,\chi_<', \nu, \xi, \eta]\bigg\},\end{equation}
where 
\begin{equation}S_{eff}[\chi_<,\chi_<', \nu, \xi, \eta]= 
Re S_{\L_c}[\chi_<,\chi_<']- \int d^4x\Big[R_-(x) \nu (x) + 
Q_-(x) \xi(x) + \chi_-(x)\eta(x)\Big].\end{equation}

From this effective action it is easy to derive the stochastic
field equation for the system
\begin{equation}
{\partial S_{eff}[\chi_<,\chi_<', \nu, \xi, \eta]\over
\partial\chi_<}\vert_{\chi_<=\chi_<'}=0\,\,\,\, .
\end{equation}
 It is given by
\begin{eqnarray}
&& \Box\chi_<+{1\over 6}\chi_<^3
+ {1\over 12}\lambda^2\chi_<^2(x)\int d^4y\theta(x_0-y_0)
ReG_{++}^{\L_c}(x,y)\chi_<^3(y)\nn\\
&&+{1\over 4}\l^2\chi_<(x)\int d^4y\theta (x_0-y_0)ImG_{++}^{\L_c 2}
(x,y)\chi_<^2(y)+{1\over 6}\l^2\int d^4y\theta(x_0-y_0)ReG_{++}^{\L_c 3}
(x,y)\chi_<(y)\nn\\
&& = {3\over 2} \nu(x) \chi_<^2(x) + \xi(x) 
\chi_<(x) + \ha \eta(x)\,\,\, .
\label{sle}
\end{eqnarray}
The field equation is real and causal, as expected.
We see that it contains multiplicative and additive 
{\it colored} noise. The non linear
coupling between modes makes the Langevin equation 
much more complicated than the usually assumed white noise equation
(see Section III C).

Greiner and M\"uller \cite{GreMul} obtained a similar stochastic equation
in flat spacetime, for a thermal environment. They analyzed in detail
the dissipative terms in the Langevin equation. In particular they
 found explicit expressions for  momentum dependent 
dissipation coefficients using a Markovian approximation for 
the soft modes.
   
\section{Renormalization Group from CTP CGEA}

\subsection{Towards a nonperturbative evaluation of the CTP CGEA: 
The exact RG equation}

The main drawback of the results presented in the previous sections 
is that we have evaluated the CTP CGEA only perturbatively. As is well known,
several applications, in particular the analysis of phase transitions
in the early universe and condensed matter physics require
nonperturbative calculations.  In this section
we will derive an exact evolution equation for the dependence of
the CGEA on the coarse graining scale \cite{DalMaz}. 
In order to simplify the
notation in this section we will denote by $\L_0$ the 
ultraviolet cutoff and by $\L$ the coarse graining scale.
The CTP CGEA interpolates between the bare theory at $\Lambda=\Lambda_0$
and the physical theory at the scale $\L$. We will now consider a
$\lambda \phi^4$ field theory in Minkowski spacetime. 

To derive such an  exact evolution equation, we 
follow the approach of  Wegner and Houghton \cite{weghou}
which is designed for Euclidean spacetime. Therefore we will invoke
the {\it Euclidean} average effective action of Wetterich et al 
 \cite{wette1,liao1,polsch,hasen,morris1,boni}.
(As already mentioned, the main difference between both actions
is that the Euclidean action averages the field
over a space-time volume, while our CTP CGEA 
averages the field over a spatial volume, and is therefore more
adept to study non-equilibrium conditions.)

The Euclidean CGEA is defined by
\begin{equation}
e^{- S_{\Lambda}(\phi)} \equiv \int 
\prod_{\Lambda_0>q > \Lambda} 
{\cal D}[\phi({ q})] \; e^{- S_{cl}[\phi]}
\end{equation} 
By considering an infinitesimal variation $\L\rightarrow
\L-\delta\L$ it is possible to obtain an evolution equation 
for $S_{\L}$, which reads
\begin{equation}
\L{\partial S_\L\over\partial\L}=
-{\L\over 2\delta\L} \int^{\prime }{d^4q\over (2\pi)^4}\ln \left( 
\frac{\delta ^{2}S_\L}{\delta \phi _{q}\delta \phi _{-q}}\right)
-\int^{\prime }{d^4q\over (2\pi)^4}\frac{\delta S_\L}{\delta 
\phi _{q}}\frac{\delta S_\L}{\delta
\phi _{-q}}\left( \frac{\delta ^{2}S_\L}{\delta \phi _{q}\delta \phi _{-q}}%
\right) ^{-1}\label{weghou1}
\end{equation}
(in the original equation of Wegner and Houghton there are additional
terms coming from a rescaling of the modes after the coarse graining).
The prime in the momenta integrals means that integration is restricted to 
the shell $\Lambda > q > \Lambda - \delta \Lambda$

A typical approximation taken to solve this evolution equation is to assume
that $S_\L$ is of the form
\begin{equation}
S_\L [\phi ]=\int d^4x \left[\frac{1}{2}\left( \partial _{\mu }\phi \right)
^{2}+V_\L(\phi )\right]  \label{derexp}
\end{equation}
Eq. (\ref{weghou1}) reduces in this case to an evolution equation
for the effective potential $V_\L(\phi)$. We will discuss this 
kind of approximation for the CTP CGEA in the next section.

Now we  start the CTP calculation by 
writing the CGEA for a scale $\Lambda - \delta \Lambda$, namely
\begin{equation}
e^{i S_{\Lambda - \delta \Lambda}(\phi_+,\phi_-)} \equiv \int 
\prod_{\Lambda_0>|{\vec q}| > \Lambda-\delta \Lambda}
 {\cal D}[\phi_+({\vec q},t)] 
{\cal D}[\phi_-({\vec q},t)] \; e^{i S_{cl}[\phi_+,\phi_-]}
\end{equation} 
The modes to be integrated can be split into two parts: one within the 
shell $\Lambda > |{\vec q}| > \Lambda - \delta \Lambda$ and another containing
modes with $\Lambda_0 > |{\vec q}| > \Lambda$. Expanding the action in 
powers of the modes within the shell, one obtains
\begin{eqnarray}
e^{i S_{\Lambda - \delta \Lambda}(\phi_+,\phi_-)} &=&
e^{i S_{\Lambda}(\phi_+,\phi_-)} \times \nonumber \\
&&\int \prod_{\Lambda >|{\vec q}| > \Lambda-\delta \Lambda} {\cal D}[\phi_+] 
{\cal D}[\phi_-] \; e^{i(S_1+S_2+ S_3)} 
\; e^{\frac{i}{2} \int^{'} \frac{d^3q}{(2\pi)^3} \int dt 
\frac{d}{dt} ( \phi_a(-{\vec q},t) \dot{\phi}_b({\vec q},t) g_{ab} )} 
\label{exp}
\end{eqnarray} 
where 
\begin{eqnarray}
S_1 &=& \int dt \int^{'} \frac{d^3q}{(2\pi)^3} \; \phi_a({\vec q},t) \;
\frac{ \partial S_{\Lambda} }{ \partial \phi_a (-{\vec q},t) } 
\nonumber \\
S_2 &=& \frac{1}{2} \int dt \; dt' \int^{'} \frac{d^3q}{(2\pi)^3} 
\; \phi_a({\vec q},t) \; \frac{\partial^2 S_{\Lambda}}
{\partial \phi_a(-{\vec q},t) \phi_b({\vec q},t')} \; 
\phi_b({\vec q},t')\,\, . 
\end{eqnarray}
In taking the functional derivatives of $S_{\Lambda}$ 
(which contains modes whose wave vectors satisfy $|{\vec q}| < \Lambda$) 
the modes within the shell are set to zero. We use the notation
\begin{eqnarray}
\phi_a({\vec q},t) = \left( \begin{array}{c}
                \phi_+({\vec q},t) \\ \phi_-({\vec q},t)
                \end{array}
         \right) & \;  ; \;& 
g_{ab} = \left( \begin{array}{cc}
                1 & 0 \\ 0 & -1
                 \end{array} \right)
\end{eqnarray}

The $S_3$ term is cubic in the modes within the shell and,
as in the Euclidean case, it  does not contribute 
in the limit $\delta \Lambda \rightarrow 0$ (basically, 
this is because one is doing a one loop calculation for the shell modes). The 
functional integrals over the shell modes have the CTP boundary conditions. 
A comment about the last exponential factor 
in Eq.(\ref{exp}) is in order. Usually one discards 
it because it is a surface term, but in the CTP formalism it must be kept 
since the boundary conditions are that $\phi_+({\vec q},T) = 
\phi_-({\vec q},T)$ with $T \rightarrow \infty$ for the modes ${\vec q}$ 
within the shell. 

In order to evaluate the functional integrals we split 
the field as $\phi_a={\bar \phi}_a + \varphi_a$ and impose the boundary 
conditions on the ``classical" fields 
${\bar \phi}_{\pm}$, i.e. they vanish in the past $-T$ (negative and positive
frequencies respectively) and match in the Cauchy surface at time $T$.
The fluctuations $\varphi_a$ vanish both in the past and in the future. 
The classical fields are solutions to
\begin{equation}
(- \frac{d^2}{dt^2} - q^2) g_{ab} \bar\phi_b({\vec q},t) + 
\int dt' 
\frac{\partial^2 S_{int}}
{\partial \varphi_a(-{\vec q},t) \partial 
\varphi_b({\vec q},t')} \bar\phi_b({\vec q},t') = 0
\label{mod1}
\end{equation}
where we have split the CGEA as 
$S_{\Lambda}(\phi_{\pm}) = S_{kin}(\phi_{\pm})+S_{int}(\phi_{\pm})$ with
\begin{equation}
S_{kin} =  \int d^4x  \left[ \frac{1}{2}  (\partial_{\mu} \phi_+)^2 + 
\frac{i \epsilon}{2} \phi_+^2 \right] -
\int d^4x  \left[ \frac{1}{2} (\partial_{\mu} \phi_-)^2 -
\frac{i \epsilon}{2} \phi_-^2 \right]
\end{equation}
As before, in the 
functional derivatives the modes within the shell are set to zero.

Let $h_a$ be solutions to Eq.(\ref{mod1}), vanishing in the past and
satisfying an arbitrary normalization in the future, and let $\phi(\vec{q})$
be the common value of the fields taken in the future. We can then write
\begin{equation}
\bar\phi_a({\vec q},t) = \phi({\vec q}) 
\frac{h_a({\vec q},t)}{h_a({\vec q},T)}
\end{equation}
We first integrate over the common value $\phi(\vec{q})$ and then proceed with
the functional integration over the fluctuations $\varphi_a$ (both are
Gaussian integrals with ``source'' terms). One finally gets

\begin{eqnarray}
\Lambda \frac{\partial S_{\Lambda}}{\partial \Lambda} &=& 
- \frac{i \Lambda}
{2 \delta \Lambda} \int^{'} \frac{d^3q}{(2\pi)^3} 
\ln \left( \frac{\dot{h}_{+}({\vec q},T)}{h_{+}({\vec q},T)} -
\frac{\dot{h}_{-}({\vec q},T)}
{h_{-}({\vec q},T)} \right) + \nonumber \\
&& \frac{\Lambda}{2 \delta \Lambda}  \int^{'} \frac{d^3q}{(2\pi)^3} 
\left( \frac{\dot{h}_{+}({\vec q},T)}{h_{+}({\vec q},T)} -
\frac{\dot{h}_{-}({\vec q},T)}
{h_{-}({\vec q},T)} \right)^{-1} \, 
\left( \int dt \frac{h_a(\vec{q},t)}{h_a(\vec{q},T)} 
\frac{\partial S_{\Lambda}}{\partial \varphi_a(-{\vec q},t)} \right)^{2} 
- \nonumber \\
&& \frac{i \Lambda}{2 \delta \Lambda} \ln det'(A_{ab}) + \nonumber \\
&& \frac{\Lambda}{2 \delta \Lambda} \int dt \; dt' \int^{'} \frac{d^3q}
{(2\pi)^3} \frac{\partial S_{\Lambda}}{\partial 
\varphi_a({\vec q},t)} A_{ab}^{-1}(-{\vec q},t;{\vec q},t') 
\frac{\partial S_{\Lambda}}{\partial \varphi_b({\vec q},t')} 
\label{exact}
\end{eqnarray}

The $2\times 2$ matrix $A_{ab}$ has the following elements
\begin{eqnarray}
A_{++}(-{\vec q},t;{\vec q\, '},t') &=& 
        (- \frac{d^2}{dt^2} - q^2 + i \epsilon) \delta(t-t') 
        \delta^3({\vec q}+{\vec q\,'}) + 
        \frac{\partial^2 S_{int}}{\partial \varphi_+(-{\vec q},t) 
        \partial \varphi_+({\vec q\,'},t')} \nonumber \\
A_{--}(-{\vec q},t;{\vec q\,'},t') &=& 
        (\frac{d^2}{dt^2} + q^2 + i \epsilon) \delta(t-t') 
        \delta^3({\vec q}+{\vec q\,'}) + 
        \frac{\partial^2 S_{int}}{\partial \varphi_-(-{\vec q},t) 
        \partial \varphi_-({\vec q\,'},t')} \nonumber \\
A_{+-}(-{\vec q},t;{\vec q\,'},t')&=& A_{-+}({\vec q\,'},t';{-\vec q},t)=
 \frac{\partial^2 S_{int}}{\partial \varphi_+(-{\vec q},t) 
        \partial \varphi_-({\vec q\,'},t')}
\end{eqnarray}
The primed determinant must be calculated as the 
product of the eigenvalues of $A_{ab}$ in 
a space of functions with wave vectors within the 
shell ($\Lambda -\delta \Lambda < |{\vec q}| < \Lambda$) 
and satisfying null conditions both in the past 
and in the future. Similar conditions are to be used to evaluate the 
inverse $A_{ab}^{-1}$.  

The equation (\ref{exact}) is exact in the sense 
that no perturbative approximation has 
so far been used. It is similar 
to its Euclidean counterpart (\ref{weghou1}), but involves 
two fields and CTP boundary conditions. It contains {\it all} the 
information of the influence of the short wavelength modes on the long 
wavelength ones, and should be the starting point for a non-perturbative 
analysis of decoherence, dissipation, domain formation and out of 
equilibrium evolution. 
 

\subsection{ Derivative expansion}

The exact renormalization group equation is too complex to be solved 
without taking some approximation. The usual 
ones are expansions in the number of powers of the fields 
(see Ref. \cite{morris2} for a detailed analysis) or in derivatives 
of them \cite{morris3,enrique,tetradis}. In the following we 
shall make use of the derivative 
expansion approach.

We will prove that, within this approach, the 
exact RG Eq.(\ref{exact}) admits a solution of the form
\begin{equation}
S_{\Lambda}(\phi_+,\phi_-) = S_{\Lambda}(\phi_+) - 
S_{\Lambda}(\phi_-)
\label{ans}
\end{equation} 
Clearly this is not the most general form that can be 
imagined for the coarse grained action because contributions 
involving mixing of both fields are not taken into account. 
The main drawback of this approach is therefore that 
we miss the stochastic 
aspects of the theory, since there can be no noise
terms. However, 
the proposed form for the CGEA will be enough for 
studying the renormalization group flow of real time field theories. 

The great technical advantage of the  form Eq. (\ref{ans}) 
is that the second  functional  derivative of the action has no crossed 
terms, leading to a diagonal  matrix $A_{ab}$ 
whose determinant is easily computed as the product of two determinants,
one for $A_{++}$ and one for $A_{--}$. Following Ref. \cite{cole} one 
can express both $det' A_{++}$ and $det' A_{--}$ as the 
product over momenta of a constant (momenta independent) times the
mode $h({\vec q},T)$ evaluated at the final time $T$. Therefore the last 
term of the exact RG equation can be written as
\begin{equation}
\ln det'(A_{ab}) = \ln [det'(A_{++}) det'(A_{--})] = 
\int^{'} \frac{d^3q}{(2\pi)^3} \ln( h_{+}({\vec q},T) h_{-}({\vec q},T) )
\end{equation}
The first and the third terms can then be cast in the form of a 
single logarithm, and we arrive at
\begin{eqnarray}
\Lambda \frac{\partial S_{\Lambda}}{\partial \Lambda} &=& 
- \frac{i \Lambda} {2 \delta \Lambda} \int^{'} \frac{d^3q}{(2\pi)^3} 
\ln({h_{-}({\vec q},T) \dot{h}_{+}({\vec q},T) - h_{+}}({\vec q},T) 
\dot{h}_{-}({\vec q},T)) 
\nonumber + \\
&&  \frac{\Lambda}{2 \delta \Lambda}  \int^{'} \frac{d^3q}{(2\pi)^3} 
\left( \frac{\dot{h}_{+}({\vec q},T)}{h_{+}({\vec q},T)} -
\frac{\dot{h}_{-}({\vec q},T)}
{h_{-}({\vec q},T)} \right)^{-1} \, 
\left( \int dt \frac{h_a(\vec{q},t)}{h_a(\vec{q},T)} 
\frac{\partial S_{\Lambda}}{\partial \varphi_a(-{\vec q},t)} \right)^{2} 
+ \nonumber \\
&&  \frac{\Lambda}{2 \delta \Lambda} \int dt \; dt' \int^{'} \frac{d^3q}
{(2\pi)^3} \frac{\partial S_{\Lambda}}{\partial 
\varphi_a({\vec q},t)} A_{ab}^{-1}(-{\vec q},t;{\vec q},t') 
\frac{\partial S_{\Lambda}}{\partial \varphi_b({\vec q},t')} 
\label{Slambda}
\end{eqnarray}
Note that the equations for the two modes $h_+$ and $h_-$ (Eq. (\ref{mod1})) 
simplify considerably, since the two equations are 
decoupled.
What we still have to prove is that the proposed form for the 
action makes the r.h.s. of the exact RG equation split in the same form. 

Next we perform a derivative expansion of the interaction term. As our 
coarse graining explicitly breaks Lorentz invariance, 
we allow different coefficients for the 
temporal and spatial derivatives, namely
\begin{equation}
S_{int}(\phi_{\pm}) = \int d^4x [ - V_{\Lambda}(\phi_{\pm}) + 
\frac{1}{2} Z_{\Lambda}(\phi_{\pm})  \dot{\phi}^2_{\pm} -
\frac{1}{2} Y_{\Lambda}(\phi_{\pm})( {\vec \nabla} \phi_{\pm})^2 + \ldots ]
\end{equation}
We expand the fields around a time dependent background: $\phi_{\pm} = 
\phi_{\pm}(t) + \varphi_{\pm}({\vec x},t)$ and Fourier transform in 
space. 
We shall solve the 
Eq. (\ref{mod1}) for the modes to zeroth order in the 
inhomogeneities, i.e. 
we equate terms 
in the equations for $h_{\pm}$ that are
independent of $\varphi_{\pm}$'s.
Since the first functional derivative of the CGEA (S') is 
linear in the inhomogeneities $\varphi_{\pm}$, we 
put $S'=0$ and keep the $\varphi_{\pm}$-independent contributions to 
$S_{int}''$. After a little algebra and functional derivations, we get
\begin{eqnarray}
\frac{\partial^2 S_{int}}{\partial \varphi({\vec q},t) \partial 
\varphi(-{\vec q}',t')}
&=& [ -V'' - \frac{1}{2} \dot{\phi}^2 Z'' - Y q^2 -  Z' \dot{\phi}
\frac{d}{dt} - Z \frac{d^2}{dt^2} - \ddot{\phi} Z' + \ldots ] 
\times \nonumber \\
&&\delta(t-t') \delta^3({\vec q}-{\vec q}')
\end{eqnarray}
where the primes denote derivation with respect to the field and the 
ellipsis denote terms linear in the fluctuations. 
In this expression and hereafter we omit (unless explicitly stated) 
the ${\pm}$ subscripts in the background 
fields $\phi_{\pm}(t)$, in the 
potential $V_{\Lambda}(\phi_{\pm}(t))$, and in the wave 
function factors $Z_{\Lambda}(\phi_{\pm}(t))$ and 
$Y_{\Lambda}(\phi_{\pm}(t))$. Note that  
the effective mass of the modes depends on the time-dependent background
$\phi(t)$.
The equations of motion for the modes $h_a$ become localized and take the 
form of
harmonic oscillators with variable frequency and a damping term.  
The boundary conditions to be imposed are the aforementioned CTP ones. 

If one defines new modes as
$f({\vec q},t) = (1+Z_{\Lambda})^{1/2} h({\vec q},t)$, the damping 
terms cancel out and the new modes are harmonic oscillators with frequency 
\begin{equation}
w_{q}^2(t)= q^2 \, \frac{1+Y_{\Lambda}}{1+Z_{\Lambda}} + 
\frac{V_{\Lambda}''}{1+Z_{\Lambda}} + 
\frac{1}{4} \frac{Z_{\Lambda}'^2}{(1+Z_{\Lambda})^{2}} \, \dot{\phi}^2 +
\frac{1}{2} \frac{Z_{\Lambda}'}{1+Z_{\Lambda}} \,  \ddot{\phi}
\end{equation}
Using an adiabatic expansion for the modes,
\begin{equation}
h_{\pm}({\vec q},t) = (1+Z_{\Lambda})^{-1/2} 
\frac{1}{\sqrt{2 W_{\pm}({\vec q},t)}} e^{\pm i 
\int_{-T}^{t} W_{\pm}({\vec q},t') dt'}
\end{equation}
we can easily evaluate the logarithmic term in the r.h.s. of the exact RGE
(\ref{Slambda}), which is the only term that survives in the 
approximation we are working. The RGE (\ref{Slambda}) reduces to  
\begin{eqnarray}
&&\Lambda \int dt 
\left\{
\left[
- \frac{d V_{\Lambda}(\phi_+)}{d \Lambda} + \frac{1}{2} 
\frac{d Z_{\Lambda}(\phi_+)}{d \Lambda} \dot{\phi}_+^2 
\right]
-
\left[
- \frac{d V_{\Lambda}(\phi_-)}{d \Lambda} + \frac{1}{2} 
\frac{d Z_{\Lambda}(\phi_-)}{d \Lambda} \dot{\phi}_-^2 
\right]
\right\} = \nonumber \\
&& \frac{ \Lambda}{2 \delta
\Lambda} \int^{'} \frac{d^3q}{(2\pi)^3}  \int 
[W_{+}({\vec q},t)-W_{-}
({\vec q},t)]dt
\end{eqnarray}
In the adiabatic expansion, the $W$'s read
\begin{equation}
W^2=A_{\Lambda} + B_{\Lambda}
\dot{\phi}^2(t)+ C_{\Lambda} \ddot{\phi}(t)
\end{equation} 
where the coefficients are 
\begin{eqnarray}
A_{\Lambda} &=& \Lambda^2 \frac{1+Y_{\Lambda}}{1+Z_{\Lambda}} +
     \frac{V_{\Lambda}''}{1+Z_{\Lambda}} \nonumber \\
B_{\Lambda} &=& \frac{Z_{\Lambda}'^2}{4(1+Z_{\Lambda})^2} + 
     \frac{5 A_{\Lambda}'^2}{16 A_{\Lambda}^2} - \frac{A_{\Lambda}''}
{4 A_{\Lambda}} \nonumber \\
C_{\Lambda} &=& \frac{Z_{\Lambda}'}{2(1+Z_{\Lambda})} 
- \frac{A_{\Lambda}'}{4 A_{\Lambda}} 
\end{eqnarray}
Integrating by parts we get 
\begin{equation}
\int dt \left\{
- \Lambda \frac{dV_{\Lambda}}{d\Lambda} + 
\frac{1}{2} \Lambda \frac{dZ_{\Lambda}}{d\Lambda} {\dot \phi}^2 \right\}
=\frac{\Lambda^3}{4 \pi^2} \int dt 
\left\{
\sqrt{A_{\Lambda}} +
\frac{1}{2} \dot{\phi}^2 
\left[
\frac{B_{\Lambda}}{\sqrt{A_{\Lambda}}} - 
\left( \frac{C_{\Lambda}}{\sqrt{A_{\Lambda}}} \right)'
\right]
\right\}
\end{equation}
Therefore the dependence of the potential and the wave function
renormalization on the infrared scale is given by
\begin{eqnarray}
\Lambda \frac{dV_{\Lambda}}{d\Lambda} &=& -\frac{\Lambda^3}{4 \pi^2} 
\sqrt{\Lambda^2 \frac{1+Y_{\Lambda}}
{1+Z_{\Lambda}} + \frac{V_{\Lambda}''}{1+Z_{\Lambda}}} \nonumber \\
\Lambda \frac{dZ_{\Lambda}}{d\Lambda} &=& \frac{\Lambda^3}{4 \pi^2} 
\left[
\frac{B_{\Lambda}}{\sqrt{A_{\Lambda}}} - 
\left( \frac{C_{\Lambda}}{\sqrt{A_{\Lambda}}} \right)'
\right]
\label{flowing}
\end{eqnarray}
These equations are valid both for the $\phi_+$ field 
and for the $\phi_-$ field.

The above equations  describe the flow
of the coarse grained action with the infrared scale in the derivative 
expansion 
of the exact CTP renormalization group equation. It is
interesting to note that the higher derivative terms
modify the differential equation for the effective
potential. 

We have obtained two equations for the three independent 
unknown functions $V_{\Lambda}$, $Z_{\Lambda}$ and $Y_{\Lambda}$. 
In order to find an additional relation between the 
spatial and temporal wave funtion renormalization 
functions $Z_{\Lambda}$ and $Y_{\Lambda}$, it is 
necessary to write the exact RGE up to quadratic order in the 
inhomogeneities. 
We will not present this long calculation here. For simplicity,
we will assume that $Z_{\Lambda}$ and $Y_{\Lambda}$ are small numbers, 
and therefore we will set them to zero on the r.h.s. of Eq.(\ref{flowing}). 
This assumption is partially confirmed by numerical 
calculations \cite{DalMaz}. 
Note that in this 
approximation we recover the RG improved equation proposed in 
Ref. \cite{strick} for the coarse grained effective potential
\begin{equation}
\Lambda \frac{dV_{\Lambda}}{d\Lambda} =  -\frac{\Lambda^3}{4 \pi^2} 
\sqrt{\Lambda^2 
+ V_{\Lambda}''}\,\,\, .
\end{equation}

There are other  points which are worth noting. First,
when we substitute $V_{\Lambda}$, $Z_{\Lambda}$
and $Y_{\Lambda}$ by their classical values 
$V_{\Lambda}=V \, , \, Z_{\Lambda}=Y_{\Lambda}=0$, on the
r.h.s. of both equations we obtain the one loop
evolution equations \cite{DalMaz}. Second, while in the one loop
approximation it is possible to take the limit $\Lambda_{0}\rightarrow\infty$
(the infinities can be absorbed into the bare mass and coupling constant),
in this non-perturbative calculation it is not possible to 
renormalize the theory (as is the case for Hartree, Gaussian and $1/N$ approximations).
For these reasons we keep $\Lambda_{0}$ as a large
(compared with the mass) but finite number. 

Once the functions $V_{\Lambda}$ and $Z_{\Lambda}$ are known,
one can write the effective dynamical equations for the coarse grained 
field.

\vskip 1cm 
{\Large{\bf PART FOUR:  Renormalization Group in Semiclassical Gravity}}

\section{Renormalization Group and Stochastic Semiclassical Gravity }
    
In this Section we will describe a different relation between the
RG and the CTP CGEA. We will consider the backreaction of quantum matter 
fields on the spacetime geometry, assumed classical. To analyze this 
problem, we will use the formulation of quantum open
systems and compute the CTP CGEA using an expansion in powers of the spacetime
curvature and discuss its relation with the RG equations. 

The usual approach to analyze backreaction 
in semiclassical gravity is based on the use
of the Semiclassical Einstein Equations (SEEs) \cite{birrel}
\begin{equation}
{1\over{8\pi G}}\left[R_{\mu\nu}-\frac{1}{2}Rg_{\mu\nu}\right]- 
\alpha H^{(1)}_{\mu\nu} - \beta H^{(2)}_{\mu\nu}= T_{\mu\nu}^{clas}+ 
<T_{\mu\nu}>.
\label{see}
\end{equation}
In the SEEs, which 
can be derived
from the {\it real part} of the  CTP CGEA
\cite{CH87,CH89,Jordan},
the effect of quantum matter fields is taken into account
by including as a source the quantum mean value of the 
energy-momentum tensor. The terms proportional to
\begin{equation} 
H^{(1)}_{\mu\nu} = \left[4 R^{;\mu\nu}-4g^{\mu\nu}\Box R\right]+ 
O(R^2),\end{equation} and 
\begin{equation} H^{(2)}_{\mu\nu}=\left[4R^{\mu\alpha;\nu}{}_{;\alpha}-2\Box 
R^{\mu\nu}- g^{\mu\nu}\Box R\right]+ O(R^2),\end{equation}
come from terms quadratic in the curvature  in the gravitational action,
which are
needed to renormalize the theory.

These equations cannot provide a full description
of the problem \cite{HuPhysica}, since they do not take into account
the fluctuations of the energy momentum tensor  around
its mean value. The fluctuations can be incorporated by  
including an additional stochastic term \cite{CH94,camver0,camver}
on the right hand side of Eq. \ref{see}. This noise-term can be derived
from the {\it imaginary part} of the CTP CGEA, in the 
same way we proceeded for the soft modes of the scalar field (see Section
VII.B). When it is incorporated in the SEE, one obtains the  
``Einstein-Langevin Equations" (ELEs), which include
both the dissipative and diffusive effects of the quantum matter
on the geometry of spacetime \cite{CH94}, in complete analogy with 
what happens in quantum Brownian motion typical of  quantum
open systems. For a review of the semiclassical stochastic gravity program
based on the ELEs, see \cite{stogra}.

The ELEs have been 
derived for arbitrary small metric perturbations conformally coupled
to a massless quantum scalar field in a spatially flat background 
\cite{camver}, and, in a cosmological setting, for a massive field 
in a spatially flat Friedmann-Robertson-Walker universe \cite{humatacz}, 
and in a Bianchi type-I spacetime\cite{husinha}. Further ellaboration
of its physical meaning can be found in the papers by Verdaguer and Roura
\cite{Enric}. In ref. \cite{CRV} it is proven that the ELE may be used 
to compute certain quantum averages, even in conditions 
where there is no decoherence.
   
Here we present the derivation of the ELEs based on the renormalization
group equations of Lombardo and Mazzitelli \cite{lombmazz97}. 
Using a 
covariant expansion in powers of the curvature
we will see that, for massless quantum fields, the ELEs are
determined to leading order
by the running couplings of the theory. 

This example will show
an interesting connection between the ``usual" renormalization 
group and the different versions of the effective action. Indeed, 
in a naive ``Wilsonian" approach, some quantum effects can be taken
into account by replacing the coupling constants of the theory by their 
running counterparts. However, the (usual) running of coupling
constants is defined in momentum space, and a careful implementation
of the Wilsonian approach leads to a nonlocal effective action
in configuration space. We will compute explicitly this non local
effective action first in its Euclidean version, and show that it
coincides with calculations based on resummations of the Schwinger
DeWitt expansion. Using general relations between Euclidean propagators
and CTP propagators, we will be able to obtain the CTP CGEA from the
Euclidean effective action. From the CTP CGEA we will derive the 
ELEs, thus showing that they are a consequence of usual renormalization
group equations of the theory. As an application, we will compute
the leading quantum corrections to the Newtonian potential. 

Consider a quantum scalar field on a classical, {\it Euclidean} curved 
background.
The classical action is given by
\begin{equation}
S=S_{grav} + S_{matter},
\label{sclas}
\end{equation}
where
\begin{equation}
S_{grav}=-\int d^4x\sqrt g\left [{1\over 16\pi G_0}(R-2\Lambda_0)
+\alpha_0 R^2 +\beta_0 R_{\mu\nu}R^{\mu\nu}\right ],\label{bare}
\end{equation}
and
\begin{equation}
S_{matter}={1\over 2}\int d^4x\sqrt g [\partial_{\mu}\phi\partial^{\mu}\phi
+m^2\phi^2+\xi R\phi^2].
\end{equation}
Here $\xi$ is the coupling to the curvature. $G_0$, $\Lambda_0$, 
and the dimensionless 
constants $\alpha_0$ and $\beta_0$ are {\it bare} constants.

The effective action for this theory is a complicated, non local
object. It is defined by integrating out the quantum scalar field, that is
\begin{equation} e^{-S_{eff}} = N \int {\cal D}\phi e^{-S[g_{\mu\nu}, 
\phi]},
\end{equation} 
where $N$ is a normalization constant. It is in general 
not possible to find a closed form for it. If we compute it using a covariant 
expansion 
in powers of the curvature, the different terms must be constructed
with the Riemann tensor and its derivatives 
$\nabla \nabla ...{\cal R}$. Using integration by parts and the Gauss
Bonnet identity,  the effective action can be
written only in terms of the  Ricci tensor $R_{\mu\nu}$  and $\Box R_{\mu\nu}$
\cite{vilkoqt}.These arguments suggest that
the effective action must have the general form 
\begin{eqnarray} S_{eff} &=& -\int d^4x\sqrt g\left 
[{1\over 16\pi G}R +\alpha R^2 +\beta R_{\mu\nu}R^{\mu\nu}\right ]\nonumber\\ 
&+&\frac{1}{32\pi^2} \int d^4x\sqrt g\left[F_0 R +  RF_1(\Box)R+R_{\mu\nu} 
F_2(\Box)R^{\mu\nu}+...\right], \label{seff} 
\end{eqnarray} 
where the ellipsis 
denote terms cubic in the curvature. For simplicity, in the above equation and 
in what follows we will  omit the cosmological constant term. Note that the 
bare constants in Eq. (\ref{bare}) have been replaced by dressed couplings in 
Eq. (\ref{seff}). The expansion is adequate for weak gravitational fields,
i.e. $\nabla\nabla {\cal R}\ll {\cal R}^2$.

Up to this order, all the information about the effect of the quantum
field is encoded in the constant $F_0$ and in the form 
factors $F_1$ and $F_2$.
The form factors are, in general, non-local two point functions constructed
with the d'Alambertian and the parameters
$\xi$ and $m^2$.  $F_0$, $F_1$, and $F_2$ also depend on an energy
scale $\mu$, introduced by the regularization method.

The dressed coupling constants depend on the energy scale
$\mu$ according to the RG equations.
Using minimal substraction these equations read
\cite{buchetal92}
\begin{eqnarray}
\mu{dG\over d\mu}&=&
 \frac{G^2 m^2}{\pi} \left(\xi -  \frac{1}{6}  \right),\label{rge1} \\
\mu{d\alpha\over d\mu}&=&
 -\frac{1}{32 \pi^2}
 \left[
 \left( \frac{1}{6} - \xi \right)^2 -\frac{1}{90}
 \right], \label{rge2}\\
\mu{d\beta\over d\mu}&=&
 -\frac{1}{960 \pi^2}. \label{rge3}
\end{eqnarray}
The dependence of $F_0$, $F_1$ and $F_2$ on $\mu$ is such that
the full equation is $\mu$-independent. For example, 
from Eqs. (\ref{seff}) and (\ref{rge1}), 
we see that $F_0=m^2\ln\left({m^2\over\mu^2}\right)(\xi-{1\over 6})+const$.

When the scalar field is massless, this information is
enough to fix completely the form factors. Indeed, as the
$F_i,\,\, i=1,2$ are  dimensionless two point functions, 
by simple dimensional analysis we obtain $F_i(\Box,\mu^2,\xi)=
F_i({\Box\over\mu^2},\xi)$. Inserting this into
Eq. (\ref{seff}), using Eqs. (\ref{rge2}) and (\ref{rge3}), 
and the fact  that $S_{eff}$ 
must be independent of
$\mu$, we obtain
\begin{eqnarray}
F_1(\Box)&=&{1\over 2}\left[(\xi-{1\over 6})^2-{1\over 90}\right ]
\ln \left [{-\Box
\over\mu^2}\right] + const,\nonumber\\                   
F_2(\Box)&=&{1\over 60} 
\ln \left [{-\Box
\over\mu^2}\right] + const.
\label{m=0}
\end{eqnarray}

The final result for the effective action has a clear interpretation:
it is just the classical action in which the coupling
constants  $\alpha$ and $\beta$ have been replaced by nonlocal 
two point functions 
that take into 
account their running  in {\it configuration
space}.

For a massive field, the situation is more complex because there
is an additional dimensional parameter. The form factors also
depend on ${m^2\over\mu^2}$ and the $\mu$-independence of the
effective action is not enough to fix the form of them.
They have 
already been computed 
in the literature \cite{barvi,avra} 
\begin{equation}
F_i(\Box)=\int_0^1 d\gamma \chi_i(\xi, \gamma) \ln \left 
[{m^2-{1\over 4}(1-\gamma^2)\Box
\over\mu^2}\right]\label{fgrav},\end{equation}
where 
\begin{eqnarray}
\chi_1(\xi, \gamma)&=&{1\over 2} \left [ \xi^2-{1\over 2}\xi(1-\gamma^2)
+{1\over 48}(3-6 \gamma^2-\gamma^4)\right ],\nonumber\\
\chi_2(\xi, \gamma)&=&{1\over 12}\gamma^4.
\label{chi}
\end{eqnarray}
These equations can be obtained through a covariant
perturbation expansion \cite{barvi}, or by a resummation
of the Schwinger DeWitt expansion \cite{avra}. 
Of course these form factors coincide with our previous Eq. (\ref{m=0})
in the massless case. 
For the sake of completness we will consider in what follows the massive case,
although it is clear that only in the massless case the form factors are
determined by the RG equations.
 
In order to clarify the meaning of the two point functions
appearing in Eqs. (\ref{m=0}) and (\ref{fgrav}), it is
useful to introduce the following integral representation 
\begin{equation}
\ln\left[{m^2-{1\over 4}(1-\gamma^2)\Box\over\mu^2}\right]
=\left[\ln
{(1-\gamma^2)\over 4}+\int_0^{\infty}dz\left({1\over z+\mu^2}- 
G_E^{(z)}\right)\right],\label{rep}
\end{equation}
so the logarithm of the d'Alambertian is written in 
terms of the massive Euclidean
propagator $G_E^{(z)} = (z+{4 m^2\over (1-\gamma^2)}-\Box )^{-1}$.
This representation will also be useful to construct the CTP
version of the effective action.

Up to here we considered the Euclidean effective action. What about the
in-out and CTP CGEA? Of course one can compute them from first
principles using the covariant 
expansion, and indeed there are some calculations
in the literature for the CTP CGEA \cite{Flan}.
However, in order to emphazise the relation with the RG equations,
we will construct the CTP CGEA from its Euclidean counterpart.

Replacing the Euclidean propagator by the Feynman one 
in the integral representation Eq. (\ref{rep}), one
obtains the usual {\it in-out} effective action. As we already pointed
out, the effective equations derived from this action are neither real
nor causal because they are equations for {\it in-out} matrix
elements and not for mean values.

The CTP CGEA
can be written as
\begin{equation} e^{i S_{eff}[g^+,g^-]} = N e^{i(S_{grav}[g^+]-S_{grav}[g^-])} 
\int {\cal D} \phi^+{\cal D} \phi^- e^{i(S_{matter}[g^+,\phi^+]- 
S_{matter}[g^-,\phi^-])}, 
\label{ctpeff}\end{equation} 
and the field equations are obtained from taking the 
variation of this action with respect to the $g_{\mu\nu}^+$ metric, and then 
setting $g_{\mu\nu}^+=g_{\mu\nu}^-$. 

In an alternative, and more concise notation, we can write this effective
action as
\cite{mottola}
\begin{equation} e^{i S_{eff}^{{\cal C}}[g]} = N e^{i S_{grav}^{{\cal C}}[g]} 
\int {\cal D} \phi e^{i S_{matter}^{{\cal 
C}}[g,\phi]},\label{neweff}\end{equation} where we have introduced the CTP 
complex temporal path ${\cal C}$, going from minus to plus infinity $\cal C_+$ 
and backwards $\cal C_-$, with a decreasing (infinitesimal) imaginary part. 
Time integration over the contour ${\cal C}$ is defined by $\int_{{\cal C}} dt 
=\int _{{\cal C_+}} dt -\int_{{\cal C_-}} dt$. The field $\phi$  appearing in 
Eq. (\ref{neweff}) is related to those in Eq. (\ref{ctpeff}) by $\phi(t,\vec x) = 
\phi_{\pm}(t,\vec x)$ if $t$ belongs to ${\cal C}_{\pm}$. 
The same applies to  $g_{\mu\nu}$. 

This equation is 
useful because it has the 
structure of the usual {\it in-out} or the Euclidean effective action. Feynman
rules are therefore the ordinary ones, replacing the Euclidean propagator by
\begin{eqnarray} G(x,y) = \left\{\begin{array}{ll} 
G_F(x,y)=i \langle 0, in\vert T \phi (x) \phi(y)\vert 0, in\rangle,& ~t, t' ~ 
\mbox{both on} ~{\cal C}_+ \\ G_D(x,y)=-i \langle 0, in\vert 
{\tilde T}\phi (x) 
\phi(y)\vert 0, in\rangle , & ~t, t' ~ \mbox{both on} 
~{\cal C}_-\\ G_+(x,y)=-

i \langle 0, in\vert \phi (x) \phi(y)\vert 0, in\rangle,    &~t ~\mbox{on}~ 
{\cal C}_-, t'  ~\mbox{on} ~{\cal C}_+\\ G_-(x,y)=i \langle 0, 
in\vert \phi (y) 
\phi(x)\vert 0, in\rangle, & ~ t  ~\mbox{on} ~ {\cal C}_+, t'~  \mbox{on}~ 
{\cal C}_-\end{array}\right. \label{prop} \end{eqnarray} 
Introducing Riemann normal coordinates, we can write, up to lowest order in 
the curvature 
 
\begin{equation} G_F(x,y)= \int {d^4p\over{(2 \pi)^4}} {e^{ip(x-
y)}\over{p^2+m^2-i\epsilon}}= G_D^*(x,y),\end{equation} 

\begin{equation} 
G_{\pm}(x,y)=\mp \int {d^4p\over{(2 \pi)^4}} e^{ip(x-y)}2 \pi i 
\delta(p^2-m^2)\theta(\pm p^0).
\label{gpm}
\end{equation}

All of the preceeding formulation of the effective action is valid for  
any field theory. In our particular case, we must replace the 
Euclidean propagator $G_E^{(z)}$ in Eq. (\ref{rep}) by the propagator 
$G(x,y)$ of Eq. (\ref{prop}) with a mass given by  ${4m^2\over{1-\gamma^2}}
+z$. After integration in $z$ we obtain
\begin{eqnarray}
\ln\left[{{4 m^2\over{(1-\gamma^2)}}- \Box\over\mu^2}\right]_{CTP}
= \left\{\begin{array}{ll}
 \int {d^4p\over{(2 \pi)^4}}e^{ip(x-y)}  \ln\left({(1-\gamma^2)
(p^2-i\epsilon)+4m^2\over{\mu^2}}\right) & ~t, t' ~ \mbox{both on} ~{\cal C}_+
\\
\int {d^4p\over{(2 \pi)^4}}e^{ip(x-y)}  \ln\left({(1-\gamma^2)
(p^2+i\epsilon)+4m^2\over{\mu^2}}\right) & ~t, t' ~ 
\mbox{both on} ~{\cal C}_-\\
\int {d^4p\over{(2 \pi)^4}}e^{ip(x-y)}2 \pi i \theta (p^0)\theta 
\left(-p^2 -{4m^2\over{1-\gamma^2}}\right)    &~t ~\mbox{on}~ {\cal C}_-, t'  
~\mbox{on} ~{\cal C}_+\\
-\int {d^4p\over{(2 \pi)^4}}e^{ip(x-y)}2 \pi i \theta (-p^0)\theta 
\left(-p^2 -{4m^2\over{1-\gamma^2}}\right) & ~ t  ~\mbox{on} ~ 
{\cal C}_+, t'~  
\mbox{on}~ {\cal C}_-\end{array}\right.\end{eqnarray}

With the expression for the CTP logarithm of the d'Alambertian we can calculate
explicitly the CTP effective action. Using the previous 
notation with $g_{\mu\nu}^+$ and $g_{\mu\nu}^-$ the CTP effective 
action reads 
\begin{eqnarray}
S_{eff}&&[g^+,g^-] = S^r_{grav}[g^+] - S^r_{grav}[g^-] \nonumber \\
&&+{i\over{8 \pi^2}}\int d^4x \int d^4y \Delta(x) \Delta(y)N_1(x,y)
- {1\over{8 \pi^2}}\int d^4x \int d^4y \Delta(x)\Sigma(y)D_1(x,y)\nonumber\\
&&+{i\over{8 \pi^2}}\int d^4x\int d^4y\Delta_{\mu\nu}(x)
\Delta^{\mu\nu}(y)N_2(x,y)
-{1\over{8 \pi^2}}\int d^4x\int d^4y \Delta_{\mu\nu}(x)\Sigma^{\mu\nu}(y)
D_2(x,y),
\label{seffctp}
\end{eqnarray}
where $\Delta = {{R^+ - R^-}\over{2}}$, $\Sigma = {{R^+ + R^-}\over{2}}$, 
$\Delta_{\mu\nu} = {{R^+_{\mu\nu} - R^-_{\mu\nu}}\over{2}}$, 
$\Sigma_{\mu\nu} = {{R^+_{\mu\nu}
 + R^-_{\mu\nu}}\over{2}}$. The classical gravitational
action $S^r_{grav}$ contains the dressed, $\mu$-dependent coupling 
constants and we absorbed $F_0$ into the gravitational constant $G$.

The real and imaginary parts of $S_{eff}$ can be associated with 
the dissipation and noise, 
respectively.  The dissipation 
$D_i$ and 
noise $N_i$ kernels are given by
\begin{equation} D_i(x,y) = \int_0^1 d\gamma \chi_i(\xi, \gamma)\int 
{d^4p\over{(2 \pi)^4}} \cos[p (x-y)]\ln\left|{{(1-\gamma^2)p^2 + 
4 m^2}\over{\mu^2}}\right|,\end{equation}

\begin{equation} N_i(x,y) = \int_0^1 d\gamma \chi_i(\xi, \gamma)\int 
{d^4p\over{(2 \pi)^4}} \cos[p(x-y)] \theta\left(- p^2 - {4 m^2\over{1-
\gamma^2}}\right). \end{equation} 

It is important to note that the imaginary part of this effective action must 
be positive definite. To make this point  explicit, one can write 
the imaginary 
part in terms of the Weyl tensor $C_{\mu\nu\alpha\beta}$ and the scalar 
curvature $R$ by means of the following relation: 
$C_{\mu\nu\alpha\beta}C^{\mu\nu\alpha\beta}=2R_{\mu\nu}R^{\mu\nu} - 2/3 R^2$. 
It is not difficult to show that the scalar and tensor contributions to the 
imaginary part of the effective action are both positive.

In order to derive the ELEs we proceed as in Section VII B.
One can regard the imaginary part of the CTP CGEA 
as 
coming from two classical stochastic sources $\eta(x)$ and 
$\eta^{\mu\nu\alpha\beta}(x)$, where the last tensor has the symmetries of the 
Weyl tensor. In fact, we can write the 
imaginary part as 
\begin{eqnarray} 
\int {\cal D}\eta(x)&& \int {\cal 
D}\eta^{\mu\nu\alpha\beta}(x) P[\eta, 
\eta^{\mu\nu\alpha\beta}]\exp\left(i\left\{ \Delta(x) \eta(x) + 
\Delta_{\mu\nu\alpha\beta} \eta^{\mu\nu\alpha\beta}\right\}\right) \nonumber 
\\ &&= \exp\left\{-\int d^4x\int d^4y \left[\Delta(x) {\tilde N}(x-y) 
\Delta(y) + \Delta_{\mu\nu\alpha\beta}(x) N_2(x-y) 
\Delta^{\mu\nu\alpha\beta}(y)\right]\right\}, 
\end{eqnarray} 
where ${\tilde N}(x,y) = N_1(x,y)+ \frac{1}{3} N_2(x,y)$, 
and $\Delta_{\mu\nu\alpha\beta} =\frac{1}{2}{ C^+_{\mu\nu\alpha\beta} - C^-_{\mu\nu\alpha\beta}}$. 
The Gaussian functional probability distribution 
$P[\eta, \eta^{\mu\nu\alpha\beta}]$ is given by 
\begin{eqnarray} 
P[\eta, \eta^{\mu\nu\alpha\beta}] = A\,\, && 
\exp\left\{-{1\over{2}} \int d^4x\int d^4y \eta(x) \left[ {\tilde N}(x, 
y)\right]^{-1} \eta(y)\right\} \nonumber \\ &&\times\exp\left\{- 
{1\over{2}}\int d^4x\int d^4y \eta_{\mu\nu\alpha\beta}(x) \left[ N_2(x, 
y)\right]^{-1} \eta^{\mu\nu\alpha\beta}(y)\right\},
\label{noise}
\end{eqnarray} 
with $A$ a normalization factor. 

Therefore, the CTP CGEA can be written as
\begin{equation} \exp\{iS_{eff}\}= \int {\cal D}\eta {\cal 
D}\eta_{\mu\nu\alpha\beta} P[\eta ,\eta_{\mu\nu\alpha\beta}] \exp\left\{i 
A_{eff}[\Delta, \Delta_{\mu\nu\alpha\beta}, \Sigma, \Sigma_{\mu\nu}, \eta, 
\eta_{\mu\nu\alpha\beta}]\right\},\end{equation} where 
\begin{equation} A_{eff} = Re 
S_{eff} + \int d^4x [\Delta(x) \eta(x) + \Delta_{\mu\nu\alpha\beta}(x) 
\eta^{\mu\nu\alpha\beta}].\end{equation} 
The field equations $\left.{\delta A_{eff}\over{\delta g^+_{\mu\nu}}} 
\right|_{g^+_{\mu\nu} = g^-_{\mu\nu}} = 0$, the Einstein-Langevin equations,
are
\begin{eqnarray}
&&{1\over 8 \pi G}\left( R^{\mu\nu}- 
{1\over{2}} g^{\mu\nu} R \right)
- \tilde{\alpha}H^{(1)}_{\mu\nu}
- \tilde{\beta}H^{(2)}_{\mu\nu}\nonumber \\
&&= -{1\over{32 \pi^2}}\int d^4y   D_1(x,y) H_{\mu\nu}^{(1)}(y)
-{1\over{32 \pi^2}}\int d^4y  D_2(x,y)H_{\mu\nu}^{(2)}(y)\nonumber \\ 
&&+ g^{\mu\nu}\Box \eta - \eta^{;\mu\nu}
+2 \eta^{\mu\alpha\nu\beta}{}{}{}{}_{;\alpha\beta}~~,
\label{ele}
\end{eqnarray}
where $\tilde{\alpha}$ and $\tilde{\beta}$ differ from 
$\alpha$ and $\beta$ by $\xi$-dependent finite constants. 
Eq. (\ref{ele}) is the main result of this section.
The r.h.s. consists of the mean value of the 
energy-momentum tensor for the scalar field plus a stochastic correction
characterized by the two point correlation functions
\begin{eqnarray}
<\eta (x)\eta (y)>&=& {\tilde N}(x,y)\nonumber\\
<\eta_{\mu\nu\alpha\beta} (x)\eta_{\rho\sigma\lambda\tau} (y)>&=& 
T_{\mu\nu\alpha\beta\rho\sigma\lambda\tau} N_2(x,y) ,
\label{corrfunct} 
\end{eqnarray} 
where the tensor $T_{\mu\nu\alpha\beta\rho\sigma\lambda\tau}$ 
is a linear combination of four-metric products in such a way that the r.h.s 
of Eq. (\ref{corrfunct}) keeps the Weyl's symmetries. 
The scalar-noise kernel is given by 
\begin{equation} 
{\tilde N}(x,y)={1\over{2}}\int_0^1 d\gamma \left[\left(\xi-
{(1-\gamma^2)\over{4}}\right)^2-{\gamma^4\over{36}}\right]\int {d^4p\over{(2 
\pi)^4}} cos[p(x-y)]\theta\left(-p^2-{4m^2\over{1-
\gamma^2}}\right).
\end{equation} 
In the massless case ${\tilde N}$ is proportional to 
$(\xi - 1/6)^2$, and vanishes for conformal coupling. Therefore this term is 
present when the quantum fields are massive and/or when the coupling is not 
conformal. This is to be expected, since the imaginary part of the CTP 
CGEA signifies particle creation. For massless, 
conformally coupled quantum fields, particle creation takes place only when the
spacetime is not conformally flat. Therefore in this case the only 
contribution to the imaginary part of the CTP CGEA is proportional 
to the square 
of the Weyl tensor. When the fields are massive and/or non-conformally 
coupled, particle creation takes place even when the Weyl tensor vanishes. 
This is why an additional contribution proportional to $R^2$ appears in the 
imaginary part of the effective action. 

From Eq. (\ref{ele}) we can define the effective energy-momentum tensor 
\begin{equation} 
T_{\mu\nu}^{eff}=<T_{\mu\nu}>+T_{\mu\nu}^{stoch}= <T_{\mu\nu}>+ 
g^{\mu\nu}\Box \eta - \eta^{;\mu\nu}
+2 \eta^{\mu\alpha\nu\beta}{}{}{}{}_{;\alpha\beta}~~,\end{equation} where 
$<T_{\mu\nu}>$ is the quantum expectation value of the energy-
momentum tensor of the quantum field and $T_{\mu\nu}^{stoch}$ 
is the contribution of 
the stochastic force, which in turn has contributions from the scalar and 
tensor noises. In the massless-conformal case the scalar-noise kernel 
vanishes, and $(T_\mu{}^\mu)^{stoch}= 0$, because the noise-source 
$\eta^{\mu\nu\alpha\beta}$ has vanishing trace. This means that there is no 
stochastic correction to the trace anomaly \cite{camver}. 

To summarize, we have obtained the ELEs using a covariant expansion
in powers of the curvature. Our results are valid for quantum
scalar fields with arbitrary mass and coupling to the curvature
$\xi$.
In the massless case, still for arbitrary $\xi$, we have shown that
it is possible to obtain the noise and dissipation kernels using 
only dimensional analysis, the running of the coupling constants 
and the relation between the Euclidean and CTP CGEA. From this 
point of view, we can therefore 
conclude that the RG equations already contain information about
the dissipation and noise kernels. 

The results of this section can be 
easily generalized to any field theory with massless quantum fields
in gravitational or Yang Mills backgrounds.

\section{ Renormalization Group and Quantum 
Corrections to the Newtonian Potential}

Two major areas of application for the stochastic and semiclassical gravity
programs are the study of the final states of black holes  and the physics of the 
early Universe. These are enormously complicated problems.
We will describe here one of the simplest applications of
semiclassical gravity, the calculation of quantum corrections to the 
Newtonian potential. Although a very simple example, it will shed light on
important conceptual issues, in particular on the relation between
the quantum corrections and the renormalization group.

Duff \cite{Duff73} computed the corrections 
to the Newtonian potential
produced by the vacuum 
polarization of gravitons. His result is, schematically,
\begin{equation}
V_{eff}(r)=-{G M\over r}(1 + a {G\over r^2})
\label{veff}
\end{equation}
This result has been rederived more recently by a number of authors
\cite{Don94,HamLiu,MuzVok}. 
It was also shown that corrections of this form 
are the leading quantum corrections for massless quantum fields. 
The constant $a$ depends on the number and spin of massless fields.
The first derivation based in the SEE has been presented in Ref.
\cite{DalMaz94}. Recently there is a renewed interest in this 
type of corrections since it is relevant to relating two
different developments in quantum gravity: Maldacena's Ads/CFT
correspondence and the Randall- Sundrum alternative to 
compactification, see for example \cite{Duff00}.

There is  an alternative,
intuitive ``Wilsonian" way of taking into account, at least partially, the 
quantum effects: just modify the classical potential by replacing the Newton 
constant by its running counterpart
\begin{equation}
V(r)=-{G(\mu = {1\over r})M\over r}
\label{newtonwilson}
\end{equation}
where $G(\mu)$ is the solution to the renormalization group equations in the 
theory considered.
As can be seen from Eqs. (\ref{rge1}) and (\ref{veff}) this argument 
does not reproduce
the leading quantum correction for massless fields.

We will now show that the result (\ref{veff}) can be derived from the ELEs, 
and are therefore a 
consequence of the RG equations for the parameters $\alpha$ and
$\beta$. For simplicity, and to make contact with previous works,
we will solve the ELEs without including the noise source. Solutions
to the ELEs including this source can be found in \cite{martinver}.

In the static, weak field approximation 
$g_{\mu\nu} = \eta_{\mu\nu} + h_{\mu\nu}$, we have
\begin{eqnarray}
R_{\mu\nu} &=& - \frac{1}{2} \Box h_{\mu\nu} \\
R &=& - \frac{1}{2} \Box h \\
H_{\mu\nu}^{(1)} &=& (-2 \partial_{\mu} \partial_{\nu} + 2 \eta_{\mu\nu}
\Box) \Box h \\
H_{\mu\nu}^{(2)} &=& (- \partial_{\mu} \partial_{\nu} + \frac{1}{2} 
\eta_{\mu\nu} \Box) \Box h + \Box \Box h_{\mu\nu} , 
\end{eqnarray}
where $h=\eta^{\mu\nu} h_{\mu\nu}$ and we assumed the Lorentz gauge
condition $(h_{\mu\nu} - \frac{1}{2} h \eta_{\mu\nu})^{; \nu}=0$.

Including a point particle source with $T_{\m\n}=\delta_\m^0
\delta_\n^0\delta^3(\bf x)$, the linearized field equations become,
\be
\left[
- \frac{1}{16 \pi G} + \frac{m^2}{32 \pi^2} (\xi-\frac{1}{6}) 
\ln \frac{m^2}{\mu^2} \right] \Box {\bar{h}_{\mu\nu}} 
- \alpha H_{\mu\nu}^{(1)} -
\beta H_{\mu\nu}^{(2)} = T_{\mu\nu} + \langle T_{\mu\nu} \rangle ,
\label{htecho}
\te
with ${\bar h}_{\mu\nu}=h_{\mu\nu} - \frac{1}{2} h \eta_{\mu\nu}$.
We define the quantum corrected Newtonian potential as 
$V(r)=-\ha h_{00}$.

The trace of the field
equations is
\be
\left[
 \frac{1}{16 \pi G} -  \frac{m^2}{32 \pi^2} (\xi-\frac{1}{6}) 
\ln \frac{m^2}{\mu^2} \right] \nabla^2 h - 2(3\alpha+\beta) \nabla^2 \nabla^2
h = T_{\mu}^{\mu} + \langle T_{\mu}^{\mu} \rangle ,
\label{eqtraza}
\te
where, to first order in ${m^2\over\nabla^2}$,
\be
\langle T_{\mu}^{\mu} \rangle = -\frac{1}{32 \pi^2}
\left[
3 (\xi -\frac{1}{6})^2 \ln(-\frac{\nabla^2}{\mu^2}) \nabla^2 \nabla^2 
- 6 m^2 (\xi^2 - \frac{1}{36}) \ln(-\frac{\nabla^2}{m^2}) \nabla^2
\right] h .
\label{traza}
\te

We shall solve Eq. (\ref{eqtraza}) perturbatively
$h=h^{(0)} + h^{(1)}$. The classical contribution $h^0$
satisfies
\begin{eqnarray}
( \nabla^{2} - \sigma^{-2} \nabla^{2} \nabla^{2} ) h^{(0)} = 
- 16 \pi G M \delta^{3}({\bf{x}}) & ~~~~ & 
\sigma^{-2}=32 \pi G (3\alpha+\beta) .
\label{eqhzero}
\end{eqnarray}
The time independent and spherically symmetric solution is
\be
h^{(0)} =  \frac{4 G M}{r} (1 - e^{- \sigma r}) 
.
\label{hzero}
\te
The equation for the first quantum correction is
\be
( \nabla^{2} - \sigma^{-2} \nabla^{2} \nabla^{2} ) h^{(1)} = 
{\cal D}(\nabla^2) h^{(0)} ,
\label{h1}
\te
where
\be
{\cal D}(\nabla^2) = 
-\frac{3 G}{2 \pi} (\xi-\frac{1}{6})^2 \ln(-\frac{\nabla^2}{\mu^2})
\nabla^2 \nabla^2 + \frac{G m^2}{\pi} 
\left[
\frac{1}{2} (\xi -\frac{1}{6})  \ln(\frac{m^2}{\mu^2}) + 
3 (\xi^2 -\frac{1}{36})  \ln(-\frac{\nabla^2}{m^2}) \nabla^2
\right] .
\te

To find a solution to this equation, we will consider
the limit $\sigma r\rightarrow\infty$  
(we are interested in long distance quantum corrections). In this limit
$h^{(0)} = 4 G M \left(\frac{1}{r} + 4 \pi \sigma^{-2} 
\delta^{3}({\bf{x}}) \right)$ and the nonlocal operator
in the rhs of Eq.(\ref{h1}) can be easily evaluated. After
a long calculation we
obtain
\be
h^{(1)} = - \frac{24 G^2 M m^2}{\pi} (\xi^2-\frac{1}{36}) 
 \frac{ \ln{ \frac{r}{r_{0}} } }{r}  -
\frac{12 G^2 M}{\pi} (\xi-\frac{1}{6})^2 \frac{1}{r^3} + \ldots .
\te
The dots denote corrections
at the origin, which are proportional
to $\delta^3({\bf{x}})$ and its derivatives.
We have not included them
because our quantum corrections are not accurate near the origin. Indeed,
we have derived the modified Einstein equations under the assumptions
$\nabla \nabla {\cal R} \gg {\cal R}^2$ and $m^2 {\cal R} \ll
\nabla \nabla R$. Both conditions
are satisfied for the $\frac{G M}{r}$ potential if $G M \ll r \ll m^{-1}$,
so the origin is excluded.

A similar analysis can be carried out for $\bar h_{00}$. We omit the 
details \cite{DalvitPhD}.
The final answer for the Newtonian potential is
\be
V(r) = -\frac{1}{2} h_{00} =
- \frac{G M}{r} 
\left\{
1 + \frac{2G}{\pi} \left[\frac{1}{2} (\xi-\frac{1}{6})^2 +
\frac{1}{90} \right]
\frac{1}{r^2} + \frac{2 m^2 G}{\pi} (\xi^2 + \frac{1}{12})
\ln \frac{r}{r_0}
\right\} .
\label{VCONF1}
\te

From Eq. (\ref{VCONF1}) 
we see that there are two different 
terms in the quantum correction. The term containing the logarithm
is
qualitatively what we expected from `Wilsonian' arguments (see Eq.
\ref{newtonwilson}).
However, the
coefficient is not exactly
the same as the one derived from the renormalization group equation
(\ref{rge1}), unless $\xi=0$.
Besides the running of $G$, we have obtained
the {\it leading} quantum $r^{-3}$ correction, which is a consequence of the 
running of $\alpha$ and $\beta$ in configuration space.

\section{Renormalization Group Theory for Nonequilibrium Systems}

\subsection{Summary Remarks}

We first give a schematic summary of this report and then discuss some
general issues concerning RG theory for NEq systems.

The CGEA contains all the information
about the influence of the  environment on the system. The CTP
version of the CGEA is suitable 
for the analysis of the dynamical evolution of the system under the
influence of the environment, and takes into account effects of 
renormalization, dissipation and noise.

In this review we have described several applications of the 
{\it in-out} and {\it in-in} effective action in semiclassical
gravity and cosmology, paying particular
attention to the relation with the RG equations. The {\it Euclidean}
averaged effective action is reviewed by Wetterich and his collaborators 
in an article in this conference.

We have described a perturbative calculation of the 
CTP CGEA in RW spacetimes.
We have shown that the CTP CGEA is useful to derive 
the master and Langevin equations in quantum field theory. 
The imaginary part
of the CTP CGEA produces diffusive terms in the master equation.
These terms can induce the reduced density matrix to become
diagonal as it evolves, and are therefore crucial for decoherence and quantum 
to classical transition of the system. We applied this formalism
to investigate the decoherence of mean fields due to quantum fluctuations relevant 
for theories of structure formation in inflationary models.

We adapted the Wegner and Houghton {\it Euclidean} approach
to the {\it in-in} CGEA in order to obtain an exact renormalization group 
equation for the dependence of $S_\L$ on the coarse graining scale.
The exact equation is extremely complicated.
We have solved it using a  derivative expansion.
In this approximation, the CTP CGEA ``decouples'' [i.e. it is
of the form $S_{\Lambda}(\phi_+,\phi_-) 
= S_{\Lambda}(\phi_+) - 
S_{\Lambda}(\phi_-)$ ] and contains 
neither dissipative nor noise terms. Previous results on the RG 
improved effective potential are recovered under this approximation.

Efforts to find solutions  beyond the adiabatic approximation 
are under  investigation. We believe that this
equation (or a simplified version of it) 
will play an essential role in the development of a 
renormalization group
theory for nonequilibrium systems.
We expect that, as soon as we decrease the scale from $\Lambda_{0}$,
dissipative and noise terms will grow: the CGEA
will develop an imaginary part (related to noise)
and a real part containing interactions between
the $\phi_{\pm}$ fields (dissipation).
This can be easily checked both in the one loop approximation and from the
exact RGE. Indeed, 
we have seen that the one loop CTP CGEA
is in 
general non-real. On the
other hand, the real and imaginary parts of the CGEA are not decoupled
in Eq. (\ref{exact}), and a non-vanishing real part at $\Lambda=\Lambda_0$
will induce an imaginary part at lower scales. 

One should be
able to  find a non-perturbative, $\L$-dependent fluctuation-dissipation
relation and RG equations for the coupling constants of the theory
that include the noise effects.

Finally, in the context of semiclassical gravity we described an interesting
relation between the usual RG equations (running coupling constants) and the 
CTP CGEA. We have seen that one can take into account backreaction
effects of quantum fields on the spacetime metric using a ``Wilsonian''
effective action in which the parameters of the theory are 
replaced by their running counterparts  
(this form of the effective action suggests itself by demanding
the theory to be independent of the scale $\mu$ introduced 
by dimensional regularization).
As the running is local in momentum
space, it becomes nonlocal in configuration space, and thus one 
obtains an Euclidean,  nonlocal effective action for the spacetime metric
previously obtained by using different approaches. This Euclidean
action can be transformed into the CTP CGEA taking into account
formal analogies between
their definitions. The final result, which contains dissipation and 
noise effects, can be viewed as a consequence of the running couplings
of the theory.


\subsection{Towards a Nonequilibrium Renormalization Group Theory}

There are three important aspects in the construction of  a renormalization group theory for 
nonequilibrium  systems: a) The idea behind the introduction
of the renormalization group (RG) description of critical phenomena characterized by
the running of the interaction parameters of the theory;  b) The scale characteristic
of the dynamical interaction absent in static critical phenomena;
c) How  noise and dissipation in the stochastic equations describing
the effective dynamics are reflected in  the renormalization group equations
governing the parameters of the theory. 
Point a) is discussed in all theories of static critical phenomena. We approached this
problem stressing the coarse graining and backreaction aspects. Point b) is discussed
in dynamical critical phenomena (see e.g., \cite{ZinnJustin}). We note, however, in 
the literature the noise in the dynamical Ginzburg-Landau equation is usually introduced by hand, and what is at issue is how the system behaves towards the critical 
point --  neither is the origin of noise accounted for nor is
the dissipative aspect of the system dynamics being incorporated  in the flow.
The first part of  Point c), i.e., how dissipation and noise appear in 
an  open system is discussed in this report. 
(For a discussion of how noise is identified in interacting quantum fields
in both a prescribed open system and an effectively open system, see,
\cite{Banff,stobol}.)
The crucial remaining issues are how they affect the RG flow and how they
manifest in the approach to critical points. We will make some general
comments here, as research on nonequilibrium (NEq) RG is still in its 
infant stage, its application to self-organized criticality, driven-diffusive systems 
and turbulence notwithstanding.

It is perhaps helpful to reflect upon the basic ideas and procedures
behind the introduction of the RG -- i.e., coarse graining and scaling.
When a system has a certain degree of regularity (homogeneity,
periodicity)  depicted by some symmetry, one can choose to
represent it with an equivalent coarse-grained description.
Examples are decimation in real time RG applied to 
a lattice, where, e.g., in the Ising model  every other spin is
eliminated and the effective bond strength between remaining
spins is doubled. This Kadanoff-Migdal procedure
produces a new Hamiltonian, which, upon iteration, can 
transform an original system into
a simpler one. In the momentum space description, this procedure
transcribes (runs) the ultraviolet (short range) behavior of the system 
to its infrared (long-ranged) domain, and hence is useful for
the description of critical phenomena, as the behavior of
the system near the critical point is dominated by the appearance
of long range order. Whether this procedure, which is one of
many possible coarse-graining schemes, can faithfully depict or
capture the essential physics depends on the compatibility of  the  procedure
with  the system and on the properties of the system near the critical point. 
If the coarse-graining respects the symmetry of the system,
(e.g., for  an anisotropic medium, use a different decimation grading
in different directions commensurate with the symmetry of the medium) 
and if the system possesses some scaling properties near  the critical point,
then the transformed problem via the RG would preserve the same
critical behavior as the original problem. Otherwise it fails.
The  applicability of the RG idea thus depends crucially on the choice
of a coarse-graining procedure and the use of scaling concepts.
Leaving scaling aside for now, which has more to do with the properties of
the particular systems of interest than with the procedure for
accessing relevant information about the system,  our approach
to nonequilibrium processes and NEqRG theory starts 
from examining closely the coarse-graining procedure. 

\subsubsection{RG procedures in the light of open system concepts}

As we stated clearly at the beginning of this report, the
necessary  steps to capture the essense of
a physical system with a simplified depiction lies in :
1) distinguish the system from the environment; 
2) coarse grain the environment and 
3) measure how the coarse-grained environment influences
the system in  providing  an effective kinematics or dynamics of
the reduced system.  Let us examine the RG procedure in the light
of the open systems scheme of nonequilibrium statistical mechanics. 
The first step consists of a) separating the order parameter field $\Phi (x,t)$ into 
two parts, $ \Phi = {\Phi_S} + { \Phi_E}$, where 
${\Phi_S }$ and ${\Phi_E }$ are respectively our  system and environment.
We assume that ${\Phi_S }$ contains the lower $k$ wave modes and
${\Phi_E }$ the higher $k$ modes. In critical phenomena, 
$ \Phi_S: \mid \vec{k} \mid < \Lambda /s,~~~~~~
\Phi_E: \Lambda/s < \mid \vec{k} \mid < \Lambda$.
Here $\Lambda $ is the ultraviolet cutoff and $ s > 1$ is the
coarse-graining parameter which gives the fraction of total $k$ modes counted 
in the environment. Then b)  one `integrates out' the short wavelength
sector, which  amounts to finding out the effect or backreaction of the coarse-grained 
environment (the short wavelength modes) on the system. This is most
succinctly and forcefully executed by the use of coarse-grained
effective action \cite{cgea}, as illustrated in the examples of this report.
The last step in the RG procedure is c) to
introduce a rescaling of the k-space and a rescaled field, 
e.g. $ \vec{k}' = s \vec{k}, \; \; \; \phi_S '(\vec{k}',t) = s^{- \frac{d+2}{2}} 
{\phi_S} (\vec{k},t)$ (d being the spatial dimension).
In identifying (or rather, demanding that they are equivalent, up to a certain order
in the perturbation expansion) the new action in terms of the rescaled
variables with the old action, one can obtain a set of differential
RG, or  the Wegner- Houghton equation.
At the critical point itself, this effective theory containing
only the low frequency modes is insensitive to further coarse graining, 
and therefore is described by a fixed point of the RG flow.

For dynamical systems, in addition to  the scaling transformation,
an artificial device of  the RG theory depicting the running of the 
coupling constants of the system towards the infrared region,
we find a real time-dependence depicting  the dynamics of the system.
\footnote{In treating dynamical systems, we may also coarse grain the system 
according to some time scale, obtaining the effective theory for slow 
modes after coarse graining the fast modes. Sometimes what is slow 
compared to what is fast may not be taken at face value, but has
to be determined relevant to the dynamics and symmetry of the system.
(Viewing eternal inflation as `static' and `slow- roll' as dynamic is an example
we showed.)}

Now that we have identified the steps of the RG procedure in
parallel to the treatment of open systems we can ask a few questions
to illuminate the relevant issues. It is easy to see where dissipation
and noise occur in the open system.  
But the appearance or nonappearance of noise and dissipation in the RG equations for 
nonequilibrium dynamics is not a simple issue.
That is why we deem it necessary to first provide a clear and thorough
discussion of the conceptual and technical basis for addressing the origin and nature of noise 
and dissipation in nonequilibrium field theory, as this report attempts to do, 
before we move on to investigate the related issues in RG theory for nonequilibrium processes. When the RG procedure is viewed 
in the light of an open system theory --- in effect the modes of interest in the
system after the Kadanoff transformation form an open system, 
and the techniques of nonequilibrium
field theory are the natural language to address these problems ---
the long wavelength / slow modes dynamics will display, in general, both
noise and dissipation.  
Indeed, the possibility  that a renormalization group flow may acquire
stochastic features for nonequilibrium processes 
is even clearer if we think of the RG as encoding the process of eliminating
irrelevant degrees of freedom from our description of a system \cite{Ma}.
These elimination processes lead as a rule to dissipation and noise, as 
was made
manifest in  the influence action and the CTP-CGEA approaches shown. 

But why, one may ask, is it  that we don't usually talk about 
dissipation in the system or  noise in the environment in a RG 
theory. A simple answer is that RG running is different from real-time
dynamics, and dissipation in the dynamics of the effective
system does not show up as dissipation runs. Another simple answer
is that the bulk of RG research has been focused on
equilibrium, stationary properties rather than the nonequilibrium
dynamics\cite{dcf}. \footnote{
As for whether it has to do with equilibrium versus nonequilibrium conditions, 
we know that even under equilibrium conditions  there is always  noise and dissipation,
as can be seen from linear reponse theory. They are related by the fluctuation-dissipation 
theorem. Indeed it is the precise balance of these processes which sustains the equilibrium. }
A major task in NEqRG is to make explicit this point,
 i.e., where does dissipation in the open system (Langevin) dynamics 
show up in the RG equations? We can offer only some speculative observations:

\subsubsection{Stochastic RG Equations}

A)  For apparently closed (yet effectively open)
systems described by the Boltzmann dynamics, this issue is more involved.
We learned that when the hierarchy of correlation functions are simply truncated
(with no slaving) \cite{stobol}, the equation of motion for the finite set of low-order correlation 
functions is unitary, such as is the case in Vlasov dynamics, which is obviously
non-equilibrium, yet no dissipation or noise appear as such.  A similar case is the n-loop
effective action in ordinary quantum field theory. If we view the separation between
the classical background field and the quantum fluctuation field as between a 
system and an environment, then the ordinary n-loop effective action is only one 
special case of the coarse-grained effective action, with quantum fluctuations being
integrated or coarse-grained away. This  results in  radiative corrections 
to the bare mass and charge as we learned from renormalization theory,
which,  in the nonequilibrium language, amounts to  backreaction of the environment 
on the system. We would expect to see dissipation in the effective dynamics of
the background field if we follow the general arguments of NEq field theory, but of course
one never talks about dissipation or noise in the equations of motion derived from an n-loop 
effective action. So what is missing? The reason is similar for both cases: the mean field or background
field is the lowest order of the hierarchy of correlation functions in the 
Schwinger-Dyson equations. Any loop expansion entails a truncation of the
hierarchy and,  just like Vlasov dynamics in classical mechanics, the equation of
motion from the effective action is unitary. So again it is not just the dynamics
which prompts the appearance of dissipation -- the causal factorization condition which 
we call slaving  (as in the molecular chaos assumption of Boltzmann) is responsible
for it.   

For example the circumstances whereby
the  ordinary RG equations are derived could be similar to the Vlasov or the loop 
approximation examples mentioned above, in that the correlation functions
are truncated from the hierarchy with a factorization condition  (rather than 
from slaving, where the correlation noise arises). This results in the coupling constants
being modified only partially, in a way similar to the renormalized mass or charge in 
one-loop (quantum) field theory  or the (classical) Vlasov dynamics following the 
averaged potential which replaces the  particle interactions.
In other words,  the  backreaction of the coarse-grained modes is not taken into account
fully. We see this, e.g. for the $\phi^4$ theory: the rescaled theory differs from 
the original theory in the  $\phi^6$ and higher order terms. By identifying
the two theories in such a way prescribed by the ordinary RG theory procedure,
we assumed that the error of ignoring the higher order corrections are inmaterial in 
our range of consideration. As such it is a good description of the theory 
only  for systems which scale
near the critical point, because there the higher order correlation functions can be ignored
or  incorporated in a way simply related to that of the lower order ones. 
Away from the critical point this is not true and this procedure would give bad results.
That is the criterion whereby one can assert the equivalence of the RG transformed theory 
and the original one. 

 Thinking  about the relation between the general nonequilibrium dynamics of an 
open system  and that depicted by the RG transformation in this light, we can see that the former 
(say, starting from  a nonlocal Langevin equation governing the system arising from 
the backreaction of  another subsystem or environment via a Zwangzig-Mori projection 
operator) keeps track of  much more information about the environment  
than that in the
RG treatment of the critical regime (as slaving being more than truncation).  
Although this is not sufficient for a proper description of the effective dynamics
as dissipation and fluctuations carry important information 
about the system and the environment, 
it is sufficient for the purpose of depicting the (static) critical phenomena of the system in the
critical region, as scaling behavior simplifies the relation between the correlation functions. 
That is why in actual RG calculations one can assume that the cutoff is in effect infinitely far 
above the scales of interest.  This advantage
near the critical point will still come to the aid of dynamical critical phenomena,
but when the full  dynamics goes beyond  an  approach to the equilibrium
(which is manifest in say, critical slowing down) or when one is interested in the  
behavior of the system
farther away from the critical point, then more of the dissipative and
noise attributes which are always present in general circumstances will have to be included
in a nonequilibrium RG theory. The art in the design of such a theory for NEq processes
is to find out what additional information (noise and dissipation)  need be kept, or what degree of coarse-graining is sufficient for the RG procedure which can still capture 
the essence of the critical behavior for  dynamical systems. 

B) One place where one can indeed see the `disappearance' of the effects 
of noise from the RG equations is when an adiabatic or quasilocal approximation is introduced
on the mode function dynamics. In \cite{DalMaz}  Dalvit and Mazzitelli show
from the Wegener - Houghton  equation  derived from  the CTP effective action that, 
for static long wavelength fields, the usual RG is recovered. In view of this result it is clear that an understanding of the running of noise and dissipation requires consideration
of dynamical long wavelength fields. However, if these fields are slowly
varying, they may be considered as actually linear or quadratic in the space
- time coordinates, in which case the functional integration of the high
momentum modes may be carried out exactly, at least to one loop order
\cite{HuOC84,qlea}. 

Alternatively, we may consider that the low momentum modes are following a
prescribed, though not necessarily slow, evolution, and compute the back
reaction of the high momentum modes keeping track of non adiabatic effects.
Such an analysis is carried out in \cite{eft}(See also  \cite{BoyDeV}), 
where the fast modes are
represented by a massive heavy field interacting with slow modes, described
as a light field. As expected, noise and dissipation display significant
running (they increase exponentially),  as the gap between the characteristic
frequencies of the heavy and light fields narrows. This result suggests that
a nonequilibrium RG should include, besides the usual interactions, the
running of suitable parameters describing both noise and dissipation.

Therefore the  challenge is to construct a RG flow where the dissipative 
and noisy
characteristics of the dynamics is ingrained therein.
There is a precedent for this problem in studies of RG flow in the Navier - Stokes 
equation \cite{RGhydro}, where one is
interested, for example, in the running of the viscosity. In this kind of
research, the starting point is generally a Langevin type equation, whose
solutions are given through functional representations following the methods of Zinn-Justin, e.g., \cite{ZinnJustin} (see also \cite{Carmen}).  
We would like to point out that even in this context,
closed time - path techniques offer an alternative way to obtain a
functional representation of the solutions of the Langevin equation \cite{DecHydro},
which is in several aspects simpler than the more familiar, forward time path, one.

C) Another lead can be found in  the stochastic correlation function ${\bf G}$
introduced in a quantum field theoretical derivation of the Boltzmann-Langevin equation
\cite{stobol}.
Its expectation value reproduces the usual propagators
(Green functions), while its fluctuations  account for the 
quantum fluctuations in the binary product of (operator) fields. 
The dynamical equation for ${\bf G} $ takes the form of an
explicitly stochastic Dyson equation. In the kinetic limit,
the fluctuations in ${\bf G}$ become the classical fluctuations in the
one particle distribution function, and the dynamical equation for ${\bf G}$'s
Wigner transform becomes the Boltzmann - Langevin equation. (Each of these
results has an interest of its own. A priori, there is no simple reason why the
fluctuations derived from quantum field theory should have a physical
meaning corresponding to a phenomenological entropy flux and Einstein's
relation.)
The studies of the fluctuating character of these field theoretic Green functions
also suggest new avenues in the
development of  RG theory. For example, we are used to fixing the
ambiguities of renormalization theory by demanding certain Green functions
to take on given values under certain conditions (conditions which should
resemble the physical situation of interest as much as possible, as
stressed  by O'Connor and Stephens \cite{OCSte}). If the Green functions
themselves are to be regarded as fluctuating, then the same ought to hold
for the renormalized coupling constants defined from them, and for the
renormalization group (RG) equations describing their scale dependence.
The notion that Green functions (and indeed, higher correlations as well)
may or even ought to be seen as possessing fluctuating characters (when
placed in the larger context of the whole hierarchy) with clearly
discernable physical meanings is likely to have an impact on the way we
perceive the statistical properties of field theory.

To end, we note that while the application of renormalization group methods to stochastic
equations is presented in well-known monographs\cite{ZinnJustin}, 
our proposal
here goes beyond these results in at least two ways. First, in our approach
the noise is not put in by hand or brought in from outside (e.g., the
environment of an open system), as in the usual Langevin equation approach,
but it follows from the (quantum) dynamics of the system itself. Actually,
the possibility of learning about the system from the noise properties
(whether it is white or coloured, additive or multiplicative, etc.) -- {\it %
unraveling} the noise, or treating noise {\it creatively}-- is a subtext in
our program. Second, our result suggests that stochasticity may, or should,
not only appear at the level of equations of motion, but also the level of
the RG equations, as they describe the running of `constants' which are
themselves fluctuating. This we feel is the first task in the construction of
a RG theory for NEq processes.\\

{\bf Acknowledgements} 
We wish to thank the organizers of the RG2000 meeting in Taxco (Mexico), 
January 1999 for their warm hospitality,  especially Denjoe O'Connor  and Chris Stephens, 
with whom we enjoyed many  close discussions over the years. 
We also enjoy the exchanges with  David Huse, Michael Fisher and 
Jean Zinn-Justin during the meeting on the role of noise in
nonequilibrium renormalization group theory.
EC, FDM are supported in part by CONICET, UBA,
Fundaci\'on Antorchas and Agencia Nacional de Promoci\'on
Cient\'\i fica y Tecnol\'ogica. 
BLH is supported in part by NSF grant PHY98-00967 and
their collaboration is supported in part by NSF grant INT95-09847.


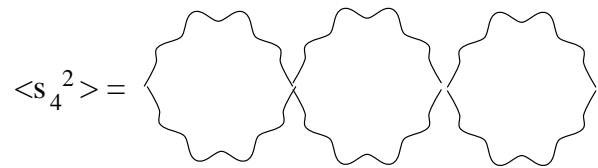
\end{document}